\documentclass[aps,prl,twocolumn,superscriptaddress]{revtex4-2}

\usepackage{graphicx}

\usepackage{amssymb}
\usepackage{amsmath}
\usepackage{verbatim}
\usepackage{tabularx}

\usepackage{algorithmicx}
\usepackage{algpseudocode}

\usepackage{mathtools}  
\usepackage{array}

\begin{document}

\title{Memory-Dominated Quantum Criticality as a Universal Route to High-Temperature Superconductivity}

\author{Byung Gyu Chae}

\affiliation{Electronics and Telecommunications Research Institute, 218 Gajeong-ro, Yuseong-gu, Daejeon 34129, Republic of Korea
\\ bgchae@etri.re.kr}

\begin{abstract}
Understanding the dynamical origin of high-temperature superconductivity
remains a central challenge in strongly correlated quantum matter.
Near quantum criticality, diverging correlation times reorganize the
infrared dynamics into a scale-free continuum of collective relaxation
processes.
We show that the infrared behavior of interacting electrons is
generically controlled by the relaxation-rate spectrum of the underlying
many-body dynamics.
Starting from a microscopic fermionic theory, we derive that the
Cooper-channel kernel admits a universal spectral representation in terms
of the time-scale density of states (TDOS) of collective decay modes,
without invoking a specific bosonic mediator.
The superconducting instability follows directly from the
vanishing of the quadratic kernel via a standard ladder resummation and
Thouless criterion, with the pairing interaction determined entirely by
the infrared structure of the relaxation spectrum.
A finite TDOS at vanishing relaxation rate produces a memory-dominated
regime characterized by long-time kernels $K(t)\sim 1/t$ and logarithmic
enhancement of the retarded pairing interaction, leading to a BCS-like
exponential transition scale set by infrared spectral weight.
More generally, infrared-singular spectra generate power-law response and
algebraic enhancement of the transition scale.
The same relaxation spectrum controls normal-state dynamics, giving rise
to long-time correlations, non-Markovian response, and strange-metal
behavior.
These results identify the spectral organization of relaxation modes as a
universal organizing principle of quantum critical matter and establish
memory-dominated criticality as a natural mechanism for enhanced pairing.
\end{abstract}

\maketitle

\section{I. Introduction}
High--temperature superconductivity remains one of the central open
problems in strongly correlated quantum matter
\cite{1,2,3,4,5,6,7}.
Despite decades of theoretical and experimental effort, no universal
consensus has emerged regarding the microscopic origin of robust pairing
far above conventional phonon scales.
A recurring feature among many candidate materials is the close proximity
of superconductivity to quantum critical regimes
\cite{8,9,10,11,12}.
Near a quantum critical point, the divergence of the correlation time,
\(
\tau_\xi \sim \xi^z \to \infty
\),
eliminates any intrinsic infrared time scale
\cite{13,14}.
As a consequence, temporal correlations become scale-free and exhibit
power-law behavior rather than simple exponential relaxation.
Such dynamics cannot generally be represented by a finite set of isolated
collective excitations, but instead suggests the emergence of a dense
continuum of relaxation processes extending toward arbitrarily long
time scales.

Within the conventional theoretical paradigm, quantum critical dynamics
is typically modeled in terms of overdamped bosonic collective modes
coupled to a fermionic continuum, producing an approximately universal
Ohmic Landau damping kernel of the form \( |\omega| \)
\cite{15,16}.
This Hertz--Millis description has proven highly successful in weakly
correlated systems.
However, its underlying assumptions become increasingly restrictive in
strongly correlated regimes, where the infrared dynamics may no longer
be reducible to a small number of collective modes subject to local
Markovian damping.
Indeed, a growing body of experimental observations in cuprate
superconductors and strange metals indicates anomalous temporal
correlations, non-Markovian response, and \(1/f\)-like noise
\cite{17,18,19,20}.
These observations point toward an infrared dynamical spectrum that
becomes increasingly dense and broadly distributed near criticality.

A complementary microscopic route to superconductivity begins from
interacting fermions and derives an effective collective theory of the
pairing order parameter
\cite{21,22}.
In particular, the Ginzburg--Landau functional may be obtained through
a Hubbard--Stratonovich decoupling of the Cooper channel followed by an
integration over fermionic degrees of freedom
\cite{23,24}.
Within this framework, the quadratic kernel of the order-parameter
theory is determined by the Cooper-channel susceptibility, which is
usually expanded analytically in momentum and frequency to obtain a
local low-energy effective action.

However, this construction implicitly assumes that the infrared dynamics
admits a regular expansion in terms of a finite set of well-defined
collective modes.
Such assumptions become increasingly questionable in strongly correlated
quantum critical regimes, where quasiparticle coherence is suppressed
and the infrared spectrum itself becomes dense and scale-free.

A related lesson from strongly interacting quantum fluids is that
macroscopic coherence and long-range order can emerge from collective
many-body organization beyond a simple weakly interacting particle
description \cite{25,26,27}.
Near quantum criticality, strong interactions and critical slowing down
invalidate descriptions based solely on coherent single-particle
propagation and instead promote collective infrared dynamics extending
over broad temporal and spatial scales.
This suggests that collective condensation is fundamentally an infrared
organizational phenomenon of many-body dynamics, governed by the
accumulation and dynamical synchronization of low-energy collective
degrees of freedom.
The central issue is therefore how such infrared collective modes
reorganize and accumulate to produce emergent macroscopic coherence and
long-range order.

At a fundamental level, the superconducting instability in conventional
BCS theory originates from the logarithmic infrared accumulation of
successive low-energy Cooper scattering processes near the Fermi surface.
In renormalization-group language, this corresponds to the marginal
infrared growth of the Cooper channel generated by the shell integral
\(
\int d\omega/\omega
\).
The essential mechanism is therefore not tied specifically to phonons
themselves, but to the infrared enhancement produced by the accumulation
of low-energy pairing shells.
In weakly correlated metals, this accumulation is organized by coherent
quasiparticle excitations and gives rise to the familiar logarithmic
Cooper kernel and exponentially suppressed transition scale.

The central question addressed in the present work is whether an
analogous infrared instability can emerge more generally in strongly
dissipative quantum critical electronic systems where sharply defined
quasiparticles cease to exist.
From a dynamical perspective, the breakdown of quasiparticle coherence
may be viewed as a reorganization of the infrared spectrum itself.
Coarse-grained collective degrees of freedom behave effectively as an
open subsystem \cite{28,29}, whose dynamics is characterized not by a
finite set of isolated collective modes, but by a continuous
distribution of relaxation rates.
The absence of an intrinsic time scale at criticality implies that this
distribution cannot remain discrete in the infrared, but instead
develops a dense continuum extending toward vanishing relaxation rates.

This motivates the introduction of the time--scale density of states
(TDOS) \cite{30,31,32}, defined as the density of collective relaxation
rates governing the infrared dynamics.
Importantly, the TDOS should not be viewed as an additional phenomenological
assumption, but as a natural consequence of scale-invariant critical
dynamics.
The infrared behavior is therefore governed not solely by excitation
energies, but by the spectral organization of slow collective relaxation
processes.

In this work, we show that infrared collective dynamics is generically
controlled by the TDOS of correlated decay channels.
Starting from a microscopic fermionic theory, we derive that the
Cooper-channel kernel admits an exact spectral representation in terms
of the relaxation-rate density, without assuming \emph{a priori} a
finite set of soft modes.
The superconducting instability then follows from a standard ladder
resummation and Thouless criterion.
Crucially, within this framework the Cooper channel is reorganized not
by quasiparticle energy shells alone, but by the infrared accumulation
of slow relaxation modes encoded in the relaxation spectrum itself.

We demonstrate that the infrared scaling of the relaxation-rate density,
$\rho(\lambda)\sim \lambda^\alpha$, defines universality classes of
dynamical critical behavior.
In particular, a flat TDOS ($\alpha=0$) produces a marginal nonanalytic
response with logarithmic frequency dependence and long-time memory,
$K(t)\sim 1/t$, while infrared-singular spectra ($\alpha<0$) generate
power-law response and algebraic enhancement of pairing.

This memory-dominated regime represents a qualitatively different
organization principle of superconducting infrared dynamics.
Rather than being governed primarily by quasiparticle coherence or by
the exchange of a specific bosonic mediator, pairing is controlled by
the collective accumulation of slow relaxation processes.
In this sense, the present framework may be viewed as a dynamical
generalization of the infrared Cooper instability itself, extended from
quasiparticle energy shells to a continuum of collective relaxation
modes.

Figure~1 summarizes the resulting dynamical universality map of correlated
quantum matter.
The infrared structure of the relaxation spectrum determines the form of
the dynamical kernel and controls physical consequences including pairing
enhancement, superconducting domes, Uemura scaling, and anomalous temporal
response \cite{1,33,34,35}.
Charge, spin, and pairing channels reorganize within a common slow-mode
reservoir, with superconductivity corresponding to condensation in the
particle--particle sector of a shared infrared spectral continuum.

Our results identify the relaxation-rate spectrum as a fundamental
organizing principle of quantum critical dynamics and establish
memory-dominated criticality as a natural dynamical environment for
enhanced pairing.
The remainder of this paper is organized as follows.
In Section~II we derive the exact TDOS representation of the collective
susceptibility and classify infrared universality classes.
In Sections~III and IV we analyze the resulting memory-dominated pairing
kernel and its implications for superconducting instabilities.
In Section~V we discuss experimental consequences, including superconducting
domes, Uemura scaling, and anomalous dynamical response.
In Section~VI we conclude with a discussion and outlook.

\begin{figure*}[t]
\centering
\includegraphics[width=0.85\textwidth, trim=0.5cm 14.8cm 0cm 0cm]{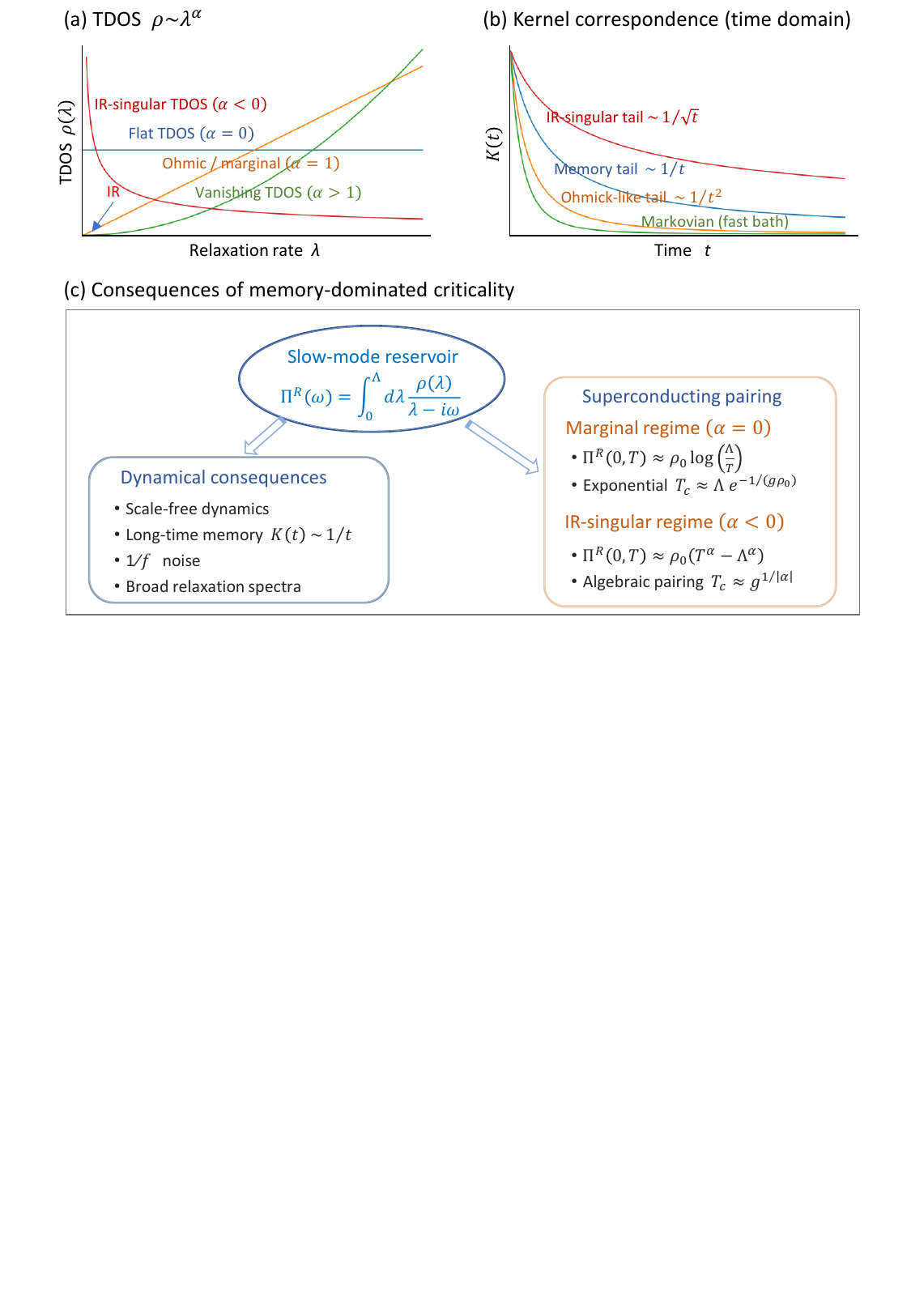}
\caption{Dynamical universality map of correlated quantum matter.
Infrared behavior is organized by the spectrum of relaxation rates
encoded in the time--scale density of states.
(a) Infrared TDOS $\rho(\lambda)\sim \lambda^\alpha$ defining distinct
dynamical universality classes: sparse spectra ($\alpha>1$),
Ohmic damping ($\alpha=1$), marginal flat TDOS ($\alpha=0$),
and infrared-singular spectra ($\alpha<0$) with an accumulation of slow modes.
(b) Corresponding time-domain memory kernels $K(t)$ obtained via Laplace transform:
rapid decay for sparse spectra, $K(t)\sim 1/t^2$ (Ohmic),
$K(t)\sim 1/t$ (flat TDOS), and slower scale-free decay for $\alpha<0$.
(c) Physical consequences of TDOS-controlled dynamics.
A finite density of slow modes produces non-Markovian response and
infrared enhancement of pairing:
logarithmic for $\alpha=0$ (BCS-like $T_c$) and algebraic for $\alpha<0$.
}
\label{fig:fig1}
\end{figure*}

\section{II. Emergence of a Slow-Mode Reservoir from Correlated Dynamical Spectra}

In this section we establish the central organizing principle of the
infrared dynamics.
We focus on strongly correlated electronic systems in the vicinity of
quantum critical regimes, where conventional quasiparticle
descriptions break down.
In such regimes, the long-time dynamics is not governed by
a finite set of quasiparticle excitations, but instead by a continuous
spectrum of relaxation processes.

We first demonstrate that scale-free critical dynamics necessarily
implies the emergence of a continuous distribution of relaxation rates.
We then derive microscopically, starting from an interacting fermionic
theory, that the Cooper-channel kernel acquires a relaxational spectral
representation.
Finally, we classify the resulting infrared behavior in terms of
universality classes determined by the low-energy scaling of the
relaxation-rate spectrum.

\subsection{A. Critical scaling and emergence of a continuous relaxation spectrum}

The infrared structure of strongly correlated quantum matter is governed
by the general properties of its critical dynamics \cite{13,14}.
Near a continuous phase transition, the correlation length diverges as
\begin{equation}
\xi \sim |g-g_c|^{-\nu},
\end{equation}
and dynamical scaling relates the corresponding correlation time to
the length scale through
\begin{equation}
\tau_\xi \sim \xi^z \sim |g-g_c|^{-z\nu}.
\end{equation}
At criticality, this implies
\begin{equation}
\tau_\xi \to \infty,
\end{equation}
which expresses the phenomenon of critical slowing down.

The divergence of the correlation time eliminates any intrinsic time
scale in the infrared theory.
As a result, temporal correlation functions cannot generically decay
as a single exponential at long times.
Instead, scale invariance requires a power-law form,
\begin{equation}
E(t) \equiv \langle \phi(t)\phi(0)\rangle \sim t^{-p},
\qquad t \to \infty,
\end{equation}
with an exponent determined by the generic class.

A single relaxation rate produces an exponential decay,
\begin{equation}
E(t) \sim e^{-\lambda t},
\end{equation}
and any finite superposition of such modes,
\begin{equation}
E(t) = \sum_i a_i e^{-\lambda_i t},
\end{equation}
remains dominated at long times by the smallest $\lambda_i$ and therefore
also decays exponentially.
Such forms are incompatible with the scale-free behavior above.

It follows that a power-law temporal decay cannot arise from a finite
set of dynamical modes.
Instead, it requires a continuous superposition of decay channels,
\begin{equation}
E(t)
=
\int_0^\infty d\lambda\;
\rho(\lambda)\, e^{-\lambda t},
\label{eq:laplace_rep_main}
\end{equation}
where $\rho(\lambda)$ is the density of relaxation rates.
Equation~(\ref{eq:laplace_rep_main}) should be understood as a general
spectral representation of dissipative dynamics.
Importantly, it is not introduced as a phenomenological assumption,
but follows directly from the requirement of scale-free infrared
behavior.

The long-time behavior of $E(t)$ is controlled by the small-$\lambda$
region of this spectrum.
Using the Laplace transform identity
\begin{equation}
\int_0^\infty d\lambda\;
\lambda^{\alpha} e^{-\lambda t}
=
\Gamma(1+\alpha)\, t^{-1-\alpha},
\end{equation}
one finds that a power-law decay $E(t)\sim t^{-p}$ implies
\begin{equation}
\rho(\lambda)
\sim
\lambda^{p-1},
\qquad \lambda \to 0.
\end{equation}
Thus, the infrared scaling of the TDOS directly
encodes the dynamical universality class.

At a classical critical point, such scale-free behavior is realized only
at a fine-tuned temperature.
By contrast, in quantum critical systems, the divergence of the correlation
time at zero temperature extends to finite temperatures, giving rise to an
extended scaling regime.
This regime, often referred to as the quantum critical fan, is characterized
by the condition
\begin{equation}
T \gg |g-g_c|^{z\nu},
\end{equation}
under which the intrinsic correlation time $\tau_\xi$ exceeds the thermal
time scale $1/T$.
As a result, the infrared dynamics is governed not by proximity to a specific
phase, but by the scale-invariant critical theory itself.

Within the TDOS framework, this finite-temperature scaling regime admits a
simple interpretation.
Temperature acts as an effective infrared cutoff of the relaxation spectrum,
so that only modes with relaxation rates $\lambda \gtrsim T$ contribute
to the long-time dynamics.
Consequently, the quantum critical fan corresponds to a regime in which
the continuous relaxation spectrum is truncated in the infrared at a scale
set by temperature.

This perspective implies a qualitative shift relative to conventional
approaches.
Rather than describing infrared dynamics in terms of a small number of
collective modes, the relevant degrees of freedom are organized by a
continuous distribution of relaxation rates.

In weakly interacting systems, the density of slow modes typically
vanishes as $\lambda \to 0$, reflecting the presence of well-defined
quasiparticles and isolated collective excitations.
By contrast, in strongly correlated critical regimes, the relaxation
spectrum can acquire substantial infrared weight.
A particularly important case for the present work is the marginal
situation in which
\begin{equation}
\rho(\lambda \to 0) = \rho_0 \neq 0,
\end{equation}
corresponding to a finite density of arbitrarily slow modes.
This accumulation of slow relaxation processes provides the dynamical
foundation for the memory-dominated behavior and infrared enhancement
mechanisms developed in the following sections.

\vspace{10pt}
\noindent

This motivates the central problem addressed in this work:
to derive the relaxation-spectrum representation of collective
response functions directly from a microscopic fermionic theory,
and to establish how scale-free infrared dynamics gives rise to a
continuous spectrum of relaxation modes.

In the following subsection, we show that such a representation
emerges naturally when the fermionic propagators themselves acquire a
dissipative form.
Starting from a standard interacting fermionic action and performing a
Hubbard--Stratonovich transformation \cite{36,37}, we demonstrate that the Cooper
channel polarization kernel can be expressed in terms of a continuum
of relaxational modes.
This provides a microscopic foundation for the relaxation-spectrum
framework and connects it directly to the infrared structure of
dissipative fermionic systems.

\subsection{B. Microscopic origin of the relaxational Cooper kernel}
\label{subsec:microscopic_relaxational_kernel}

The relaxational spectral representation introduced above should not
be viewed as an ad hoc phenomenological assumption.
Rather, it emerges microscopically from the Cooper-channel projection
of dissipative fermionic dynamics once the infrared regime is dominated
by finite damping rather than by sharply propagating quasiparticles.

We begin from a generic interacting fermionic action in imaginary time,
\begin{equation}
\begin{aligned}
&S[\bar c,c]
=
\int_0^\beta d\tau
\sum_{k,\sigma}
\bar c_{k\sigma}(\tau)
\bigl(
\partial_\tau+\xi_k
\bigr)
c_{k\sigma}(\tau)
\\
&\qquad
-
g
\int_0^\beta d\tau
\sum_{k,k',q}
\bar c_{k+q,\uparrow}(\tau)
\bar c_{-k,\downarrow}(\tau)
c_{-k',\downarrow}(\tau)
c_{k'+q,\uparrow}(\tau),
\end{aligned}
\label{eq:micro_action_pairing_revised}
\end{equation}
where \(g\) denotes the effective interaction strength in the
particle--particle (Cooper) channel.
Microscopically, such a coupling may arise from short-range electronic
correlations, but its specific origin is not essential for the present
infrared analysis.

Introducing the Cooper pair bilinear,
$B_q(\tau)
\equiv
\sum_k
c_{-k,\downarrow}(\tau)\,
c_{k+q,\uparrow}(\tau)$,
the quartic interaction is rewritten as
\(
-g\,B_q^\dagger B_q
\).
Using the Gaussian Hubbard--Stratonovich identity,
\begin{equation}
\begin{aligned}
&
\exp
\Biggl[
g
\int_0^\beta d\tau\;
B_q^\dagger(\tau)B_q(\tau)
\Biggr]
\propto
\int
\mathcal D\Delta^\ast
\mathcal D\Delta\;
\\
&\times \exp
\Biggl[
-
\int_0^\beta d\tau
\sum_q
\left(
\frac{|\Delta_q(\tau)|^2}{g}
-
\Delta_q^\ast(\tau)\,
B_q(\tau) - \rm{h.c.}
\right)
\Biggr].
\end{aligned}
\label{eq:HS_identity_pairing_revised}
\end{equation}
The bosonic field \(\Delta_q(\tau)\) is initially an auxiliary
collective field introduced for exact decoupling of the four-fermion
interaction.
Its physical interpretation as the superconducting order parameter
emerges only after the fermions are integrated out and the saddle-point
condition is imposed.
At that stage it satisfies
\begin{equation}
\Delta_q(\tau)
=
g
\sum_k
\left\langle
c_{k+q,\uparrow}(\tau)\,
c_{-k,\downarrow}(\tau)
\right\rangle,
\end{equation}
which is the usual gap equation in the Cooper channel.

The partition function becomes
\begin{equation}
Z
=
\int
\mathcal D\bar c\,
\mathcal Dc\,
\mathcal D\Delta^\ast\,
\mathcal D\Delta\;
e^{-S[\bar c,c,\Delta]},
\end{equation}
with
\begin{equation}
\begin{aligned}
S[\bar c,c,\Delta]
=
&
\int_0^\beta d\tau
\sum_q
\frac{|\Delta_q|^2}{g}
\\
&
+
\int_0^\beta d\tau
\sum_k
\Psi_k^\dagger(\tau)
\,
\hat{\mathcal G}^{-1}[\Delta]
\,
\Psi_k(\tau),
\end{aligned}
\end{equation}
where
\(
\Psi_k=(c_{k\uparrow},\,\bar c_{-k\downarrow})^T
\)
is the Nambu spinor and
\begin{equation}
\hat{\mathcal G}^{-1}[\Delta]
=
\begin{pmatrix}
\partial_\tau+\xi_k & \Delta
\\
\Delta^\ast & \partial_\tau-\xi_k
\end{pmatrix}.
\end{equation}
Integrating out the fermions yields the exact effective action
\begin{equation}
S_{\rm eff}[\Delta]
=
\int_0^\beta d\tau
\sum_q
\frac{|\Delta_q|^2}{g}
-
\mathrm{Tr}
\ln
\hat{\mathcal G}^{-1}[\Delta].
\label{eq:Seff_logdet_revised}
\end{equation}

Expanding around the normal state \(\Delta=0\) to quadratic order gives
\begin{equation}
S_2[\Delta]
=
\sum_{q,\Omega_n}
\Delta^\ast(q,\Omega_n)
\,
\Gamma_2(q,\Omega_n)
\,
\Delta(q,\Omega_n),
\end{equation}
with
\begin{equation}
\Gamma_2(q,\Omega_n)
=
\frac{1}{g}
-
\Pi(q,\Omega_n),
\label{eq:Gamma2_general_revised}
\end{equation}
where the microscopic Cooper bubble is
\begin{equation}
\Pi(q,\Omega_n)
=
T
\sum_{\omega_n}
\int
\frac{d^dk}{(2\pi)^d}
\;
G(k+q,i\omega_n+i\Omega_n)\,
G(-k,-i\omega_n).
\label{eq:Cooper_bubble_general_revised}
\end{equation}

Up to this point, the construction follows the standard
Hubbard--Stratonovich formulation.
The essential difference in the present theory lies in the infrared
structure of the fermionic propagator entering the Cooper kernel.

In strongly interacting metallic systems, scattering processes broaden
the single-particle pole and generate fermionic excitations with finite
lifetimes.
Instead of sharply propagating quasiparticles, the infrared fermionic
response is characterized by a distribution of relaxation scales.
The Cooper bubble is therefore constructed with respect to the dressed
normal-state propagator.
A convenient Matsubara representation is
\begin{equation}
G(k,i\omega_n)
=
\frac{1}{
i\omega_n
-
\xi_k
+
i\gamma_k\,\mathrm{sgn}(\omega_n)
},
\label{eq:damped_matsubara_green}
\end{equation}
where \(\gamma_k>0\) denotes the fermionic relaxation rate.

This form preserves the quasiparticle pole structure while including a
finite lifetime.
Importantly, both the dispersive part \(\xi_k\) and the damping scale
\(\gamma_k\) remain present.
The key point is that, after projection to the Cooper channel, the
infrared pair response reorganizes into an effective relaxational
kernel governed by the pair relaxation rate.

To make this structure explicit, it is convenient to insert the
spectral representation of the damped fermionic propagator directly
into the microscopic Matsubara Cooper bubble:
\begin{equation}
\begin{aligned}
\Pi(q,\Omega_n)
=
&T\sum_{\omega_n}
\int
\frac{d^dk}{(2\pi)^d}
\\
&\times \int
d\varepsilon\,
d\varepsilon'
\;
\frac{A_{k+q}(\varepsilon)}
{i\omega_n+i\Omega_n-\varepsilon}
\frac{A_{-k}(\varepsilon')}
{-i\omega_n-\varepsilon'},
\end{aligned}
\label{eq:cooper_bubble_spectral_matsubara}
\end{equation}
where \begin{equation}
A_k(\varepsilon)
=
-\frac{1}{\pi}
\mathrm{Im}\,G_k^R(\varepsilon)
=
\frac{1}{\pi}
\frac{\gamma_k}{
(\varepsilon-\xi_k)^2+\gamma_k^2}
\label{eq:lorentzian_spectral_function}
\end{equation}
is the Lorentzian spectral function of the broadened fermion.

Performing the Matsubara sum and then analytically continuing
\(i\Omega_n\to \Omega+i0^+\),
the retarded Cooper kernel becomes
\begin{equation}
\begin{aligned}
\Pi^R(q,\Omega)
=
&\int
\frac{d^dk}{(2\pi)^d}
\int
d\varepsilon\,
d\varepsilon'
\;
\\
&
\times
A_{k+q}(\varepsilon)\,
A_{-k}(\varepsilon')
\,
\frac{
1-n_F(\varepsilon)-n_F(\varepsilon')
}{
\Omega+i0^+-\varepsilon-\varepsilon'
}.
\end{aligned}
\label{eq:real_axis_cooper_kernel}
\end{equation}

The central microscopic result is that the energy convolution
determining the retarded denominator can be evaluated exactly.
Defining
\begin{equation}
\xi_1\equiv\xi_{k+q},
\quad
\gamma_1\equiv\gamma_{k+q},
\quad
\xi_2\equiv\xi_{-k},
\quad
\gamma_2\equiv\gamma_{-k},
\end{equation}
one finds
\begin{equation}
\begin{aligned}
\int
d\varepsilon\,
d\varepsilon'
\;
A_1(\varepsilon)\,
A_2(\varepsilon')
\,
&
\frac{1}{
\Omega+i0^+-\varepsilon-\varepsilon'
}
\\
&=
\frac{1}{
\Omega-\xi_1-\xi_2+i(\gamma_1+\gamma_2)
}.
\end{aligned}
\label{eq:exact_lorentzian_convolution}
\end{equation}

This result follows from the Cauchy-transform identity
\begin{equation}
\int_{-\infty}^{\infty}
d\varepsilon\;
\frac{1}{\pi}
\frac{\gamma}{
(\varepsilon-\xi)^2+\gamma^2
}
\,
\frac{1}{w-\varepsilon}
=
\frac{1}{w-\xi+i\gamma},
\quad
\mathrm{Im}\,w>0,
\label{eq:cauchy_transform_identity_main}
\end{equation}
applied successively to the \(\varepsilon\) and \(\varepsilon'\)
integrals.
Because Lorentzian (Cauchy) distributions are closed under convolution,
the dispersive parts add and the damping rates add:
the pair mode inherits the center \(\xi_1+\xi_2\) and the total width
\(\gamma_1+\gamma_2\).

For a time-reversed Cooper pair in the long-wavelength limit,
\begin{equation}
\xi_{-k}=\xi_k,
\qquad
\gamma_{-k}=\gamma_k,
\label{eq:TR_pair_condition}
\end{equation}
so that Eq.~(\ref{eq:exact_lorentzian_convolution}) reduces to
\begin{equation}
\Pi_k^R(\Omega)
\propto
\frac{1}{
\Omega
-
2\xi_k
+
i\,2\gamma_k
}.
\label{eq:pair_kernel_exact_pair}
\end{equation}

Near quantum criticality, the divergence of the correlation time
eliminates any intrinsic infrared time scale and reorganizes the
collective dynamics into a broad continuum of relaxation processes.
As a result, the low-energy Cooper response is governed not primarily
by sharply propagating quasiparticles, but by overdamped infrared
modes characterized by finite relaxation rates.

The dominant contribution to the Cooper instability therefore arises
from the overdamped infrared shell satisfying
\begin{equation}
|\xi_k|
\lesssim
\gamma_k,
\label{eq:overdamped_shell_condition}
\end{equation}
where the residual dispersive energy becomes subleading compared with
the fermionic damping scale.
Within this regime, Eq.~(\ref{eq:pair_kernel_exact_pair}) reduces to
an effective relaxational kernel controlled by the pair relaxation
rate
\begin{equation}
\lambda_k
\equiv
2\gamma_k,
\end{equation}
so that
\begin{equation}
\Pi_k^R(\Omega)
\propto
\frac{1}{\lambda_k-i\Omega}.
\label{eq:pair_single_rate_frequency_revised}
\end{equation}

Critical slowing down then causes these relaxation scales to accumulate
toward the infrared, generating the dense continuum of slow collective
modes described by the TDOS.

The relaxational form therefore emerges only within the infrared
critical shell defined by
Eq.~(\ref{eq:overdamped_shell_condition}), where damping dominates
over residual dispersive propagation.
Higher-energy sectors away from the overdamped shell retain their full
microscopic dispersive structure and contribute primarily to the
overall normalization and effective interaction parameters.
The emergence of the relaxational denominator is therefore not an
assumption of purely dissipative dynamics, but the infrared consequence
of projecting broadened fermionic propagators onto the overdamped
Cooper sector near quantum criticality.

It is important to emphasize that the relaxational denominator
is fixed directly by the Lorentzian convolution itself.
The thermal factor
\(
1-n_F(\varepsilon)-n_F(\varepsilon')
\)
does not modify the analytic pole structure of the Cooper kernel,
but instead acts only as a smooth occupation weight multiplying the
relaxational response.

This situation is analogous to the standard finite-temperature BCS
kernel, where thermal occupation factors regularize the infrared
singularity through the smooth crossover function
\(
\tanh(\xi/2T)
\).
Likewise, in the present broadened critical regime, the occupation
dependence enters through the spectrally broadened convolution
\(
A_k(\varepsilon)n_F(\varepsilon)
\),
which smooths the infrared weight over an energy scale set by
\(
\max(\gamma_k,T)
\).

Accordingly, the finite-temperature Cooper kernel takes the form
\begin{equation}
\Pi^R(\Omega;T)
\simeq
\int
\frac{d^dk}{(2\pi)^d}
\;
\frac{
W_k(T)
}{
\lambda_k-i\Omega
},
\end{equation}
where \(W_k(T)\) is a smooth thermal occupation weight generated by
the spectral convolution.
In the overdamped infrared regime, this weight varies smoothly compared
to the broad continuum of relaxation rates, so the leading infrared
behavior is governed primarily by the accumulation of small
\(\lambda_k\).

At the scaling level, the Cooper kernel may therefore be represented as
\begin{equation}
\Pi^R(\Omega)
\sim
\int
\frac{d^dk}{(2\pi)^d}
\;
\frac{1}{
\lambda_k-i\Omega
}.
\label{eq:momentum_integrated_kernel}
\end{equation}
Introducing the time-scale density of states (TDOS),
\begin{equation}
\rho(\lambda)
=
\int
\frac{d^dk}{(2\pi)^d}
\;
\delta(\lambda-2\gamma_k),
\label{eq:TDOS_definition_weighted}
\end{equation}
the kernel reorganizes into the spectral representation
\begin{equation}
\Pi^R(\Omega)
=
\int_0^\Lambda
d\lambda
\;
\frac{
\rho(\lambda)
}{
\lambda-i\Omega
}.
\label{eq:TDOS_kernel_revised}
\end{equation}

Thus the infrared Cooper problem is organized not by sharply defined
quasiparticle energies, but by the spectrum of relaxation rates.
More generally, the relaxation rates \(\lambda\) may be interpreted as
the eigenvalue spectrum of the effective Liouvillian governing the
infrared dissipative dynamics.
In this sense, the TDOS represents the density of Liouvillian decay
modes projected onto the Cooper channel.
A formal operator-level derivation of this interpretation is presented
in Appendix~A.

The superconducting instability is therefore controlled by the
accumulation of slow dissipative pair modes, rather than by a narrow
energy shell around a sharply defined Fermi surface.

After analytic continuation, we use \(\omega\) interchangeably with the
real external frequency variable of the collective response.
Equation~(\ref{eq:pair_single_rate_frequency_revised}) shows that the
relaxational Cooper kernel is not a phenomenological assumption, but
emerges directly from the infrared convolution structure of
dissipative fermionic dynamics.

At the same time, the spatial dependence of the collective kernel is
not an independent phenomenological addition.
The relaxational denominator
\(
(2\gamma_k-i\omega)^{-1}
\)
is obtained in the static and uniform Cooper channel
(\(\mathbf q=0\)),
where the common dispersive shift of the time-reversed pair cancels
exactly inside the frequency convolution.
For finite center-of-mass momentum \(\mathbf q\neq0\), however, this
exact cancellation no longer holds, since the two fermionic lines carry
momenta
\(
\mathbf k+\mathbf q/2
\)
and
\(
-\mathbf k+\mathbf q/2
\).
Expanding the Cooper bubble in the long-wavelength limit then produces
the standard leading spatial gradient term \(cq^2\), while the temporal
infrared structure remains governed by the relaxational kernel
\(\Pi^R(\omega)\).

To connect this microscopic result to the long-wavelength collective
theory, we write the quadratic kernel as
\begin{equation}
\Gamma_2^R(\mathbf q,\omega)
=
g^{-1}
-
\Pi^R(\mathbf q,\omega).
\end{equation}
Expanding about the static and uniform limit gives
\begin{equation}
\Gamma_2^R(\mathbf q,\omega)
\simeq
r
+
c q^2
+
\Gamma_{\rm dyn}^R(\omega),
\end{equation}
where
\begin{equation}
r
\equiv
g^{-1}
-
\Pi^R(\mathbf 0,0),
\end{equation}
and
\begin{equation}
\Gamma_{\rm dyn}^R(\omega)
\equiv
\Pi^R(\mathbf 0,0)
-
\Pi^R(\mathbf 0,\omega).
\end{equation}
Thus
\begin{equation}
\chi_R^{-1}(\mathbf q,\omega)
\equiv
\Gamma_2^R(\mathbf q,\omega)
\simeq
r+cq^2+\Gamma_{\rm dyn}^R(\omega).
\end{equation}

Absorbing the static contribution into \(r\), this reduces to the
compact form
\begin{equation}
\chi_R^{-1}(\mathbf q,\omega)
\simeq
r
+
c q^2
+
\Pi^R(\omega),
\end{equation}
with the relaxational spectral representation
\begin{equation}
\Pi^R(\omega)
=
\int_0^\Lambda
d\lambda\;
\frac{\rho(\lambda)}{\lambda-i\omega}.
\label{eq:PiR_relax_pair}
\end{equation}
This expression shows that the dynamical part of the collective
response is entirely determined by the relaxation spectrum of the
underlying degrees of freedom.

Importantly, the spectral representation above admits a direct
interpretation in the time domain.
Taking the inverse transform, one obtains a causal memory kernel
\begin{equation}
K(t)
=
\int_0^\Lambda
d\lambda\;
\rho(\lambda)\,e^{-\lambda t},
\qquad
t\ge0,
\end{equation}
which governs the non-Markovian dynamics of the collective field.

Thus, the polarization function \(\Pi^R(\omega)\) appearing in the
susceptibility is simply the frequency representation of a memory
kernel.
This establishes that the infrared dynamics is controlled not by a
small number of propagating modes, but by the accumulation of slow
relaxational processes.

The microscopic Cooper-channel formulation developed in the main text
is complemented by a dynamical construction of the infrared memory
kernel presented in Appendices~B and C.
Appendix~B derives the emergence of a scale-free nonlocal temporal
kernel by integrating out a continuum of relaxational modes within the
MSRJD formalism \cite{38,39,40}, while Appendix~C provides an exact
time-domain realization based on an Ornstein--Uhlenbeck reservoir.
Together, these constructions establish the dynamical foundation
underlying the TDOS description of collective infrared behavior.

\subsection{C. Infrared universality classes}

Having established that the collective response is governed by the
spectral representation
\begin{equation}
\chi_R^{-1}(\mathbf q,\omega)
\simeq
r + c q^2
+
\int_0^\Lambda d\lambda\;
\frac{\rho(\lambda)}{\lambda - i\omega},
\end{equation}
the infrared behavior of the system is determined entirely by the
low--$\lambda$ structure of the time--scale density of states.

For a generic scaling form
\begin{equation}
\rho(\lambda)\sim \lambda^{\alpha},
\qquad \lambda\to 0 ,
\end{equation}
the dynamical kernel inherits a corresponding infrared structure.
In the time domain, this corresponds to a memory kernel
\begin{equation}
K(t)=\int_0^\Lambda d\lambda\;\rho(\lambda)e^{-\lambda t}
\;\sim\; t^{-1-\alpha},
\end{equation}
showing that the exponent $\alpha$ directly controls the long-time
memory of the system.

The dissipative part of the kernel is
\begin{equation}
\mathrm{Im}\,\chi_R^{-1}(\mathbf q,\omega)
=
\omega\int_0^\Lambda d\lambda\,
\frac{\rho(\lambda)}{\lambda^2+\omega^2}
\;\sim\;
\mathrm{sgn}(\omega)\,|\omega|^{\alpha}.
\end{equation}
Because the response function is causal, the reactive component is not
independent but is related to the dissipative part through the
Kramers--Kronig relation.
The real part therefore inherits the same infrared scaling structure
and, depending on the TDOS exponent, produces logarithmic or
subleading corrections to the static stiffness.

It is convenient to express the long--wavelength response in the
scaling form
\begin{equation}
\chi_R^{-1}(\omega,\mathbf q)
=
r + c q^2 + \gamma |\omega|^{s},
\end{equation}
so that the frequency exponent is directly determined by the TDOS scaling, $s=\alpha$.
While this form resembles the conventional Hertz--Millis
representation, its physical interpretation is different.
In the present framework the exponent $s$ does not originate from
fermionic damping of a single order--parameter mode 
but instead reflects the infrared organization of the entire relaxation spectrum,
equivalently encoding the long-time memory kernel generated by the continuum of slow modes.

The appropriate renormalization viewpoint is therefore naturally
formulated in terms of temporal coarse--graining.
Integrating out fast relaxation processes with decay rates
$\lambda>\Lambda/b$ corresponds to eliminating short time scales
$t\lesssim b^{-1}\Lambda^{-1}$.
Under this temporal renormalization step the remaining effective
theory is characterized by a renormalized TDOS
\begin{equation}
\rho_b(\lambda)=b^{\alpha}\rho(b\lambda),
\end{equation}
so that the infrared scaling exponent $\alpha$ defines the
universality class of the relaxation spectrum.
Infrared universality classes are therefore classified directly by the
low--energy structure of the relaxation--rate density.

A particularly important case occurs when the TDOS remains finite at
the origin,
\begin{equation}
\rho(\lambda)\xrightarrow{\lambda\to0}\rho_0 .
\end{equation}
In this flat--TDOS regime the dissipative kernel simplifies to
\begin{align}
\mathrm{Im}\,\chi_R^{-1}(\omega)
=
\omega\int_0^\Lambda d\lambda\,
\frac{\rho_0}{\lambda^2+\omega^2}
\xrightarrow{|\omega|\ll\Lambda}
\frac{\pi}{2}\rho_0\,\mathrm{sgn}(\omega),
\label{eq:sgn_kernel}
\end{align}
while the reactive component develops a logarithmic infrared
singularity,
\begin{equation}
\mathrm{Re}\,\chi_R^{-1}(\omega)
=
\int_0^\Lambda d\lambda\,
\rho_0
\frac{\lambda}{\lambda^2+\omega^2}
\simeq
\rho_0\ln\frac{\Lambda}{|\omega|}.
\label{eq:log_kernel}
\end{equation}
Collective infrared dynamics is therefore governed by the marginal
non--Markovian kernel
\begin{equation}
\chi_R^{-1}(\omega,\mathbf q)
\simeq
r+cq^2+\rho_0\ln\frac{\Lambda}{|\omega|}
+i\frac{\pi}{2}\rho_0\,\mathrm{sgn}(\omega)+\cdots ,
\label{eq:marginal_kernel}
\end{equation}
which corresponds in the time domain to a generic long--memory
kernel $K(t)\sim1/t$, indicating scale-free temporal correlations
characteristic of memory-dominated critical dynamics.

This flat--TDOS fixed point therefore defines a marginal dynamical universality class
 in which the infrared behavior is governed by the accumulation of a continuum of slow decay modes.
Rather than arising from overdamped fluctuations of a single soft
order parameter, collective relaxation is controlled by the spectral
organization of the full relaxation spectrum.

\section{III. Memory--Enhanced Pairing Beyond Eliashberg Theory}

Building on the microscopic framework developed in Sec.~II,
we now analyze the structure of superconducting pairing in a regime
where infrared dynamics is governed by a continuum of relaxation modes.

In contrast to conventional approaches based on a specific bosonic
pairing mediator, we show that superconductivity can emerge directly
from the relaxation-spectrum organization of the Cooper channel.
The pairing kernel is not determined by quasiparticle exchange,
but by the infrared structure of the retarded collective response.

We first review the marginal behavior of pairing in conventional
overdamped bosonic-mediator theories.
We then formulate the Cooper channel in terms of the relaxational
kernel and derive the corresponding ladder resummation.
Finally, we analyze the infrared scaling and renormalization of the
pairing interaction, showing how the structure of the relaxation
spectrum controls the nature of the superconducting instability.

\subsection{A. Marginal pairing in overdamped bosonic-mediator theories}

As a reference point, it is useful to recall the structure of pairing
in conventional theories based on overdamped collective mediators,
such as spin-fluctuation approaches near metallic quantum criticality \cite{7,41,42}.
In such treatments one introduces an effective retarded propagator
(or susceptibility) for a particle--hole collective mode of the form
\begin{equation}
D_R^{-1}(\omega,\mathbf q)
=
r + c q^2 + \Gamma_{\rm dyn}(\omega,\mathbf q),
\end{equation}
where $\Gamma_{\rm dyn}(\omega,\mathbf q)$ denotes the dynamical
polarization kernel generated by coupling to the electronic continuum.
In many cases of interest, such as antiferromagnetic quantum criticality,
this reduces to an overdamped form
$\Gamma_{\rm dyn}(\omega)\sim \gamma |\omega|$,
reflecting Landau damping of the collective mode.

In bosonic-mediator theories \cite{21,22}, the overdamped collective mode is then
treated as an effective pairing mediator.
After momentum integration, the resulting interaction kernel entering
the Cooper channel is schematically of the form $V(\omega)\sim 1/|\omega|$,
up to model-dependent form factors and ultraviolet regularization.
If one further assumes that pairing can still be analyzed within a
quasiparticle-based ladder or gap-equation framework, the linearized
pairing equation takes the schematic form
\begin{equation}
\Delta(\omega)
=
\int_{T}^{\Lambda}
\frac{d\omega'}{|\omega'|}
\,V(\omega-\omega')\,\Delta(\omega'),
\end{equation}
where $\Lambda$ is the ultraviolet cutoff and
$T$ provides the infrared cutoff.

The corresponding infrared flow can be understood by integrating out
fast frequency modes in the shell $\omega'\in[\Lambda/b,\Lambda]$.
The leading ladder correction is then
\begin{equation}
dg
\propto
g^2
\int_{\Lambda/b}^{\Lambda}\frac{d\omega'}{\omega'},
\end{equation}
which yields
\begin{equation}
\int_{\Lambda/b}^{\Lambda}\frac{d\omega'}{\omega'}
=
\ln b .
\end{equation}
Introducing the RG scale $l=\ln b$, one obtains the standard marginal flow
\begin{equation}
\frac{dg}{dl}=C g^2 .
\end{equation}

Thus, within overdamped bosonic-mediator theories, pairing grows only
logarithmically in the infrared \cite{43,44}.
The physical reason is that the mediator itself is continuously damped
by decay into the electronic continuum, so coherence is not retained
across an extended hierarchy of time scales.

This benchmark should be distinguished sharply from the framework
developed in the present work.
Here we do not assume a well-defined quasiparticle description nor an
effective pairing glue represented by a single bosonic propagator.
Instead, after integrating out microscopic electronic degrees of freedom,
the infrared Cooper sector is organized by the collective relaxation
spectrum itself.
The relevant object is not the retarded propagator of an overdamped
order-parameter mode, but by the retarded polarization kernel generated by a
continuum of relaxational pair modes, as derived below.

\subsection{B. Cooper-channel kernel and ladder resummation from the microscopic theory}

The analysis of superconducting instability naturally follows from the
microscopic Cooper-channel kernel derived in Sec.~II.
After integrating out fermionic degrees of freedom, the quadratic kernel
(inverse propagator) of the pair field is given by
\begin{equation}
\Gamma_2(q,\Omega_n)
=
\frac{1}{g}
-
\Pi(q,\Omega_n),
\label{eq:Gamma2_sec3}
\end{equation}
where $\Pi(q,\Omega_n)$ denotes the Cooper bubble generated by fermionic loops.
After analytic continuation, we use $\omega$ to denote the real-frequency
variable of the collective response.
In the dissipative infrared regime, the Cooper kernel admits a
relaxational spectral representation of the form
in Eq.~(\ref{eq:PiR_relax_pair}).

The superconducting susceptibility is obtained by summing repeated pair
scattering processes in the Cooper channel.
At the level of linear response, this corresponds to a ladder resummation
built from the interaction $g$ and the kernel $\Pi^R(\omega)$:
\begin{equation}
\chi_\Delta^R(\omega)
=
\Pi^R(\omega)
+
\Pi^R(\omega)\,g\,\Pi^R(\omega)
+
\cdots .
\end{equation}
Summing the geometric series yields
\begin{equation}
\chi_\Delta^R(\omega)
=
\frac{\Pi^R(\omega)}{1 - g\,\Pi^R(\omega)}.
\label{eq:ladder_resum_relax}
\end{equation}

Superconductivity corresponds to a pole of the retarded pair susceptibility.
The instability condition is therefore determined by the Thouless criterion,
\begin{equation}
1 - g\,\mathrm{Re}\,\Pi^R(\omega) = 0.
\label{eq:Thouless_relax_pair}
\end{equation}
Using Eq.~(\ref{eq:PiR_relax_pair}), the real part of the Cooper kernel is
\begin{equation}
\mathrm{Re}\,\Pi^R(\omega)
=
\int_0^\Lambda d\lambda\;
\rho(\lambda)
\frac{\lambda}{\lambda^2 + \omega^2}.
\label{eq:RePiR_relax_pair}
\end{equation}

For a power-law form of the time-scale density of states,
\begin{equation}
\rho(\lambda) \sim \lambda^\alpha,
\end{equation}
the infrared behavior depends on the exponent $\alpha$.
In the infrared-singular regime $\alpha < 0$, one obtains
\begin{equation}
\mathrm{Re}\,\Pi^R(\omega)
\sim
|\omega|^{\alpha}
\int_0^\infty dx\;
\frac{x^{\alpha+1}}{x^2 + 1}
=
C_\alpha\,|\omega|^{\alpha},
\end{equation}
where $C_\alpha$ is a positive finite constant.

This result shows that the structure of the superconducting instability
is directly controlled by the infrared scaling of the relaxation spectrum.
In particular, pairing is governed not by quasiparticle coherence or
the exchange of a specific bosonic mediator, but by the organization of
collective relaxation processes.

Thus, while the ladder resummation and Thouless criterion retain their
standard formal structure, the physical content of the Cooper kernel is
fundamentally altered: the pairing interaction is controlled by the
density of relaxation rates rather than by excitation energies.

\subsection{C. Infrared scaling and renormalization of the Cooper channel}

The superconducting instability is determined by the Thouless
criterion derived in Sec.~III B.
The infrared behavior of pairing is therefore controlled by the
low-frequency scaling of the retarded kernel $\Pi^R(\omega)$,
which is in turn determined by the structure of the relaxation spectrum.

In the flat-TDOS regime ($\alpha=0$), the kernel takes the form
\begin{equation}
\Pi^R(\omega)\simeq \rho_0 \ln\frac{\Lambda}{|\omega|},
\end{equation}
so that coarse graining over a logarithmic frequency shell
$\omega\in[\Lambda/b,\Lambda]$ produces
\begin{equation}
dg \propto g^2 \int_{\Lambda/b}^{\Lambda}\frac{d\omega}{\omega}
= g^2 \ln b,
\end{equation}
leading to the marginal flow $dg/dl = C g^2$.
Thus, a flat TDOS yields a logarithmic infrared enhancement of pairing,
analogous to marginal behavior in conventional Eliashberg theories.

When the TDOS develops an infrared divergence ($\alpha<0$), the retarded
kernel acquires a power-law form,
\begin{equation}
\Pi^R(\omega)\sim |\omega|^{\alpha}.
\end{equation}
Under the scale transformation $\omega\to \omega/b$, the kernel transforms as
\begin{equation}
\Pi^R(\omega) \to b^{-\alpha}\Pi^R(\omega),,
\end{equation}
implying a scale dependence of the effective coupling
\begin{equation}
g' = b^{-\alpha} g.
\end{equation}
The corresponding RG flow is therefore
\begin{equation}
\frac{dg}{dl}
=
(-\alpha)\,g + C g^2,
\end{equation}
so that for $\alpha<0$ the interaction becomes relevant and grows
algebraically under coarse graining.

The qualitative difference from the Eliashberg regime originates from
the organization of the relaxation spectrum.
In the marginal case, each logarithmic shell contributes independently,
leading to only logarithmic enhancement.
By contrast, IR-singular spectra correspond to an accumulation of
slow modes toward vanishing relaxation rates, which reorganizes the
Cooper channel into a relevant infrared interaction.

The corresponding RG flows are illustrated schematically in Fig.~2.
The schematic curves correspond to the solutions of the RG equations
$dg/dl = C g^2$ (marginal case) and $dg/dl = (-\alpha) g$
(IR-singular case), leading to exponential growth
of the effective pairing interaction, respectively.
In the marginal regime associated with a flat TDOS ($\alpha=0$),
the pairing interaction increases only logarithmically under
coarse graining, reflecting the weak accumulation of infrared
spectral weight.
By contrast, for infrared-singular TDOS ($\alpha<0$), the coupling
acquires a finite scaling dimension and grows exponentially with
the RG scale, indicating that the Cooper channel becomes a relevant
infrared instability.
The figure thus provides a direct visualization of how the infrared
structure of the relaxation spectrum controls the qualitative nature
of pairing enhancement.

Superconductivity thus emerges as an instability of the retarded
collective response controlled by the TDOS,
with logarithmic enhancement for flat spectra and
power-law scaling of the kernel in the presence of an
IR-singular slow-mode density.

\vspace{6pt}
The superconducting instability is thus fully determined at the level
of the quadratic Cooper kernel through the Thouless condition.
For completeness, we note that the consistency of the present
relaxation-spectrum formulation with a Wilsonian renormalization-group
framework is established in Appendix~D, where fluctuation corrections
to the effective mass and interaction are analyzed.

In that formulation, the RG flow is naturally organized in terms of
a shell defined by the full inverse propagator
$r + c q^2 + \Gamma_{\rm dyn}(\omega)$, reflecting the fact that
infrared dynamics is controlled by the relaxation spectrum rather
than by frequency or momentum alone.
This provides a geometrically transparent representation of the
infrared scaling in terms of equal-kernel contours.

Importantly, these fluctuation corrections do not modify the leading
pairing mechanism, which remains governed by the infrared structure
of the retarded Cooper kernel.

\begin{figure}[t]
\centering
\includegraphics[scale=0.65, trim=0.7cm 20.4cm 0cm 0cm]{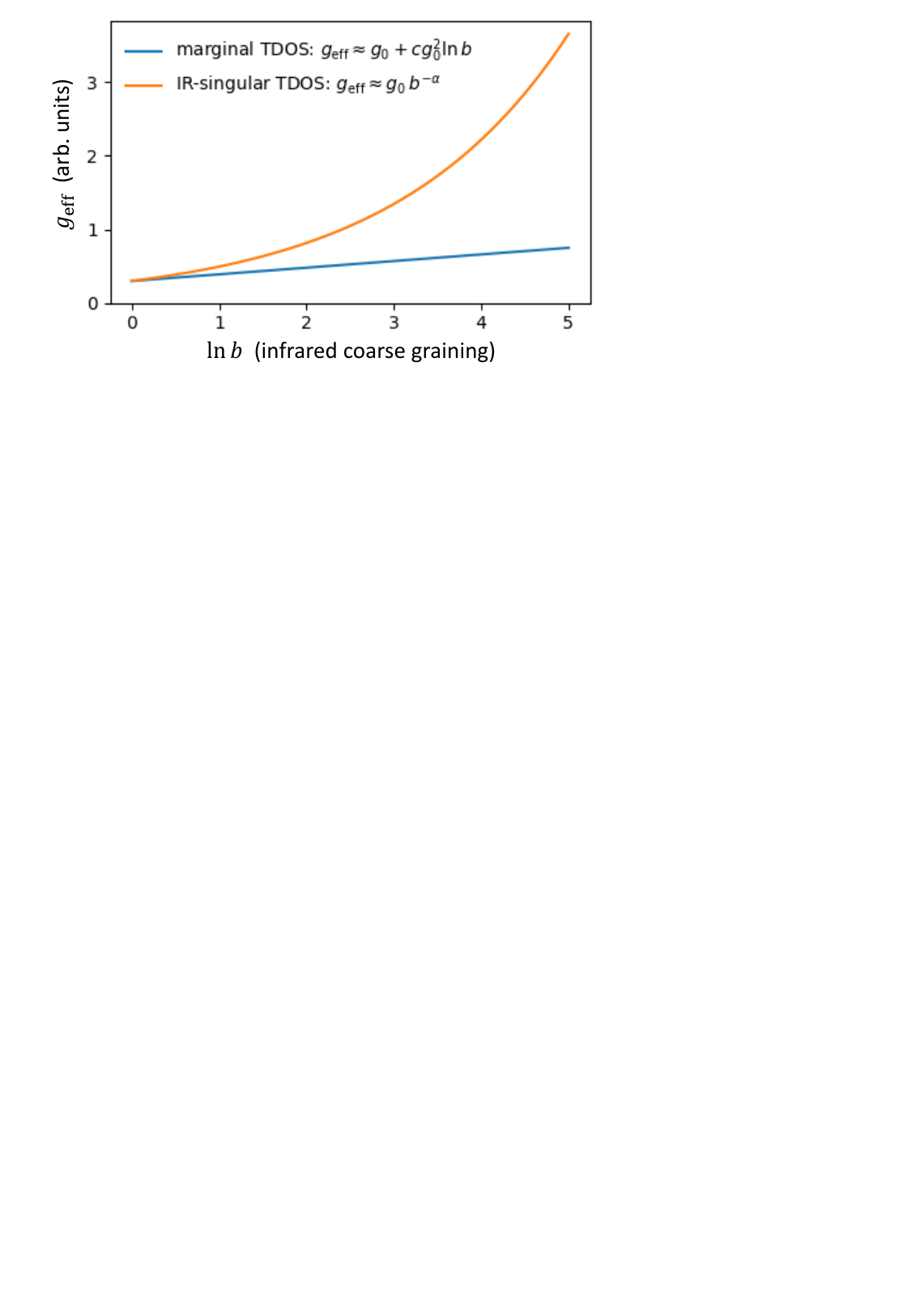}
\caption{Infrared renormalization-group flow of the Cooper-channel interaction.
Shown schematically is the flow of the dimensionless pairing coupling $g$ under
infrared coarse graining.
A flat relaxation-rate density of states (marginal TDOS, $\alpha=0$)
produces logarithmic growth,
$dg/dl\sim g^2$, corresponding to a marginal infrared enhancement of pairing (blue).
This behavior is analogous to the logarithmic scaling familiar from
conventional BCS/Eliashberg-type pairing theories.
In contrast, IR-singular relaxation spectra with enhanced infrared weight
($\alpha<0$) generate a relevant contribution,
$dg/dl=(-\alpha)g$, leading to algebraic growth of the pairing interaction (orange).
.}
\label{fig:fig2}
\end{figure}

\section{IV. Pairing from Memory--Dominated Critical Dynamics}

The previous section established the universal infrared structure of the
Cooper channel generated by the relaxation spectrum.
In particular, the slow--mode reservoir reorganizes the retarded
Cooper kernel into a singular form, leading to marginal enhancement
of pairing when the TDOS remains finite in the infrared, and to a
relevant instability only when the TDOS develops an infrared
divergence.

The remaining question is how this infrared amplification is tied to
microscopic electronic pairing tendencies in correlated materials and
how it determines the observable superconducting transition scale.
In this section we connect the intrinsic electronic pairing seed to
the memory--dominated Cooper kernel and show how their interplay produces an enhanced superconducting
instability, which acquires algebraic scaling only when the TDOS develops an infrared divergence.

In the present framework, superconductivity is not primarily controlled
by the formation of tightly bound Cooper pairs at the microscopic level.
Rather, a weak intrinsic pairing tendency is already present in the
electronic system, giving rise to fluctuating Cooper-pair correlations.

The key mechanism is the accumulation of slow collective modes in the
relaxation spectrum, which enhances both the lifetime and the overlap
of these pair fluctuations.  As a result, superconductivity emerges as
a collective dynamical instability driven by the infrared
reorganization of relaxation modes, rather than by the microscopic
formation of pairs itself.

\subsection{A. Intrinsic electronic pairing and dynamical formulation}

In strongly correlated materials, short-range electronic interactions
can naturally enhance local pairing tendencies.
For example, in systems proximate to Mott insulating phases,
superexchange processes generated by strong Coulomb repulsion favor
short-range spin-singlet correlations \cite{1,45,46}.

At low energies, electronic interactions generically generate
multiple collective channels, including the Cooper channel.
Projection onto the particle--particle sector defines an effective
pairing interaction, which serves as the intrinsic
pairing scale in the present framework.
In conventional approaches these channels are analyzed through
diagrammatic resummations, such as ladder ($t$-matrix) constructions,
typically formulated in terms of specific effective interaction
channels.
In the present framework, the same resummation structure is retained,
but the Cooper kernel is governed by the relaxation spectrum emerging
from the underlying microscopic dynamics.

While several ordering channels are in principle allowed, their
infrared relevance is not identical.  Particle--hole instabilities,
such as charge- or spin-density waves, involve bilinears of the form
$c^\dagger_{\mathbf k+\mathbf Q}c_{\mathbf k}$ and therefore require
that the ordering wave vector $\mathbf Q$ connect extended low-energy
regions of the Fermi surface.  In the absence of nesting or hot-spot
structure, only a restricted portion of momentum space contributes to
the infrared accumulation.  By contrast, the Cooper channel involves
time-reversed pairs $(\mathbf k,-\mathbf k)$ with zero total momentum.
Once a Fermi surface is present, such pairs are available over the
entire Fermi shell.  The particle--particle channel therefore provides
the largest and most uniform phase space for accumulating slow
relaxational spectral weight, making superconductivity the natural
leading instability of the memory-dominated infrared spectrum.

Within the dynamical framework developed in Sec.~III, the pairing
susceptibility assumes a universal $t$-matrix structure, as given in
Eq.~(\ref{eq:ladder_resum_relax}).
In this formulation, $g$ represents the intrinsic electronic pairing
seed, while $\Pi^R(\omega)$ denotes the retarded Cooper kernel governed
by the relaxation spectrum that emerges after integrating out
microscopic fermionic degrees of freedom.

Importantly, this relaxation spectrum is not an ad hoc assumption, but
arises naturally from the momentum-space structure of interacting
fermionic systems.
As shown in Appendix~E, the time-scale density of states follows
directly from the distribution of fermionic relaxation rates in
Eq.~(\ref{eq:TDOS_definition_weighted}),
which is formally analogous to the ordinary electronic density of
states, with the energy dispersion replaced by the relaxation-rate
dispersion.

A particularly important consequence follows from the existence of a
Fermi surface.
Because low-energy fermions occupy a finite momentum shell
$k\simeq k_F\neq0$, slow pair modes are distributed over an extended
$(d-1)$-dimensional manifold rather than concentrated at a single
point in momentum space.
For a generic overdamped dispersion
$\gamma(k)\sim |k-k_F|$, this naturally produces a finite infrared
TDOS, $\rho(\lambda\to0)\sim \rho_0$, while stronger softening can
lead to infrared-singular spectra.
The accumulation of slow modes is therefore a robust consequence of
finite-shell phase space, rather than a special microscopic
assumption.

Strongly entangled quantum many-body systems provide even stronger
realizations of such behavior. In particular, models such as
Sachdev--Ye--Kitaev-type systems and resonating valence bond states
\cite{3,47} exhibit scale-free infrared dynamics governed by a
continuum of relaxation processes, often corresponding to enhanced or
infrared-singular TDOS.
Concrete realizations and microscopic constructions of such relaxation
spectra are discussed in Appendix~F.

In weakly interacting systems the pairing instability remains
logarithmic, leading to the exponentially small transition scale of
conventional BCS theory.
Near correlated criticality, however, the relaxation spectrum becomes
dense as $\lambda\to0$, generating an extensive reservoir of slow
collective modes.
This slow-mode reservoir reorganizes the infrared dynamics by
redistributing spectral weight toward long time scales and enhances
the Cooper channel.
Thus, critical dynamics does not generate a microscopic pairing glue,
but instead amplifies the Cooper-channel instability through infrared
spectral reorganization.

\vspace{10pt}
\noindent
\emph{Phase-space origin of the TDOS scaling.}
The time--scale density of states admits a direct microscopic
representation in terms of the momentum-space distribution of
fermionic relaxation rates,
\begin{equation}
\rho(\lambda)
=
\int \frac{d^d k}{(2\pi)^d}\;
\delta\!\bigl(\lambda-2\gamma_{\mathbf k}\bigr),
\end{equation}
which counts the number of states contributing to a given decay rate.
For an isotropic dispersion $\gamma_{\mathbf k}=\gamma(k)$, this may be
evaluated as
\begin{equation}
\rho(\lambda)
=
\frac{S_{d-1}}{(2\pi)^d}
\sum_i
\frac{k_i^{\,d-1}}{|2\gamma'(k_i)|},
\qquad
\lambda = 2\gamma(k_i),
\label{eq:TDOS_general_main}
\end{equation}
where $S_{d-1}$ is the surface area of the $(d-1)$-sphere.

To determine the infrared behavior, consider a power-law form near a
soft manifold,
\begin{equation}
\gamma(k)\sim a\,|k-k_0|^n.
\end{equation}
When the slow modes are distributed near a finite momentum shell
($k_0\neq 0$), one has $k_i\simeq k_0$ for $\lambda\to 0$, so that the
phase-space factor remains finite, $k_i^{d-1}\simeq k_0^{d-1}$.
In this case,
\begin{equation}
\rho(\lambda)\sim \lambda^{-(n-1)/n}.
\label{eq:TDOS_shell_main}
\end{equation}
By contrast, when the slow modes are centered at $k=0$, the phase-space
factor vanishes as $k^{d-1}$, yielding
\begin{equation}
\rho(\lambda)\sim \lambda^{d/n-1}.
\label{eq:TDOS_origin_main}
\end{equation}

This distinction has a clear physical origin: in the finite-shell case,
the slow modes are distributed over an extended momentum manifold, so
that their number does not diminish as $\lambda\to 0$, allowing for a
uniform or enhanced accumulation of slow modes.
In contrast, when the slow modes collapse toward $k=0$, the available
phase-space volume itself vanishes, suppressing the density of slow
modes.
As a result, flat or infrared-singular TDOS arises naturally in systems
with a finite momentum shell, providing the microscopic basis for the
memory-dominated enhancement of pairing.

\subsection{B. Transition scale from infrared spectral structure}

The superconducting transition is determined by the Thouless condition
in the static, uniform limit.
Equivalently, it can be formulated as the vanishing of the
temperature-dependent quadratic kernel (inverse propagator),
\begin{equation}
\Gamma_2^R(\mathbf 0,0;T)
=
\frac{1}{g}-\Pi^R(0;T),
\end{equation}
so that the transition temperature is defined by
\begin{equation}
r(T)\equiv \frac{1}{g}-\Pi^R(0;T),
\qquad
r(T_c)=0.
\end{equation}

In the preceding microscopic derivation, the Lorentzian convolution was
used to identify the infrared analytic structure of the Cooper kernel.
This step fixes the relaxational denominator
\((\lambda-i\Omega)^{-1}\).
The finite-temperature occupation factors do not modify this pole
structure, but instead regularize the static infrared singularity
through a smooth thermal crossover, analogous to the way the factor
\(\tanh(\xi/2T)\) regularizes the logarithmic Cooper singularity in the
conventional BCS gap equation.

Accordingly, the finite-temperature static kernel should more generally
be written as
\begin{equation}
\Pi^R(0;T)
\simeq
\int_0^\Lambda d\lambda\;
\frac{\rho(\lambda)}{\lambda}\,
F\!\left(\frac{\lambda}{T}\right),
\label{eq:PiR_static_thermal_function}
\end{equation}
where \(F(x)\) is a smooth thermal crossover function generated by the
finite-temperature occupation factors.
It satisfies
\begin{equation}
F(x)\to0
\quad (x\ll1),
\qquad
F(x)\to1
\quad (x\gg1),
\end{equation}
so that finite temperature suppresses the contribution of sufficiently
slow relaxation modes.

The appearance of the thermal infrared scale should not be interpreted
as an identification of the relaxation rate with a fermionic
Matsubara frequency.
The relaxation rate \(\lambda\) characterizes the decay time of
infrared collective modes, whereas Matsubara frequencies arise from the
imaginary-time Fourier decomposition of finite-temperature correlation
functions.
The relation
\begin{equation}
\lambda_T\sim T
\end{equation}
should therefore be understood only at the scaling level.

Physically, finite temperature introduces the thermal time scale
\begin{equation}
\tau_T\sim\frac{1}{T},
\end{equation}
beyond which temporal correlations cannot be coherently maintained.
Consequently, collective relaxation modes with decay rates
\begin{equation}
\lambda\lesssim\tau_T^{-1}\sim T
\end{equation}
are effectively suppressed by thermal decoherence.
The scale \(\lambda_T\) therefore represents an infrared crossover
scale generated by finite-temperature dynamics, rather than a
microscopic Matsubara pole.

For the purpose of extracting the leading infrared scaling, the smooth
thermal regularization may therefore be represented by the effective
infrared scale \(\lambda_T\), giving
\begin{equation}
\Pi^R(0;T)
\sim
\int_{\lambda_T}^{\Lambda} d\lambda\;
\frac{\rho(\lambda)}{\lambda},
\qquad
\lambda_T\sim T .
\label{eq:PiR_static_cutoff}
\end{equation}

\vspace{6pt}
\noindent
\emph{Flat TDOS ($\alpha=0$, marginal regime).}
For a finite infrared density of states,
\begin{equation}
\rho(\lambda\to0)=\rho_0,
\end{equation}
the finite-temperature static kernel takes the scaling form
\begin{equation}
\Pi^R(0;T)
\sim
\rho_0
\int_{\lambda_T}^{\Lambda}
\frac{d\lambda}{\lambda},
\qquad
\lambda_T\sim T,
\end{equation}
where the thermal scale $\lambda_T$ represents the smooth infrared
regularization generated by finite-temperature occupation effects.
The leading infrared contribution is therefore
\begin{equation}
\Pi^R(0;T)
\sim
\rho_0
\ln\frac{\Lambda}{\lambda_T},
\qquad
\lambda_T\sim T.
\end{equation}

The transition condition $r(T_c)=0$ then gives
\begin{equation}
\frac{1}{g}
=
\rho_0
\ln\frac{\Lambda}{\lambda_{T_c}}
+
O(1),
\end{equation}
where the additive constant depends on the detailed form of the
finite-temperature crossover function.
Since $\lambda_{T_c}\propto T_c$ up to a nonuniversal numerical factor,
the transition temperature takes the scaling form
\begin{equation}
T_c
=
C\,\Lambda
\exp\!\left(-\frac{1}{g\, \rho_0}\right),
\label{eq:Tc_flat_final}
\end{equation}
where $C$ is a nonuniversal constant determined by the precise
finite-temperature regularization of the infrared kernel.

Thus, a flat TDOS produces logarithmic enhancement of the static
retarded kernel, leading to a BCS-like exponentially generated
transition scale despite the presence of strong infrared fluctuations.
This corresponds to a marginal dynamical regime in which long-time
memory is present but does not render the pairing interaction
infrared-relevant.

\vspace{6pt}
\noindent
\emph{Infrared-divergent TDOS ($\alpha<0$, IR-singular regime).}
When the relaxation spectrum develops an infrared divergence,
\begin{equation}
\rho(\lambda)\sim \lambda^{\alpha},
\qquad
\alpha<0,
\end{equation}
the static kernel scales as
\begin{equation}
\Pi^R(0;T)
\sim
\int_{\lambda_T}^{\Lambda}
d\lambda\;
\lambda^{\alpha-1}
\sim
T^{\alpha},
\qquad
(T\ll\Lambda),
\end{equation}
again with $\lambda_T\sim T$ representing the effective thermal
infrared scale.

The transition condition $r(T_c)=0$ then gives
\begin{equation}
1-g\,C_\alpha T_c^{-|\alpha|}=0,
\end{equation}
where $C_\alpha$ is a positive nonuniversal constant.
Solving for the transition temperature yields
\begin{equation}
T_c
\sim
(g\,C_\alpha)^{1/|\alpha|}
\sim
g^{1/|\alpha|}.
\label{eq:Tc_powerlaw_final}
\end{equation}

In this regime, the accumulation of slow modes toward vanishing
relaxation rates reorganizes the Cooper channel into a relevant
infrared interaction.
As a result, the exponential suppression characteristic of
conventional pairing is replaced by an algebraic dependence on the
interaction strength.

\vspace{6pt}
\noindent
\emph{Infrared-depleted TDOS ($\alpha>0$).}
For completeness, it is useful to contrast the above infrared-enhanced
regimes with the case
\begin{equation}
\rho(\lambda)\sim \lambda^\alpha,
\qquad
\alpha>0,
\end{equation}
for which the TDOS vanishes toward the infrared.
In this regime,
\begin{equation}
\Pi^R(0;T)
\sim
\int_{\lambda_T}^{\Lambda}
d\lambda\;
\lambda^{\alpha-1}
=
\frac{1}{\alpha}
\left(
\Lambda^\alpha-T^\alpha
\right),
\end{equation}
which remains finite as $T\to0$.
The infrared logarithmic or algebraic enhancement is therefore absent,
and superconductivity no longer emerges from an infrared accumulation
instability at arbitrarily weak coupling.
Instead, a finite critical interaction strength is required to produce
pairing.

\vspace{6pt}
\noindent
\emph{Physical interpretation.}
The superconducting transition is governed by the infrared structure
of the static retarded Cooper kernel.
A finite TDOS produces the familiar logarithmic Cooper enhancement and
an exponentially generated transition scale, while an infrared-singular
TDOS modifies the kernel itself and leads to algebraic growth of the
pairing instability.
By contrast, an infrared-depleted TDOS suppresses the accumulation of
slow collective modes and eliminates the infrared enhancement mechanism.

The ladder instability derived from the retarded Cooper kernel already
corresponds to a collective infrared ordering instability of the Cooper
channel.
The resulting superconducting state is characterized by a finite
superfluid stiffness $\rho_s$, which measures the spatial and
electromagnetic rigidity of the condensate.
In strongly correlated systems, the observable transition temperature
may be reduced when this condensate rigidity becomes parametrically
small, leading to
\begin{equation}
k_B T_c \sim \min\left(\Delta,\, \rho_s\right),
\end{equation}
with $\rho_s \sim n_s/m^*$.

The central result is therefore that both the Cooper instability and
the resulting condensate rigidity are governed by the same infrared
structure of the relaxation-rate spectrum.
The low-energy TDOS acts as a unifying infrared organizational
principle linking memory-dominated critical dynamics, pairing
susceptibility, and macroscopic superconducting coherence.

\subsection{C. Phase dynamics and electromagnetic response in the superconducting state}

We now turn to the dynamical structure of the superconducting state
below the transition, including the coupling to the electromagnetic
field. In contrast to conventional treatments based on propagating
collective modes, we show that the phase dynamics inherits the
nonlocal memory structure of the Cooper kernel, leading to a
dissipative or critical response rather than a sharp quasiparticle mode.

Starting from the Hubbard--Stratonovich formulation in the Cooper channel,
the fermions can be integrated out, yielding an effective action for
the complex pair field $\Delta(\mathbf r,t)$:
\begin{equation}
S_{\rm eff}[\Delta]
=
\int d^d r\,dt\;
\Delta^*
\left[
g^{-1}
-
\Pi^R(-i\partial_t,-i\nabla)
\right]
\Delta.
\end{equation}
Expanding near the transition, the inverse pair propagator takes the form
\begin{equation}
D_R^{-1}(\mathbf q,\omega)
=
g^{-1}
-
\Pi^R(\mathbf q,\omega)
\simeq
r(T)
+
c\,\mathbf q^2
+
\Gamma_{\rm dyn}^R(\omega),
\end{equation}
where
$
r(T)
=
g^{-1}
-
\Pi^R(\mathbf 0,0;T),
\quad
\Gamma_{\rm dyn}^R(\omega)
=
\Pi^R(0)-\Pi^R(\omega)$.

Below the transition, the order parameter acquires a finite expectation
value, and we parameterize fluctuations as
\begin{equation}
\Delta(\mathbf r,t)
=
(\Delta_0+\eta(\mathbf r,t))e^{i\theta(\mathbf r,t)}.
\end{equation}
At low energies the amplitude mode $\eta$ is gapped and may be
integrated out, leaving an effective theory for the phase field.

To derive the phase dynamics, it is essential to expand the nonlocal
kernel systematically rather than identifying $|\Delta|^2$ directly
with $\theta^2$.
Using
\begin{equation}
\Delta^*(t)\Delta(t')
=
\Delta_0^2 e^{i[\theta(t')-\theta(t)]},
\end{equation}
we expand for small phase differences,
\begin{equation}
e^{i[\theta(t')-\theta(t)]}
\simeq
1
+
i\bigl(\theta(t')-\theta(t)\bigr)
-
\frac{1}{2}
\bigl(\theta(t')-\theta(t)\bigr)^2
+\cdots.
\end{equation}
Because the memory kernel depends only on the time difference and is
symmetric under $t\leftrightarrow t'$, the linear contribution vanishes
upon integration.
The leading correction is therefore quadratic in phase differences,
showing that the dynamical kernel couples to temporal phase variations
rather than directly to $\theta^2$ itself.

The resulting nonlocal phase action takes the form
\begin{equation}
S_{\theta}
=
\frac{\kappa_\theta}{2}
\int dt\,dt'\,
K(t-t')
\bigl[
\theta(t)-\theta(t')
\bigr]^2,
\end{equation}
with $\kappa_\theta\sim\Delta_0^2$.
Equivalently, in frequency space,
\begin{equation}
S_{\theta}
=
\frac{1}{2}
\sum_{\mathbf q,\omega}
\left[
\rho_s q^2
+
\mathcal K_\theta(\omega)
\right]
|\theta(\mathbf q,\omega)|^2,
\end{equation}
where
\begin{equation}
\mathcal K_\theta(\omega)
=
\kappa_\theta
\Gamma_{\rm dyn}^R(\omega),
\end{equation}
with
\begin{equation}
\Gamma_{\rm dyn}^R(\omega)
=
2\int dt\,
K(t)
\bigl[
1-\cos(\omega t)
\bigr].
\end{equation}
For a local analytic kernel,
$\Gamma_{\rm dyn}^R(\omega)\sim \omega^2$,
this reduces to the conventional time-dependent
Ginzburg--Landau form.
For a scale-free memory kernel, however,
$\Gamma_{\rm dyn}^R(\omega)$ becomes nonanalytic,
and the phase dynamics inherits the same infrared temporal structure
as the underlying Cooper polarization.

Gauge invariance requires that the phase enters only through
gauge-invariant combinations.
For the spatial sector, this gives the covariant derivative
\begin{equation}
\nabla\theta(\mathbf r,t)-2e\mathbf A(\mathbf r,t).
\end{equation}
The corresponding stiffness contribution to the phase action is
\begin{equation}
S_{\theta}^{\rm sp}
=
\frac{\rho_s}{2}
\int dt\,d^dr\,
\left(
\nabla\theta-2e\mathbf A
\right)^2 .
\end{equation}
Equivalently, in momentum space,
\begin{equation}
S_{\theta}^{\rm sp}
=
\frac{\rho_s}{2}
\sum_{\mathbf q,\Omega}
\left|
i\mathbf q\,\theta(\mathbf q,\omega)
-
2e\mathbf A(\mathbf q,\omega)
\right|^2 .
\end{equation}
Unlike the schematic form
$(\mathbf q-2e\mathbf A)^2|\theta|^2$,
this expression explicitly retains the phase--gauge-field cross term.
As a result, variation with respect to $\mathbf A$ correctly reproduces
the supercurrent response.

The temporal memory part must be written in terms of the
gauge-invariant phase difference between two times.  The appropriate
nonlocal form is
\begin{equation}
\begin{aligned}
S_{\theta}^{\rm dyn}
=
&\frac{\kappa_\theta}{2}
\int d^dr\,dt\,dt'\,
\\
& \times K(t-t') \left[
\theta(\mathbf r,t)-\theta(\mathbf r,t')
+
2e\int_{t'}^{t} ds\,\Phi(\mathbf r,s)
\right]^2 .
\end{aligned}
\end{equation}
This expression is invariant under
\[
\theta\rightarrow\theta+2e\chi,\qquad
\mathbf A\rightarrow\mathbf A+\nabla\chi,\qquad
\Phi\rightarrow\Phi-\partial_t\chi .
\]
For $\Phi=0$, it reduces to the nonlocal phase action derived above.
In the local analytic limit, the temporal part reduces to the familiar
covariant form proportional to
$(\partial_t\theta+2e\Phi)^2$.

The full gauge-invariant phase action is therefore decomposed as
\begin{equation}
S_\theta
=
S_\theta^{\rm sp}
+
S_\theta^{\rm dyn},
\end{equation}
where the first term describes the equilibrium phase stiffness and
electromagnetic response, while the second encodes the nonlocal temporal
memory inherited from the Cooper polarization kernel.

The dynamical kernel is inherited directly from the Cooper polarization.
Using the TDOS representation,
\[
\Pi^R(\omega)
=
\int_{\lambda_{\rm IR}}^\Lambda d\lambda\;
\frac{\rho(\lambda)}{\lambda-i\omega},
\]
we define the dynamical contribution to the inverse pair susceptibility as
\begin{equation}
\Gamma_{\rm dyn}^R(\omega)
=
\Pi^R(0)-\Pi^R(\omega).
\end{equation}
For a flat TDOS, $\rho(\lambda)=\rho_0$, this gives
\begin{equation}
\Gamma_{\rm dyn}^R(\omega)
=
\rho_0
\ln
\left[
\frac{\Lambda(\lambda_{\rm IR}-i\omega)}
{\lambda_{\rm IR}(\Lambda-i\omega)}
\right].
\end{equation}
In the scaling regime
$\lambda_{\rm IR}\ll |\omega|\ll \Lambda$,
this reduces to
\begin{equation}
\Gamma_{\rm dyn}^R(\omega)
\simeq
\rho_0
\ln\frac{-i\omega}{\lambda_{\rm IR}}.
\end{equation}

For an infrared-singular TDOS,
\[
\rho(\lambda)=A\lambda^\alpha,
\qquad
-1<\alpha<0,
\]
the static polarization is dominated by the infrared cutoff,
\[
\Pi^R(0;T)
\simeq
A\int_T^\Lambda d\lambda\,\lambda^{\alpha-1}
=
\frac{A}{|\alpha|}
\left(
T^\alpha-\Lambda^\alpha
\right)
\simeq
\frac{A}{|\alpha|}T^\alpha .
\]
Thus the Thouless condition yields an algebraic transition scale.
The corresponding dynamical kernel exhibits the scaling form
\[
\Gamma_{\rm dyn}^R(\omega)
=
\Pi^R(0)-\Pi^R(\omega)
\propto
(-i\omega)^\alpha
\]
within the scaling regime, up to cutoff-dependent real constants.
The complex power follows from the retarded analytic continuation
$\omega\rightarrow\omega+i0^+$, for which
\[
-i\omega
=
|\omega|
e^{-i\frac{\pi}{2}\operatorname{sgn}(\omega)}.
\]
Consequently,
\[
(-i\omega)^\alpha
=
|\omega|^\alpha
e^{-i\frac{\pi\alpha}{2}\operatorname{sgn}(\omega)},
\]
showing that the infrared kernel contains both reactive (real) and
dissipative (imaginary) contributions fixed by causal retarded
analyticity.

In conventional superconductors, the phase dynamics is governed
by an analytic low-frequency kernel $\Gamma(\omega)\sim\omega^2$.
In neutral systems, this leads to a propagating
Anderson--Bogoliubov mode.
In charged superconductors, however, coupling to the electromagnetic
field lifts this mode to the plasma frequency via the Anderson--Higgs
mechanism, so that no gapless phase mode is observed in the low-energy
spectrum \cite{23,24,48,49,50}.

In contrast, in the present case the dynamical kernel
$\Gamma_{\rm dyn}^R(\omega)$ is intrinsically nonanalytic.
As a result, the phase dynamics does not generically support a
propagating solution of the form $\omega\sim q$.
Instead, the phase response is overdamped or critical, reflecting
the underlying continuum of relaxation processes.

Despite this unconventional dynamical structure, the equilibrium
electromagnetic response remains standard.
The supercurrent is
\begin{equation}
\mathbf j
=
2e\rho_s(\nabla\theta-2e\mathbf A),
\end{equation}
which for static configurations reduces to
\begin{equation}
\mathbf j
=
-4e^2\rho_s \mathbf A,
\end{equation}
recovering the London relation and the Meissner effect.
Thus, the equilibrium properties of the superconducting state
are controlled solely by the phase stiffness, while its dynamical
response is governed by the memory kernel.

This demonstrates that superconductivity does not rely on the
presence of a well-defined propagating low-energy collective mode.
Instead, it can arise in a regime where phase fluctuations are
intrinsically dissipative or critical, reflecting a continuum of
relaxation processes encoded in the Cooper kernel.

\begin{figure}[t]
\centering
\includegraphics[scale=0.45, trim=0cm 20cm 0cm 0cm]{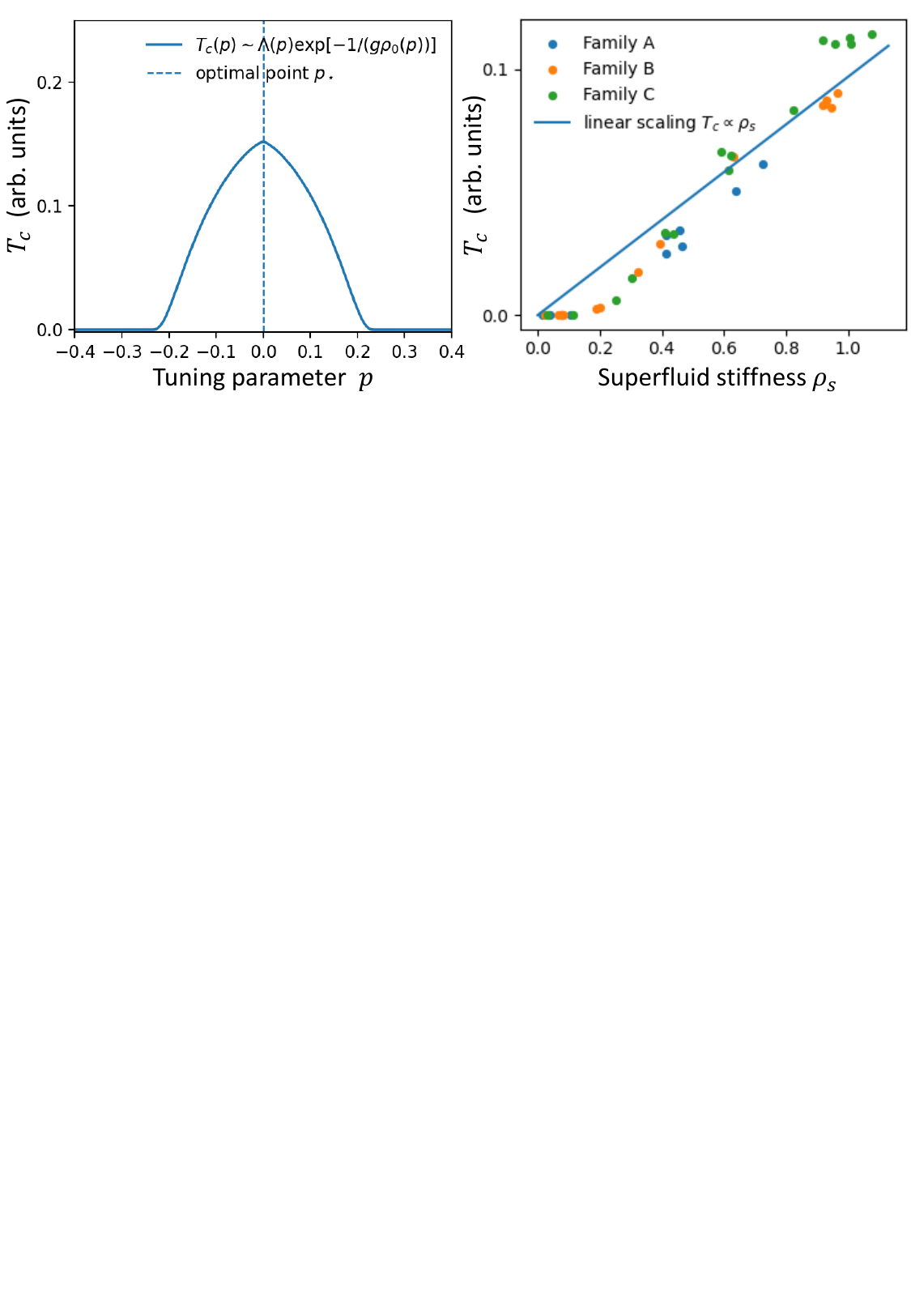}
\caption{Experimental consequences of memory--dominated critical dynamics.
(a) Schematic superconducting dome emerging from the infrared spectral
organization of the slow--mode reservoir.
The transition temperature follows
$T_c(p)\sim \Lambda(p)\exp\!\left(-\frac{1}{g\,\rho_0(p)}\right)$
and is maximal near $p_\star$, where the infrared slow--mode sector is
most fully developed.
In this regime, both the low-energy spectral weight and the infrared
spectral extent of the reservoir are enhanced, leading to a strengthened
pairing instability.
Away from optimality, the gradual weakening of the infrared slow--mode
organization suppresses superconductivity without invoking competing
orders or a specific bosonic mediator.
(b) Uemura scaling showing the proportionality between superconducting
transition temperature and superfluid stiffness, $T_c \propto \rho_s$,
in regimes where phase coherence limits the transition.
Both pairing and phase stiffness are influenced by the same infrared
spectral weight $\rho_0$, providing a common dynamical origin for the
observed scaling across material families.}
\label{fig:fig3}
\end{figure}

\section{V. Experimental consequences of memory--dominated critical dynamics}

The results of Secs.~II--IV imply that unconventional superconductivity in
strongly correlated systems is governed by a unifying organizing principle:
the infrared spectral reorganization of collective relaxation modes.
When the time--scale density of states develops a flat low--$\lambda$
sector, the system enters a memory--dominated critical regime
characterized by long--time kernels $K(t)\sim 1/t$ and a universal
infrared enhancement of the pairing susceptibility.

In this section we show that several central experimental signatures---the
superconducting dome, Uemura scaling, and anomalous long--time dynamics---follow
directly and generically from the same infrared spectral weight of the
slow--mode reservoir.
Figure~3 shows how infrared spectral reorganization of the slow--mode reservoir
naturally produces both the superconducting dome and Uemura scaling.

\subsection{A. Infrared spectral control and automatic emergence of the dome}

The analysis of Secs.~III and IV shows that the superconducting transition
temperature is governed by the infrared structure of the retarded Cooper
kernel, which in turn is controlled by the relaxation spectrum.
For a finite low--$\lambda$ TDOS, $\rho(\lambda\to0)=\rho_0$, the transition
scale follows the BCS-like form obtained in Eq.~(\ref{eq:Tc_flat_final}),
where $\Lambda$ denotes the effective infrared spectral extent of the
slow--mode reservoir.

We interpret this regime not as a finely tuned quantum critical point,
but as a self--organized dynamical state characterized by an extended
distribution of slow relaxation modes.
In this state, feedback between electronic correlations and collective
dynamics maintains a broad near-marginal sector of the TDOS,
\begin{equation}
\rho(\lambda\to0)=\rho_*(p),
\end{equation}
whose magnitude and infrared extent evolve with external control parameters
such as doping or pressure.

Let $p$ denote a tuning parameter and $p_\star$ the point at which the
slow--mode reservoir is most fully developed.
Deviations from $p_\star$ introduce a finite mass scale $r(p)$ that
progressively truncates the infrared sector of slow collective modes.
Accordingly, the effective infrared spectral extent decreases away from
optimality,
\begin{equation}
\Lambda(p)
\sim
\Lambda_0
-
c\,|p-p_\star|^{z\nu},
\end{equation}
where $\Lambda_0$ denotes the maximal infrared extent realized near
$p_\star$.

At the same time, the low-energy spectral weight evolves smoothly,
\begin{equation}
\rho_0(p)
\sim
\rho_*\,
f\!\left(\frac{|p-p_\star|}{p_0}\right),
\end{equation}
where $f$ is a monotonically decreasing crossover function satisfying
$f(0)=1$.

Substituting these dependencies into Eq.~(\ref{eq:Tc_flat_final}) yields
\begin{equation}
T_c(p)
\sim
\Lambda(p)\,
\exp\!\left(-\frac{1}{g\,\rho_0(p)}\right),
\label{eq:Tc_dome_final}
\end{equation}
which naturally produces a superconducting dome.

The superconducting dome reflects the gradual reorganization of the
infrared slow-mode reservoir across the phase diagram.
Near $p_\star$, the slow--mode reservoir is maximally extended,
leading to both a large infrared spectral weight $\rho_0$ and a large
infrared extent $\Lambda$, thereby enhancing the pairing instability.
Away from this regime, the infrared organization of slow relaxation modes
is progressively weakened, leading to a simultaneous reduction of the
infrared spectral weight and the effective infrared extent of the
slow--mode reservoir.
As a result, the pairing instability and superconducting transition
temperature are suppressed away from optimality.

In this interpretation, optimal doping does not correspond to a
fine-tuned symmetry-breaking quantum critical point.
Rather, it marks the regime in which the infrared spectral organization
of relaxation modes is most fully developed.
The superconducting dome therefore emerges naturally as a consequence of
infrared spectral reorganization, without requiring a specific bosonic
mediator or fine tuning of microscopic interactions.

\subsection{B. Phase stiffness and Uemura scaling from the slow--mode reservoir}

While Sec.~IV established that superconductivity arises from the
infrared divergence of the retarded Cooper kernel, the ordered state is
characterized not only by pairing correlations but also by its rigidity
against long-wavelength phase deformations.
This rigidity is quantified by the superfluid stiffness $\rho_s$,
which measures the energetic cost of imposing a spatial phase twist on
the condensate.

At long wavelengths, the superconducting phase sector is described by a
standard phase-only effective theory, with the corresponding ordering
scale satisfying
\begin{equation}
T_\theta \sim \rho_s.
\label{eq:Ttheta_safe}
\end{equation}
Within the present framework, however, the phase stiffness should not
be regarded as an independent mechanism separate from pairing.
The Cooper instability derived from the ladder resummation already
corresponds to a collective infrared ordering instability of the Cooper
channel, while $\rho_s$ characterizes the spatial and electromagnetic
rigidity of the resulting ordered state.

Both the Cooper instability and the condensate rigidity originate from
the same infrared spectral organization of slow collective modes.
In a memory-dominated regime, long-time phase coherence is governed by
the low-energy TDOS weight $\rho_0$, implying
\begin{equation}
\rho_s \propto \rho_0,
\end{equation}
up to material-dependent prefactors associated with quantities such as
effective mass and carrier density.
Consequently, the superconducting transition temperature naturally
tracks the superfluid density,
\begin{equation}
T_c \propto \rho_s ,
\label{eq:uemura_safe}
\end{equation}
consistent with the empirical Uemura scaling observed in cuprates and
other unconventional superconductors.

In conventional weak-coupling BCS superconductors, the superfluid
stiffness is typically much larger than the pairing scale, so that
phase fluctuations play only a minor role in determining $T_c$.
In strongly correlated systems, by contrast, both pairing enhancement
and condensate rigidity are controlled by the same infrared collective
dynamics.
The TDOS therefore acts as a unifying infrared structure linking
pairing susceptibility, phase rigidity, and macroscopic superconducting
coherence without requiring a specific microscopic bosonic pairing glue.

\subsection{C. Dynamical signatures: long--time correlations and anomalous dissipation}

A distinctive prediction of the present theory concerns dynamical response in
the normal state.
As shown in Sec.~II, a flat TDOS produces a marginal nonanalytic retarded
kernel with
\begin{equation}
\mathrm{Im}\,\chi_R^{-1}(\omega)\propto \mathrm{sgn}(\omega),
\end{equation}
corresponding in the time domain to a universal long--memory kernel
$K(t)\sim 1/t$.
This implies scale--free temporal correlations extending over broad time
windows.

Such long-time tails naturally produce \(1/f\)-type noise spectra and
anomalously slow relaxation, phenomena widely reported in strange metals and
cuprate superconductors.
Unlike conventional overdamped criticality, where correlations decay
exponentially and dynamics is effectively Markovian, the memory--dominated
regime predicts non-Markovian response with persistent temporal structure.
Because the same slow--mode reservoir controls both pairing enhancement and
dynamical response, superconducting and normal--state anomalies become two
manifestations of a single infrared spectral reorganization.
Related derivations are presented in
Appendix~G.

\medskip
\noindent
\emph{Unified infrared scaling picture.}
We emphasize that all experimental signatures discussed above---the dome,
Uemura scaling, and long--time dynamical correlations---originate from the same
infrared parameter $\rho_0$, the low--energy spectral weight of the slow--mode
reservoir.

The density of relaxation rates therefore plays a central organizational role.
It controls the infrared extent of the slow-mode reservoir, the
dynamical amplification of intrinsic Cooper-channel tendencies, the
superfluid stiffness and condensate rigidity, and the emergence of
non-Markovian long-time response.

Quantum criticality thus acts not merely as a source of fluctuations,
but as a dynamical infrared reorganizer of the collective spectrum.
Within this picture, thermodynamic ordering, phase rigidity, and
collective dynamical response all emerge from the same infrared
spectral organization encoded in the TDOS.

This unified picture yields a concrete strategy for experimental tests:
systems that exhibit enhanced low--frequency TDOS weight (or equivalently,
strong long--time correlations) are expected to display increased pairing
susceptibility, elevated $T_c$, and reduced phase stiffness consistent with
Uemura scaling.
The resulting real-frequency spectral function generated by the
memory-dominated self-energy, including its universal scaling form and infrared
structure, is derived in Appendix H.

\begin{figure}[t]
\centering
\includegraphics[scale=0.5, trim=1.0cm 17.4cm 0cm 0cm]{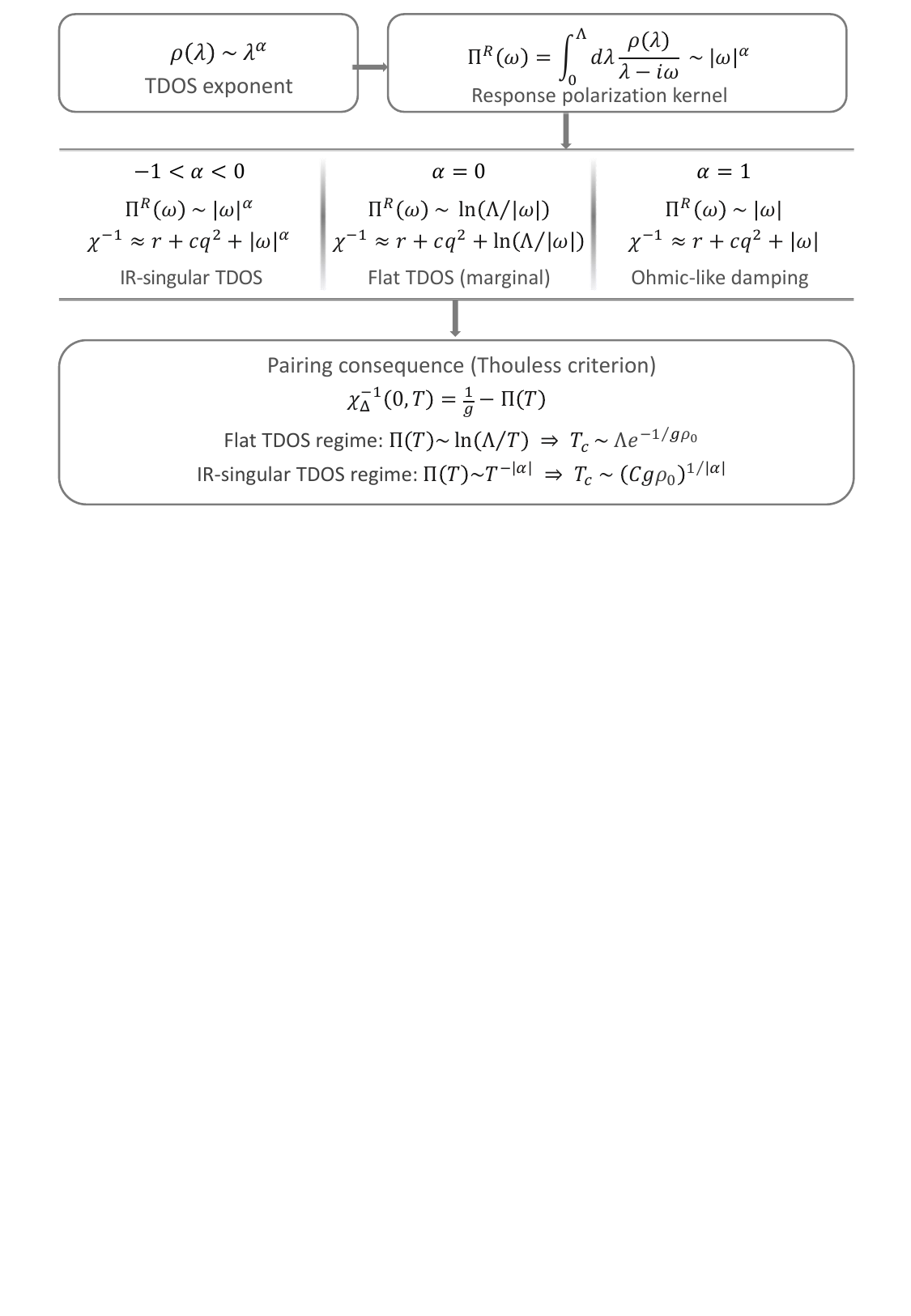}
\caption{
Dynamical classification determined by the TDOS exponent $\alpha$.
The scaling $\rho(\lambda)\sim\lambda^{\alpha}$ implies
$\Pi^R(\omega)\sim|\omega|^{\alpha}$ (with logarithmic behavior at $\alpha=0$),
and separates the system into infrared-singular ($\alpha<0$),
marginal ($\alpha=0$), Ohmic ($\alpha=1$),
and analytic ($\alpha>1$) regimes, with corresponding pairing consequences.
Here $\omega$ denotes frequency (Matsubara or real, depending on context).
}
\label{fig:fig4}
\end{figure}

\section{VI. Discussion}

The central result of this work is that infrared collective dynamics in
strongly correlated quantum matter is organized by the structure of its
relaxation-rate spectrum, rather than being described solely by the
softening of a single order parameter.
The time--scale density of states provides a quantitative measure
of this organization.
When the TDOS remains finite as $\lambda\to0$, the system enters a
memory--dominated regime characterized by long--time kernels
$K(t)\sim1/t$ and nonanalytic infrared response functions.

This viewpoint clarifies the limits of conventional coarse-grained
field theories.
Away from criticality, the relaxation spectrum is gapped in the infrared,
leading to finite memory times and justifying reduction to
local-in-time damping and a small number of collective fields.
Near correlated critical regimes, however, relaxation rates form a
dense continuum extending toward zero.
In this regime, the internal relaxation spectrum remains dynamically
relevant across a wide range of time scales and cannot be reduced
to a small set of collective modes.
The breakdown of Markovian coarse-graining is therefore not a
model-dependent artifact but a generic consequence of dynamical
criticality.

This observation highlights a structural difference from
Hertz--Millis–type approaches.
Conventional quantum critical theories assume overdamped local dynamics
of a small set of bosonic modes, with dissipation controlled by Landau
damping.
In contrast, the present framework allows for spectral condensation of
near-marginal relaxation channels.
Criticality is reinterpreted not as single-mode softening but as a
reorganization of the decay spectrum.
The TDOS therefore provides a microscopic spectral quantity governing
infrared universality, complementing phenomenological damping
descriptions.

This structure is summarized schematically in Fig.~4.
The figure illustrates how the infrared universality class of the
collective dynamics is determined by the exponent of the relaxation-rate
density of states.
For a power-law TDOS, $\rho(\lambda)\sim\lambda^\alpha$, the retarded
polarization kernel exhibits the infrared scaling
$\mathrm{Re}\,\Pi^R(\omega)\sim|\omega|^{\alpha}$ for $\alpha<1$,
with logarithmic behavior in the marginal case $\alpha=0$.
The inverse susceptibility is correspondingly given by
$\chi^{-1}(q,\omega)\simeq r+cq^2+\Pi^R(\omega)$,
so that different dynamical regimes emerge as special cases of a unified
spectral organization.
For $\alpha>1$, the density of slow modes vanishes in the infrared,
and the dynamical correction becomes subleading compared to the static
terms at low frequencies.
In this regime the collective dynamics approaches a weakly damped,
regular (analytic) limit, with the infrared response dominated by the
nondissipative contribution.
The case $\alpha=1$ corresponds to the Ohmic boundary.
In this case the kernel acquires a linear frequency dependence up to
logarithmic corrections, $\Pi^R(\omega)\sim |\omega|$,
reproducing the Hertz--Millis form
$\chi^{-1}\simeq r+cq^2+\gamma|\omega|$.
Most importantly, when the TDOS approaches a finite constant
($\alpha=0$), a dense continuum of near-marginal relaxation modes
emerges.
In this regime the polarization exhibits logarithmic infrared behavior,
$\mathrm{Re}\,\Pi^R(\omega)\sim \rho_0 \ln(\Lambda/|\omega|)$,
while the absorptive part remains finite at low frequencies.
This leads to long-time memory kernels
$K(t)\sim1/t$ and the non-Markovian infrared dynamics discussed
throughout this work.

Interestingly, the infrared response generated by a flat TDOS
is closely related to anomalous dynamical susceptibilities
that have appeared in several apparently unrelated contexts \cite{6,51}.
In particular, the resulting spectrum corresponds to a logarithmic real
part and a weakly frequency-dependent absorptive component of the susceptibility.
Similar infrared structures were introduced phenomenologically in the
marginal Fermi liquid theory of strange metals and also arise in
solvable strongly correlated spin models such as the Sachdev--Ye spin liquid.
In the present framework, however, this behavior follows directly from
the accumulation of slow relaxation modes and therefore emerges
naturally as a consequence of a finite infrared relaxation-rate
density of states.

The lower panel of Fig.~4 illustrates the consequence of this spectral
organization for superconductivity.
Within the Thouless criterion, the transition scale is controlled by
the infrared structure of the pairing kernel.
In the memory-dominated regime with a flat TDOS,
$\rho(\lambda\to0)=\rho_0$,
the retarded Cooper kernel develops logarithmic infrared enhancement,
$\Pi^R(0;T)\sim \rho_0\ln(\Lambda/T)$,
analogous to the familiar Cooper logarithm of conventional BCS theory.
The transition scale therefore retains a BCS-like exponential form,
$T_c\sim \Lambda \exp[-1/(g\rho_0)]$.
However, the infrared amplification is now governed not by coherent
fermionic quasiparticle shells alone, but by the spectral accumulation
of slow collective relaxation modes.
The effective pairing scale is therefore controlled directly by the
infrared spectral weight of the slow-mode reservoir.
Only when the TDOS becomes infrared-singular ($\alpha<0$),
$\rho(\lambda)\sim\lambda^\alpha$,
does the pairing kernel acquire algebraic infrared enhancement,
$\Pi^R(0;T)\sim T^\alpha$,
leading to power-law scaling of the transition temperature,
$T_c\sim g^{1/|\alpha|}$.
By contrast, when the TDOS becomes infrared-depleted ($\alpha>0$),
the low-energy spectral weight vanishes toward the infrared,
the static pairing kernel remains finite as $T\to0$,
and the infrared enhancement mechanism is lost.

A direct physical consequence of this spectral reorganization is the
enhancement of superconducting instabilities.
Superconductivity need not rely on finely tuned bosonic mediators.
Instead, intrinsic short-range electronic pairing tendencies are
dynamically amplified by the slow-mode reservoir.
In the marginal regime, the exponential suppression characteristic of
weak-coupling BCS theory is not removed but can be substantially reduced
due to enhanced infrared spectral weight.
Only when the TDOS becomes infrared-singular does the pairing scale
exhibit algebraic enhancement.
In this sense, superconducting order is governed by condensation in a
Cooper sector embedded within an extended slow-mode manifold rather than
by binding through a sharp bosonic excitation.

A single infrared parameter, $\rho_0$, plays a central role in both
thermodynamic and dynamical phenomena.
The superconducting dome can be interpreted as reflecting truncation
of long-time memory away from the optimal spectral reorganization point;
Uemura scaling follows from the shared spectral origin of pairing and
phase coherence; and anomalous normal-state dynamics arises from the
same non-Markovian kernel.
Quantum criticality therefore may act as a dynamical amplifier linking
transport, coherence, and pairing within a unified spectral framework.

\medskip
\noindent
\emph{Temporal scaling and dynamical criticality.}
An important conceptual aspect of the present framework concerns the
role of temporal scaling in dynamical criticality.
In conventional renormalization-group treatments of equilibrium
critical phenomena, spatial scale invariance provides the primary
structure, while temporal scaling emerges through the dynamical exponent
$z$, relating relaxation time and correlation length via
$\tau\sim\xi^z$.
The Hertz--Millis theory of quantum critical metals follows a similar
philosophy, in which spatial critical fluctuations generate an
effective dynamical scaling once fermionic degrees of freedom are
integrated out.

The present relaxation-spectrum formulation does not discard this
spatial critical structure.
Rather, it emphasizes that quantum critical slowing down reorganizes
the infrared dynamics into a broad continuum of relaxation processes.
When the relaxation-rate spectrum develops substantial low-energy
weight, the system acquires a continuum of slow modes extending over
arbitrarily long time scales.
The resulting dynamics naturally exhibits scale-free temporal behavior,
including long-time memory kernels and nonanalytic infrared response
functions.

From this viewpoint, temporal scale invariance emerges through the
infrared organization of the relaxation spectrum itself.
The low-energy TDOS therefore provides a natural dynamical language for
describing quantum critical relaxation and the associated infrared
pairing enhancement.

\medskip
\noindent
\emph{Memory--dominated dynamics and prospects for high-temperature superconductivity.}
Within the present framework, superconducting transition temperatures
are controlled not only by the microscopic pairing interaction but also
by the infrared dynamical organization of collective modes encoded in
the relaxation-rate density of states.

In the marginal regime corresponding to a flat TDOS ($\alpha=0$),
the retarded pairing kernel exhibits logarithmic infrared enhancement,
leading to a BCS-like transition scale, $T_c\sim \Lambda \exp[-1/(g\rho_0)]$,
where $\Lambda$ is an electronic cutoff scale and $\rho_0$ is the
infrared spectral weight of slow collective modes.
Because $\Lambda$ is set by electronic energy scales, 
this form can naturally accommodate transition temperatures in the range of
$10^2$~K for moderate coupling, providing a possible framework
for understanding high-$T_c$ superconductivity without relying on finely tuned bosonic mediators.

Further enhancement of the transition scale may occur if the
relaxation spectrum develops an infrared-singular structure
($\alpha<0$), corresponding to an accumulation of slow modes
toward vanishing relaxation rates.
In this regime, the pairing kernel acquires a power-law infrared
enhancement, and the transition temperature follows an algebraic
scaling with the effective interaction strength.

These considerations suggest a complementary perspective on
superconductivity design:
beyond increasing microscopic pairing interactions,
controlling the infrared organization of collective dynamics
may provide an effective route toward higher transition temperatures \cite{52}.
In this sense, dynamical phases supporting enhanced slow-mode
spectral weight represent promising candidates for realizing
superconductivity at elevated temperatures.

\section{VII. Conclusion}

We have identified dynamical spectral organization as a fundamental
principle governing infrared behavior in strongly correlated quantum matter.
By formulating collective dynamics in terms of the time--scale density
of states, we show that criticality is generically controlled not
by local Markovian dissipation but by the accumulation of slow collective
relaxation modes.

A finite infrared TDOS defines a memory--dominated universality class
characterized by long--time kernels and nonanalytic
dynamical response.
Within this regime, intrinsic electronic pairing tendencies are
dynamically enhanced, leading to logarithmic infrared amplification
of the retarded pairing kernel.
While the resulting transition scale retains a BCS-like exponential form,
it is controlled by the infrared spectral weight of slow collective modes.
Algebraic enhancement of the transition temperature arises only when the
TDOS becomes infrared-singular ($\alpha<0$), corresponding to an
accumulation of slow modes toward vanishing relaxation rates.

This mechanism provides a unified perspective on key experimental phenomena
in unconventional superconductors, including superconducting domes,
Uemura scaling of transition temperatures with superfluid stiffness,
and anomalous long--time dynamical correlations in the normal state.
These signatures arise naturally from infrared spectral reorganization,
without necessarily requiring specific bosonic mediators or material-dependent fine tuning.

More broadly, the results recast quantum criticality as a dynamical process
that enhances collective quantum phenomena through extended temporal
correlations.
Memory--dominated criticality therefore represents a universality class
that extends beyond traditional dissipative field theories, with implications
for non-Fermi liquids, glassy dynamics, and engineered quantum platforms.

In this perspective, high--temperature superconductivity can be understood
as arising from favorable infrared spectral organization in strongly
correlated matter.
Systematic exploration of TDOS-controlled dynamics may provide a pathway
toward stabilizing and enhancing collective quantum states in future
materials and devices.

\vspace{6pt}
\emph{Acknowledgements}---This work was partially supported by the Institute of Information \& Communications Technology Planning \& Evaluation (IITP) grant 
funded by the Korea government (MSIT) (IITP-RS-2025-02214780).

\vspace{3pt}
The author acknowledges the support of ChatGPT (GPT-5, OpenAI) for assistance in literature review and conceptual structuring during development.

\clearpage
\appendix

\renewcommand{\thefigure}{S\arabic{figure}}
\renewcommand{\theequation}{S\arabic{equation}}

\setcounter{figure}{0}
\setcounter{equation}{0}

\vspace*{1.5cm}
{\centering\large\bfseries Supplementary Materials\par}
\vspace{1.0cm}

\appendix

\section{Appendix A: Relaxation-rate spectra as the fundamental dynamical basis of collective dynamics}

In conventional quantum many-body theory, collective behavior is often
organized in terms of energy eigenmodes of an underlying Hamiltonian,
\begin{equation}
H|n\rangle = E_n |n\rangle,
\end{equation}
with dynamical response expressed through oscillatory Fourier components at
frequencies set by energy differences.

However, this description represents a special limit.
In generic interacting systems, particularly in the presence of dissipation,
coarse-graining, or environmental coupling, the long-time dynamics is
governed not by energy eigenmodes but by the spectrum of the dynamical
generator that controls relaxation.
This structure is most naturally formulated in terms of the Liouvillian
operator.

\vspace{10pt}
\noindent
\emph{(I) Liouvillian formulation (closed, open, and non-Markovian dynamics).}

A conceptually complete formulation of nonequilibrium quantum dynamics is given
in terms of the density matrix $\hat{\varrho}(t)$ and its dynamical generator.
For a closed system, $\hat{\varrho}(t)$ evolves by the von~Neumann equation,
\begin{equation}
\frac{d\hat{\varrho}}{dt}=-i[H,\hat{\varrho}]\equiv \mathcal L_H\hat{\varrho},
\end{equation}
where $\mathcal L_H$ is the Hamiltonian Liouvillian superoperator.
If $H|n\rangle=E_n|n\rangle$, then the operator basis $|n\rangle\langle m|$
diagonalizes $\mathcal L_H$:
\begin{equation}
\mathcal L_H\big(|n\rangle\langle m|\big)
=
-i(E_n-E_m)\,|n\rangle\langle m|,
\end{equation}
so the Liouvillian eigenvalues are
\begin{equation}
\lambda_{nm}=-i(E_n-E_m),
\end{equation}
lying purely on the imaginary axis and reproducing undamped oscillations.

For open quantum systems, the reduced density matrix typically obeys a
master equation with a dissipative contribution.
In the Markovian limit, the evolution takes Lindblad form,
\begin{equation}
\frac{d\hat{\varrho}}{dt}
=
\mathcal L\hat{\varrho}
=
-i[H,\hat{\varrho}]
+
\sum_\mu
\left(
L_\mu\hat{\varrho} L_\mu^\dagger
-\frac{1}{2}\{L_\mu^\dagger L_\mu,\hat{\varrho}\}
\right),
\label{eq:lindblad_appF}
\end{equation}
where the jump operators $L_\mu$ encode decoherence and dissipation.
The resulting Liouvillian $\mathcal L$ is generically non-Hermitian as a
superoperator and therefore possesses complex eigenvalues
\begin{equation}
\mathcal L R_\alpha=\lambda_\alpha R_\alpha,
\qquad
\mathrm{Re}\,\lambda_\alpha\le 0,
\end{equation}
with decay rates set by $-\mathrm{Re}\,\lambda_\alpha$.
The density matrix admits the modal expansion
\begin{equation}
\hat{\varrho}(t)=\hat{\varrho}_{\rm ss}+\sum_{\alpha\ne 0} c_\alpha\,e^{\lambda_\alpha t}\,R_\alpha,
\end{equation}
where $\hat{\varrho}_{\rm ss}$ is the stationary state ($\lambda_0=0$) and the
long-time approach is controlled by the Liouvillian spectral gap and any
near-marginal continuum.

More generally, when the environment or eliminated internal coordinates retain
memory, the reduced dynamics becomes non-Markovian and takes a memory-kernel
form (Nakajima--Zwanzig structure),
\begin{equation}
\frac{d\hat{\varrho}}{dt}
=
\int_0^t ds\;\mathcal K(t-s)\,\hat{\varrho}(s),
\label{eq:NZ_appF}
\end{equation}
where $\mathcal K(t)$ is a causal superoperator kernel.
In Laplace space,
\begin{equation}
\tilde{\hat{\varrho}}(s)
=
\big[s-\tilde{\mathcal K}(s)\big]^{-1}\hat{\varrho}(0),
\end{equation}
so the poles and branch cuts of $s-\tilde{\mathcal K}(s)$ define the intrinsic
relaxation spectrum.
In this sense, even for quantum systems the relevant organizing structure for
long-time dynamics is generically a \emph{relaxation-rate spectrum} (possibly
continuous), not an energy spectrum.

\vspace{10pt}
\noindent
\emph{(II) Memory kernels and continuous relaxation spectra.}

The same spectral logic appears already at the level of linear causal dynamics
for a collective coordinate $\phi(t)$:
\begin{equation}
\dot\phi(t)+\int_0^t ds\,K(t-s)\phi(s)=\eta(t).
\end{equation}
Taking the Laplace transform gives
\begin{equation}
\tilde\phi(s)=\frac{\tilde\eta(s)}{s+\tilde K(s)}.
\end{equation}
The analytic structure of $s+\tilde K(s)$ determines the intrinsic relaxation
content: isolated poles correspond to discrete decay modes, while branch cuts
encode continua of decay rates.
Whenever long-time memory is present, the causal kernel admits a spectral
representation of the form
\begin{equation}
K(t)=\int_0^\infty d\lambda\,\rho(\lambda)e^{-\lambda t},
\end{equation}
which defines the time-scale spectrum.
A flat infrared TDOS produces the universal tail $K(t)\sim 1/t$.

Thus continuous relaxation spectra arise generically from causality and memory,
providing the microscopic foundation for the time-scale density of states
introduced in the main text.

\vspace{10pt}
\noindent
\emph{(III) Energy modes as a special limit.}

The familiar Fourier-mode picture of conservative dynamics corresponds to the
singular limit in which the relevant spectrum collapses onto the imaginary axis.
In the closed-system Liouvillian case this appears as
$\lambda_{nm}=-i(E_n-E_m)$, while in the memory-kernel representation it
corresponds to eliminating all decay rates (no weight at $\mathrm{Re}\,\lambda>0$),
recovering purely oscillatory behavior.
Away from this limit, generic interacting and/or open systems naturally develop
broad spectra of relaxation rates controlling infrared dynamics.

\vspace{10pt}
\noindent
\emph{Physical implication.}
The TDOS---the infrared distribution of collective relaxation rates---
is therefore the fundamental organizing variable of collective dynamics in
correlated systems.
Energy eigenmodes describe an integrable, conservative limit, whereas
relaxation-rate spectra capture the universal structure of irreversible,
critical, and strongly interacting matter.

Memory-dominated criticality corresponds precisely to the formation of an
extensive continuum of near-marginal relaxation modes.
The resulting TDOS controls dynamical response, pairing enhancement, and
infrared universality, providing a more general dynamical foundation than energy-based descriptions.


\section{Appendix B: Gaussian MSRJD integration of a relaxation-mode reservoir and emergence of the self-energy}
\label{app:msrjd_gaussian_selfenergy}

In this appendix we derive the emergence of a nonlocal memory kernel
and the associated frequency-dependent self-energy $\Sigma(\omega)$
by integrating out a continuum of relaxational
(Ornstein--Uhlenbeck) modes within the MSRJD formalism.

The construction provides a controlled Gaussian realization of the
relaxation-spectrum representation used in the main text, in which
the infrared dynamics is governed by a continuum of decay modes rather
than by a small number of quasiparticle excitations.

\vspace{10pt}
\noindent
\emph{(I) Coupled stochastic dynamics.}

We consider a collective degree of freedom $\phi(t)$ coupled linearly to a
continuum of reservoir modes $X_\lambda(t)$ labeled by their relaxation rates
$\lambda>0$.  The coupled stochastic dynamics is
\begin{align}
\dot \phi(t)
&=
-r\,\phi(t)
+
\int d\lambda\; g(\lambda)\,X_\lambda(t)
+
\eta(t),
\label{eq:app_phi_sde}
\\
\dot X_\lambda(t)
&=
-\lambda\,X_\lambda(t)
+
g(\lambda)\,\phi(t)
+
\xi_\lambda(t).
\label{eq:app_X_sde}
\end{align}
The noises are assumed Gaussian, stationary, and white:
\begin{align}
\langle \eta(t)\eta(t')\rangle
&=
2D_\phi\,\delta(t-t'),
\label{eq:app_eta_corr}
\\
\langle \xi_\lambda(t)\,\xi_{\lambda'}(t')\rangle
&=
2D_\lambda\,\delta(\lambda-\lambda')\,\delta(t-t').
\label{eq:app_xi_corr}
\end{align}
No equilibrium fluctuation--dissipation relation is assumed; $D_\lambda$ may be
arbitrary.  Our goal is to integrate out $\{X_\lambda\}$ and obtain an exact
effective MSRJD action for $(\phi,\tilde\phi)$ containing a memory kernel.

\vspace{10pt}
\noindent
\emph{(II) MSRJD action for the coupled system.}

For a generic additive-noise Langevin equation $\dot y=f(y)+\zeta$ with
$\langle \zeta(t)\zeta(t')\rangle=2D\,\delta(t-t')$, the MSRJD functional
integral may be written (Ito convention) as
\begin{equation}
\label{eq:app_generic_msrjd}
\begin{aligned}
Z
&=
\int \mathcal Dy\,\mathcal D\tilde y\;
\\
&\exp\!\left[
-\int dt\;\tilde y(t)\big(\dot y(t)-f(y(t))\big)
+\int dt\;D\,\tilde y(t)^2
\right],
\end{aligned}
\end{equation}
up to an overall normalization independent of $y$.
Applying \eqref{eq:app_generic_msrjd} to \eqref{eq:app_phi_sde}--\eqref{eq:app_X_sde}
gives
\begin{equation}
Z
=
\int \mathcal D\phi\,\mathcal D\tilde \phi\;
\prod_\lambda \mathcal D X_\lambda\,\mathcal D\tilde X_\lambda\;
e^{-S[\phi,\tilde \phi,\{X_\lambda,\tilde X_\lambda\}]},
\label{eq:app_Z_full}
\end{equation}
with total action
\begin{equation}
S
=
S_\phi
+
\int d\lambda\; S_\lambda,
\label{eq:app_S_total}
\end{equation}
where
\begin{align}
S_\phi
&=
\int dt\;
\tilde \phi(t)\!\left[
\dot \phi(t)+r\,\phi(t)-\int d\lambda\; g(\lambda)\,X_\lambda(t)
\right]
\nonumber\\
&\quad -
\int dt\; D_\phi\,\tilde \phi(t)^2,
\label{eq:app_Sphi}
\\
S_\lambda
&=
\int dt\;
\tilde X_\lambda(t)\!\left[
\dot X_\lambda(t)+\lambda\,X_\lambda(t)-g(\lambda)\,\phi(t)
\right]
\nonumber\\
&\quad
-
\int dt\; D_\lambda\,\tilde X_\lambda(t)^2.
\label{eq:app_Slambda}
\end{align}

\vspace{10pt}
\noindent
\emph{(III) Isolating a single $\lambda$-mode contribution.}

Fix $\lambda$ and collect all terms in $S$ that contain $X_\lambda$ or
$\tilde X_\lambda$.  From \eqref{eq:app_Sphi} we extract
\begin{equation}
S_\phi \supset
-\int dt\;\tilde \phi(t)\,g(\lambda)\,X_\lambda(t).
\label{eq:app_Sphi_coupling}
\end{equation}
Combining \eqref{eq:app_Sphi_coupling} with \eqref{eq:app_Slambda} yields the
$\lambda$-sector action
\begin{equation}
\begin{aligned}
&S^{(\lambda)}[X_\lambda,\tilde X_\lambda; \phi,\tilde \phi]
=
\int dt\;
\tilde X_\lambda(t)\big(\dot X_\lambda(t)+\lambda X_\lambda(t)\big)
\\
&\quad-\int dt\;g(\lambda)\,\tilde X_\lambda(t)\,\phi(t)
-\int dt\;g(\lambda)\,\tilde \phi(t)\,X_\lambda(t)
\\
&\quad-\int dt\;D_\lambda\,\tilde X_\lambda(t)^2.
\label{eq:app_S_singlelambda_raw}
\end{aligned}
\end{equation}
In what follows, $\phi,\tilde \phi$ are treated as external sources.

\vspace{10pt}
\noindent
\emph{(IV) Functional integration over $X_\lambda$.}

Define the linear operator
\begin{equation}
\mathcal L_\lambda \equiv \partial_t+\lambda.
\label{eq:app_L_def}
\end{equation}
Then the first term in \eqref{eq:app_S_singlelambda_raw} is
\begin{equation}
\int dt\;\tilde X_\lambda(t)\,\mathcal L_\lambda X_\lambda(t)
=
\int dt\;\tilde X_\lambda(t)\big(\partial_t X_\lambda(t)+\lambda X_\lambda(t)\big).
\label{eq:app_tildeX_L_X}
\end{equation}
We now rewrite this term so that $X_\lambda$ appears without derivatives.
Using integration by parts,
\begin{equation}
\begin{aligned}
&\int dt\;\tilde X_\lambda(t)\,\partial_t X_\lambda(t)
\\
&\quad =
\big[\tilde X_\lambda(t)\,X_\lambda(t)\big]_{t_i}^{t_f}
-
\int dt\;(\partial_t \tilde X_\lambda(t))\,X_\lambda(t).
\end{aligned}
\end{equation}
Assuming boundary terms vanish (e.g.\ $t_i\to-\infty$, $t_f\to+\infty$ with
sufficient decay, or appropriate causal boundary conditions), we drop the
bracketed term and obtain
\begin{equation}
\int dt\;\tilde X_\lambda(t)\,\partial_t X_\lambda(t)
=
-\int dt\;(\partial_t \tilde X_\lambda(t))\,X_\lambda(t).
\label{eq:app_ibp2}
\end{equation}
Therefore,
\begin{align}
&\int dt\;\tilde X_\lambda(t)\,\mathcal L_\lambda X_\lambda(t)
=
-\int dt\;(\partial_t \tilde X_\lambda(t))\,X_\lambda(t)\\
&+\int dt\;\lambda\,\tilde X_\lambda(t)\,X_\lambda(t)
\nonumber=
\int dt\;X_\lambda(t)\big(-\partial_t+\lambda\big)\tilde X_\lambda(t).
\label{eq:app_Ladjoint}
\end{align}
Define the adjoint operator
\begin{equation}
\mathcal L_\lambda^\dagger \equiv -\partial_t+\lambda.
\label{eq:app_Ldagger_def}
\end{equation}
Then \eqref{eq:app_S_singlelambda_raw} becomes
\begin{equation}
\begin{aligned}
S^{(\lambda)}
&=
\int dt\;X_\lambda(t)\Big(\mathcal L_\lambda^\dagger \tilde X_\lambda(t)-g(\lambda)\tilde \phi(t)\Big)
\\
&\quad
-\int dt\;g(\lambda)\,\tilde X_\lambda(t)\,\phi(t)
-\int dt\;D_\lambda\,\tilde X_\lambda(t)^2.
\label{eq:app_S_singlelambda_rewritten}
\end{aligned}
\end{equation}
Now $X_\lambda$ enters \emph{only linearly}.  Hence its functional integral
produces a functional delta constraint:
\begin{align}
\int \mathcal D X_\lambda\;
\exp\!\left\{
-\int dt\;X_\lambda(t)\,A(t)
\right\}
&\propto
\delta[A(t)],
\label{eq:app_delta_identity}
\end{align}
where here
\begin{equation}
A(t) \equiv \mathcal L_\lambda^\dagger \tilde X_\lambda(t)-g(\lambda)\tilde \phi(t).
\label{eq:app_A_def}
\end{equation}
Therefore,
\begin{equation}
\begin{aligned}
&\int \mathcal D X_\lambda\;e^{-S^{(\lambda)}}
\propto
\int \mathcal D\tilde X_\lambda\;
\delta\!\left(\mathcal L_\lambda^\dagger \tilde X_\lambda-g(\lambda)\tilde \phi\right)\,
\\
&\quad \times 
\exp\!\left[
+\int dt\;g(\lambda)\,\tilde X_\lambda(t)\,\phi(t)
-\int dt\;D_\lambda\,\tilde X_\lambda(t)^2
\right].
\label{eq:app_after_X_integration}
\end{aligned}
\end{equation}

\noindent
The Jacobian factor associated with the functional constraint,
$|\det \mathcal L_\lambda^\dagger|^{-1}$, is independent of $\phi,\tilde \phi$ and can
therefore be absorbed into the overall normalization of $Z$.

\vspace{10pt}
\noindent
\emph{(V) Solving the delta constraint for $\tilde X_\lambda$ via Green functions.}

The constraint
\begin{equation}
\mathcal L_\lambda^\dagger \tilde X_\lambda(t) = g(\lambda)\tilde \phi(t),
\qquad
\mathcal L_\lambda^\dagger = -\partial_t+\lambda,
\label{eq:app_constraint_eq}
\end{equation}
is a linear inhomogeneous ODE for $\tilde X_\lambda(t)$.
Introduce the Green function $G_\lambda^{A}(t-t')$ solving
\begin{equation}
(-\partial_t+\lambda)\,G_\lambda^{A}(t-t')=\delta(t-t').
\label{eq:app_greenA_def}
\end{equation}

\noindent
Because the constraint is written with the \emph{adjoint} operator
$\mathcal L_\lambda^\dagger=-\partial_t+\lambda$, its Green function carries
advanced support (equivalently, the retarded kernel will reappear after
relabelling, ensuring causality of the $\phi$-sector response).

The solution with advanced support is
\begin{equation}
G_\lambda^{A}(t)=\Theta(-t)\,e^{\lambda t}.
\label{eq:app_greenA_solution}
\end{equation}
(Indeed, for $t\neq 0$, $(-\partial_t+\lambda)G_\lambda^A=0$ and the jump at
$t=0$ produces the delta distribution.)
Multiplying \eqref{eq:app_constraint_eq} by $G_\lambda^{A}$ and integrating,
we obtain
\begin{align}
\tilde X_\lambda(t)
&=
g(\lambda)\int dt'\;G_\lambda^{A}(t-t')\,\tilde \phi(t').
\label{eq:app_tildeX_solution}
\end{align}
Substituting \eqref{eq:app_tildeX_solution} into \eqref{eq:app_after_X_integration}
eliminates $\tilde X_\lambda$ entirely and yields an effective contribution to
the $(\phi,\tilde \phi)$ action.

\vspace{10pt}
\noindent
\emph{(VI) Effective action: memory kernel (self-energy) and induced colored noise.}

We now evaluate the two remaining terms in the exponent of
\eqref{eq:app_after_X_integration} using \eqref{eq:app_tildeX_solution}.
Start from
\begin{equation}
\int dt\;g(\lambda)\,\tilde X_\lambda(t)\,\phi(t).
\label{eq:app_term1_start}
\end{equation}
Insert \eqref{eq:app_tildeX_solution}:
\begin{align}
\int dt\;g(\lambda)\,\tilde X_\lambda(t)\,\phi(t)
=
\int dt\,dt'\;g(\lambda)^2\,G_\lambda^{A}(t-t')\,\tilde \phi(t')\,\phi(t).
\label{eq:app_term1_mid}
\end{align}
Relabel integration variables $(t,t')\to(t',t)$:
\begin{equation}
\int dt\,dt'\;g(\lambda)^2\,G_\lambda^{A}(t'-t)\,\tilde \phi(t)\,\phi(t').
\label{eq:app_term1_relabel}
\end{equation}
Using the identity $G_\lambda^{A}(t'-t)=G_\lambda^{R}(t-t')$ with the retarded
Green function
\begin{equation}
G_\lambda^{R}(t)\equiv \Theta(t)\,e^{-\lambda t},
\qquad
(\partial_t+\lambda)G_\lambda^{R}(t)=\delta(t),
\label{eq:app_greenR_def}
\end{equation}
we obtain
\begin{align}
\int dt\;g(\lambda)\,\tilde X_\lambda(t)\,\phi(t)
&=
\int dt\,dt'\;\tilde \phi(t)\,\underbrace{g(\lambda)^2\,G_\lambda^{R}(t-t')}_{\equiv K_\lambda(t-t')}\,\phi(t').
\label{eq:app_memory_kernel_single}
\end{align}
Thus a single OU mode contributes the causal memory kernel
\begin{equation}
K_\lambda(t-t') = g(\lambda)^2\,\Theta(t-t')\,e^{-\lambda (t-t')}.
\label{eq:app_Klambda_time}
\end{equation}

Next consider
\begin{equation}
\int dt\;D_\lambda\,\tilde X_\lambda(t)^2.
\label{eq:app_term2_start}
\end{equation}
Insert \eqref{eq:app_tildeX_solution}:
\begin{align}
&\int dt\;D_\lambda\,\tilde X_\lambda(t)^2
\nonumber\\
&=
\int dt\,dt'\,dt''\;D_\lambda\,g(\lambda)^2\,
G_\lambda^{A}(t-t')\,G_\lambda^{A}(t-t'')\,
\tilde \phi(t')\,\tilde \phi(t'').
\label{eq:app_term2_mid}
\end{align}
Define the induced (generally nonlocal) noise kernel
\begin{equation}
\mathcal N_\lambda(t',t'')
\equiv
D_\lambda\,g(\lambda)^2
\int dt\;G_\lambda^{A}(t-t')\,G_\lambda^{A}(t-t''),
\label{eq:app_Nlambda_def}
\end{equation}
so that
\begin{equation}
\int dt\;D_\lambda\,\tilde X_\lambda(t)^2
=
\int dt'\,dt''\;\tilde \phi(t')\,\mathcal N_\lambda(t',t'')\,\tilde \phi(t'').
\label{eq:app_term2_final}
\end{equation}
In equilibrium one may impose a fluctuation--dissipation relation linking
$\mathcal N_\lambda$ and $K_\lambda$, but this is not required for the
deterministic self-energy structure derived below.

Combining \eqref{eq:app_memory_kernel_single} and \eqref{eq:app_term2_final} for
all $\lambda$ yields the exact effective action
\begin{align}
&S_{\mathrm{eff}}[\phi,\tilde \phi]
=
\int dt\;\tilde \phi(t)\big(\dot \phi(t)+r \phi(t)\big)
-\int dt\;D_\phi\,\tilde \phi(t)^2
\nonumber\\
&
-\int dt\,dt'\;\tilde \phi(t)\,K(t-t')\,\phi(t')
-\int dt\,dt'\;\tilde \phi(t)\,\mathcal N(t,t')\,\tilde \phi(t'),
\label{eq:app_Seff_time}
\end{align}
where the total memory kernel is the reservoir sum
\begin{equation}
K(t-t')
=
\int d\lambda\;K_\lambda(t-t')
=
\int d\lambda\;g(\lambda)^2\,\Theta(t-t')\,e^{-\lambda (t-t')},
\label{eq:app_K_total_time}
\end{equation}
and likewise $\mathcal N=\int d\lambda\,\mathcal N_\lambda$.

\vspace{10pt}
\noindent
\emph{(VII) Frequency-domain self-energy $\Sigma(\omega)$ from the memory kernel.}

To connect \eqref{eq:app_Seff_time} to the standard response-function form, we
Fourier transform using
\begin{equation}
f(\omega)=\int_{-\infty}^{\infty} dt\;e^{i\omega t}\,f(t),
\qquad
f(t)=\int \frac{d\omega}{2\pi}\;e^{-i\omega t}\,f(\omega).
\label{eq:app_fourier_conv}
\end{equation}
The bilinear term involving $K$ becomes
\begin{align}
\int dt\,dt'\;\tilde \phi(t)\,K(t-t')\,\phi(t')
&=
\int \frac{d\omega}{2\pi}\;\tilde \phi(-\omega)\,K(\omega)\,\phi(\omega),
\label{eq:app_K_bilinear_freq}
\end{align}
where $K(\omega)$ is the Fourier transform of the causal kernel \eqref{eq:app_K_total_time}:
\begin{align}
K(\omega)
&=
\int_{-\infty}^{\infty} dt\;e^{i\omega t}\,K(t)
=
\int d\lambda\;g(\lambda)^2\int_{0}^{\infty} dt\;e^{i\omega t}e^{-\lambda t}.
\label{eq:app_Komega_start}
\end{align}
The time integral is elementary:
\begin{align}
\int_{0}^{\infty} dt\;e^{-(\lambda-i\omega)t}
&=
\left[
\frac{-1}{\lambda-i\omega}e^{-(\lambda-i\omega)t}
\right]_{0}^{\infty}
=
\frac{1}{\lambda-i\omega},
\nonumber\\
&\qquad
(\mathrm{Re}\,\lambda>0).
\label{eq:app_time_integral}
\end{align}
Substituting \eqref{eq:app_time_integral} into \eqref{eq:app_Komega_start} gives
the exact spectral representation
\begin{equation}
K(\omega)
=
\int d\lambda\;\frac{g(\lambda)^2}{\lambda-i\omega}.
\label{eq:app_Komega_spectral}
\end{equation}
In the MSRJD response sector, the retarded inverse propagator for $\phi$ is read
off from the $\tilde \phi(-\omega)\phi(\omega)$ coefficient in $S_{\mathrm{eff}}$:
\begin{equation}
G_R^{-1}(\omega)= -i\omega + r + K(\omega).
\label{eq:app_GR_inverse}
\end{equation}
It is therefore natural to define the (retarded) self-energy as
\begin{equation}
\Sigma_R(\omega)\equiv K(\omega)
=
\int d\lambda\;\frac{g(\lambda)^2}{\lambda-i\omega}.
\label{eq:app_selfenergy}
\end{equation}

\noindent
Our convention is $\Sigma_R(\omega)\equiv K(\omega)$ so that
$G_R^{-1}(\omega)=-i\omega+r+\Sigma_R(\omega)$; other sign conventions for the
self-energy appear in the literature.

Equations \eqref{eq:app_Komega_spectral}--\eqref{eq:app_selfenergy} demonstrate
that integrating out a continuum of OU relaxation modes produces a frequency-
dependent self-energy whose analytic structure is determined by the
relaxation-rate spectrum and couplings.

\vspace{10pt}
\noindent
\emph{(VIII) Connection to TDOS (relaxation-rate spectral weight).}

The spectral representation derived above shows that each relaxational
mode contributes to the retarded self-energy with pole
$(\lambda_\alpha-i\omega)^{-1}$ weighted by the square of its overlap
with the collective field.
Starting from the discrete mode expansion,
\begin{equation}
\Sigma_R(\omega)=\sum_\alpha \frac{g_\alpha^2}{\lambda_\alpha-i\omega},
\label{eq:app_sigma_discrete_modes}
\end{equation}
it is therefore natural to define the TDOS directly as the
coupling-weighted spectral distribution of relaxation rates,
\begin{equation}
\rho(\lambda)
\equiv
\sum_\alpha g_\alpha^2\,\delta(\lambda-\lambda_\alpha).
\label{eq:app_TDOS_direct_def}
\end{equation}

With this definition, the continuum form of the self-energy becomes
\begin{equation}
\Sigma_R(\omega)
=
\int d\lambda\;\frac{\rho(\lambda)}{\lambda-i\omega}.
\label{eq:app_sigma_tdos_direct}
\end{equation}
In this sense, $\rho(\lambda)$ plays a role directly analogous to the
Lehmann spectral function in Hamiltonian many-body theory.
For an operator $O$ with energy eigenstates $|n\rangle$, the usual
spectral representation involves weights
$|\langle0|O|n\rangle|^2$ multiplying poles at energies $E_n$.
Here, by contrast, the relevant poles are located at the relaxation
rates $\lambda_\alpha$, and the corresponding spectral weights are the
squared overlap amplitudes $g_\alpha^2$ between the collective field
and the relaxation eigenmodes.

Thus the TDOS should be understood not merely as a bare counting
density of modes, but as the physically observable spectral weight
distribution of the relaxation spectrum for the collective field
under consideration.
Equation~(\ref{eq:app_sigma_tdos_direct}) makes explicit, at the level
of a controlled Gaussian MSRJD integration, why the relaxation-rate
spectrum provides the natural basis for the effective dynamics and
response of the collective field.

\vspace{10pt}
\noindent
\emph{Summary of the exact Gaussian result.}
Starting from the coupled Langevin system \eqref{eq:app_phi_sde}--\eqref{eq:app_X_sde},
the MSRJD action \eqref{eq:app_Sphi}--\eqref{eq:app_Slambda} is Gaussian in the
reservoir variables.  Performing the functional integration over
$\{X_\lambda,\tilde X_\lambda\}$ yields the exact nonlocal effective action
\eqref{eq:app_Seff_time} with a causal memory kernel \eqref{eq:app_K_total_time}.
In frequency space the kernel takes the spectral form
\eqref{eq:app_Komega_spectral}, which is naturally identified with the retarded
self-energy \eqref{eq:app_selfenergy}.  This derivation is purely algebraic
and involves no approximation beyond standard boundary-condition assumptions
required for integration by parts.


\section{Appendix C: Exact emergence of a memory kernel from integrating out an OU reservoir}
\label{app:OU_memory_kernel}

In this Appendix we show that integrating out a continuum of
Ornstein--Uhlenbeck (OU) relaxation modes generates an exact
nonlocal memory kernel for the collective coordinate.
This provides a direct time-domain derivation of the relaxation-spectrum
representation used in the main text.

\vspace{10pt}
\noindent
\emph{(I) Coupled Markovian dynamics (collective field plus relaxational reservoir).}

For clarity we focus on temporal dynamics and suppress spatial dependence.
We consider a slow collective variable $\phi(t)$ linearly coupled to a
continuum of relaxational Ornstein--Uhlenbeck (OU) modes $X_\lambda(t)$
labeled by decay rate $\lambda\ge 0$:
\begin{align}
\dot\phi(t)
&=
-r\,\phi(t)
+\int_{0}^{\Lambda} d\lambda\, g(\lambda)\,X_\lambda(t)
+\eta(t),
\label{eq:phi_coupled_app}
\\
\dot X_\lambda(t)
&=
-\lambda\,X_\lambda(t)
+g(\lambda)\,\phi(t)
+\zeta_\lambda(t).
\label{eq:X_coupled_app}
\end{align}
Here $r>0$ is the bare restoring (mass) parameter, $\Lambda$ is an ultraviolet
cutoff on relaxation rates, and $\eta$ and $\zeta_\lambda$ are (possibly
independent) noise sources.
For $\lambda>0$ each reservoir coordinate is exponentially stable; the sector
$\lambda\to 0^+$ should be understood as the infrared limit of a continuous
spectrum of increasingly slow relaxational modes rather than an exactly
conserved degree of freedom.
Equations
(\ref{eq:phi_coupled_app})--(\ref{eq:X_coupled_app}) define a \emph{Markovian}
(first-order) dynamical system in the enlarged state space
$\{\phi, X_\lambda\}$.

\vspace{10pt}
\noindent
\emph{(II) Exact solution for the OU reservoir modes.}

Equation (\ref{eq:X_coupled_app}) is linear and can be solved exactly by the
integrating-factor method.
Multiply (\ref{eq:X_coupled_app}) by $e^{\lambda t}$:
\begin{equation}
e^{\lambda t}\dot X_\lambda(t)+\lambda e^{\lambda t}X_\lambda(t)
=
e^{\lambda t}\big[g(\lambda)\phi(t)+\zeta_\lambda(t)\big].
\label{eq:integrating_factor_step1_app}
\end{equation}
The left-hand side is the total derivative
\begin{equation}
\frac{d}{dt}\Big(e^{\lambda t}X_\lambda(t)\Big)
=
e^{\lambda t}\big[g(\lambda)\phi(t)+\zeta_\lambda(t)\big].
\label{eq:integrating_factor_step2_app}
\end{equation}
Integrate from an initial time $t_0$ to $t$:
\begin{equation}
e^{\lambda t}X_\lambda(t)-e^{\lambda t_0}X_\lambda(t_0)
=
\int_{t_0}^{t} ds\, e^{\lambda s}\big[g(\lambda)\phi(s)+\zeta_\lambda(s)\big].
\label{eq:integrating_factor_step3_app}
\end{equation}
Solving for $X_\lambda(t)$ yields the exact causal representation
\begin{equation}
\begin{aligned}
X_\lambda(t)
&=
e^{-\lambda (t-t_0)}X_\lambda(t_0)
\\
&\quad+
\int_{t_0}^{t} ds\, e^{-\lambda (t-s)}
\Big[g(\lambda)\phi(s)+\zeta_\lambda(s)\Big].
\end{aligned}
\label{eq:X_solution_app}
\end{equation}
This expression already shows that the reservoir coordinate at time $t$ depends
on the entire history $\{\phi(s)\}_{s<t}$.

\vspace{10pt}
\noindent
\emph{(III) Substitution into the collective equation and emergence of a memory kernel.}

Substituting (\ref{eq:X_solution_app}) into the collective equation
(\ref{eq:phi_coupled_app}) gives
\begin{align}
&\dot\phi(t)
=
-r\,\phi(t)
+\int_{0}^{\Lambda} d\lambda\, g(\lambda)\,
\nonumber\\
&\times \Bigg[
e^{-\lambda (t-t_0)}X_\lambda(t_0)
+
\int_{t_0}^{t} ds\, e^{-\lambda (t-s)}
\Big(g(\lambda)\phi(s)+\zeta_\lambda(s)\Big)
\Bigg]
\nonumber\\
&\quad +\eta(t)
\nonumber\\
\label{eq:kernel_identification_step_app}
\end{align}
This motivates the definition of the causal memory kernel
\begin{equation}
K(t-s)\equiv \Theta(t-s)\int_{0}^{\Lambda} d\lambda\, g(\lambda)^2\,e^{-\lambda (t-s)},
\label{eq:memory_kernel_def_app}
\end{equation}
so that (\ref{eq:kernel_identification_step_app}) becomes
\begin{equation}
\int_{0}^{\Lambda} d\lambda\, g(\lambda)
\int_{t_0}^{t} ds\, e^{-\lambda (t-s)}g(\lambda)\phi(s)
=
\int_{t_0}^{t} ds\, K(t-s)\,\phi(s).
\label{eq:kernel_compact_app}
\end{equation}

With this definition, Eq.~(\ref{eq:kernel_identification_step_app}) can be written as
\begin{equation}
\dot\phi(t)
=
-r\,\phi(t)
+\int_{t_0}^{t} ds\,K(t-s)\phi(s)
+\eta(t)
+\xi_{\rm eff}(t)
+J_{\rm ic}(t),
\label{eq:phi_memory_full_app}
\end{equation}
where we have collected the remaining contributions into an effective noise
term (generated by the reservoir noise) and an initial-condition transient:
\begin{align}
\xi_{\rm eff}(t)
&\equiv
\int_{0}^{\Lambda} d\lambda\, g(\lambda)
\int_{t_0}^{t} ds\, e^{-\lambda (t-s)}\zeta_\lambda(s),
\label{eq:xi_eff_def_app}
\\
J_{\rm ic}(t)
&\equiv
\int_{0}^{\Lambda} d\lambda\, g(\lambda)\,e^{-\lambda (t-t_0)}X_\lambda(t_0).
\label{eq:Jic_def_app}
\end{align}
Equation (\ref{eq:phi_memory_full_app}) is an exact \emph{generalized Langevin}
equation for $\phi(t)$.
Although the enlarged system (\ref{eq:phi_coupled_app})--(\ref{eq:X_coupled_app})
is Markovian, eliminating the hidden reservoir coordinates produces a
non-Markovian (memory) term and, generally, colored effective noise.

\vspace{6pt}
\noindent
\emph{Long-time limit.}
For $t-t_0\gg \Lambda^{-1}$, the transient $J_{\rm ic}(t)$ decays.
More precisely, decay holds provided the effective spectral weight
$g(\lambda)X_\lambda(t_0)$ is integrable near $\lambda=0$ so that the
$\lambda\to 0^+$ sector does not produce a non-decaying contribution.

\vspace{10pt}
\noindent
\emph{(IV) TDOS form and Laplace-transform structure.}

As derived in Appendix~C, integrating out the continuum of relaxation
modes produces a retarded self-energy of the form (\ref{eq:app_sigma_tdos_direct}),
where the time-scale density of states is defined as the
spectral weight distribution of relaxation eigenmodes,
\begin{equation}
\rho(\lambda)
=
\sum_\alpha g_\alpha^2\,\delta(\lambda-\lambda_\alpha).
\end{equation}

Here $g_\alpha$ represents the overlap amplitude between the collective
field and the relaxation eigenmode $\alpha$, so that $g_\alpha^2$
plays the role of the spectral weight of that mode.
In this sense the TDOS is directly analogous to the Lehmann spectral
function of Hamiltonian many-body theory, but defined for the
relaxation spectrum of the Liouvillian dynamics.

With this definition the memory kernel introduced above can be written
in the compact form
\begin{equation}
K(t)
=
\int_{0}^{\Lambda} d\lambda\;
\rho(\lambda)\,e^{-\lambda t},
\qquad t\ge0,
\label{eq:K_as_Laplace_app}
\end{equation}
showing that the memory kernel is the Laplace transform of the TDOS.
Consequently the long-time dynamics of the collective field is fully
controlled by the infrared structure of $\rho(\lambda)$.

\vspace{10pt}
\noindent
\emph{(V) Flat TDOS implies universal long-memory tail $K(t)\sim 1/t$.}

If the TDOS is finite at the origin (``flat'' TDOS),
\begin{equation}
\rho(\lambda)\xrightarrow{\lambda\to 0}\rho_0,
\label{eq:flat_TDOS_memory_app}
\end{equation}
then for times $t\gg \Lambda^{-1}$ the infrared part of
(\ref{eq:K_as_Laplace_app}) dominates and we may approximate $\rho(\lambda)\simeq
\rho_0$ over the relevant range.
In that case the kernel can be evaluated exactly:
\begin{align}
K(t)
=
\int_{0}^{\Lambda} d\lambda\, \rho_0\,e^{-\lambda t}.
\label{eq:K_flat_step1_app}
\end{align}
Perform the change of variables $u=\lambda t$ (so $d\lambda=du/t$):
\begin{align}
K(t)
&=
\rho_0\int_{0}^{\Lambda t} \frac{du}{t}\, e^{-u}
=
\frac{\rho_0}{t}\Big(1-e^{-\Lambda t}\Big).
\label{eq:K_flat_exact_app}
\end{align}
Therefore, at long times $t\gg \Lambda^{-1}$,
\begin{equation}
K(t)\simeq \frac{\rho_0}{t},
\qquad (t\gg \Lambda^{-1}),
\label{eq:K_tail_1_over_t_app}
\end{equation}
which is the universal long-memory tail discussed in the main text.

\vspace{10pt}
\noindent
\emph{(VI) Frequency-space form and connection to the retarded self-energy.}

For completeness we relate the time-domain memory kernel to the standard
frequency-space self-energy.
Taking the Fourier transform of the reservoir solution
(\ref{eq:X_solution_app}) in the stationary (long-time) limit yields
\begin{equation}
X_\lambda(\omega)=\frac{g(\lambda)}{\lambda-i\omega}\,\phi(\omega)
+\frac{1}{\lambda-i\omega}\,\zeta_\lambda(\omega),
\label{eq:X_omega_app}
\end{equation}
and substituting into (\ref{eq:phi_coupled_app}) gives
\begin{equation}
\Big[-i\omega+r+\Sigma_R(\omega)\Big]\phi(\omega)
=
\eta(\omega)+\int_0^\Lambda d\lambda\,\frac{g(\lambda)}{\lambda-i\omega}\zeta_\lambda(\omega),
\label{eq:phi_omega_app}
\end{equation}
with the retarded self-energy
\begin{equation}
\Sigma_R(\omega)=\int_0^\Lambda d\lambda\,\frac{g(\lambda)^2}{\lambda-i\omega}
=\int_0^\Lambda d\lambda\,\frac{\rho(\lambda)}{\lambda-i\omega}.
\label{eq:SigmaR_from_TDOS_app}
\end{equation}
\noindent
We use the convention $G_R^{-1}(\omega)=-i\omega+r+\Sigma_R(\omega)$; other
sign conventions for the self-energy appear in the literature.

Separating real and imaginary parts using
\begin{equation}
\frac{1}{\lambda-i\omega}
=
\frac{\lambda}{\lambda^2+\omega^2}
+i\,\frac{\omega}{\lambda^2+\omega^2},
\label{eq:decompose_app}
\end{equation}
one obtains
\begin{align}
\mathrm{Re}\,\Sigma_R(\omega)
&=
\int_0^\Lambda d\lambda\,\rho(\lambda)\frac{\lambda}{\lambda^2+\omega^2},
\\
\mathrm{Im}\,\Sigma_R(\omega)
&=
\omega\int_0^\Lambda d\lambda\,\rho(\lambda)\frac{1}{\lambda^2+\omega^2}.
\end{align}

For a flat TDOS $\rho(\lambda)\simeq\rho_0$ and $|\omega|\ll\Lambda$,
\begin{align}
\mathrm{Im}\,\Sigma_R(\omega)
&\simeq
\omega\rho_0\int_0^\Lambda \frac{d\lambda}{\lambda^2+\omega^2}
=
\omega\rho_0\left[\frac{1}{|\omega|}\arctan\!\Big(\frac{\Lambda}{|\omega|}\Big)\right]
\nonumber\\
&\xrightarrow{|\omega|\ll\Lambda}
\frac{\pi}{2}\rho_0\,\mathrm{sgn}(\omega),
\label{eq:ImSigma_sgn_app}
\end{align}
and
\begin{align}
\mathrm{Re}\,\Sigma_R(\omega)
&\simeq
\rho_0\int_0^\Lambda d\lambda\,\frac{\lambda}{\lambda^2+\omega^2}
=
\frac{\rho_0}{2}\ln\!\left(\frac{\Lambda^2+\omega^2}{\omega^2}\right)
\nonumber\\
&
\simeq
\rho_0\ln\!\left(\frac{\Lambda}{|\omega|}\right),
\quad (|\omega|\ll\Lambda),
\label{eq:ReSigma_log_app}
\end{align}
which reproduces the logarithmic reactive singularity discussed in the main
text.

\vspace{10pt}
\noindent
\emph{Summary.}
Equations (\ref{eq:phi_memory_full_app})--(\ref{eq:K_as_Laplace_app}) show that
integrating out a continuum of OU relaxation modes yields an \emph{exact}
generalized Langevin equation for the collective coordinate with a causal
memory kernel $K(t)$.
The kernel is the Laplace transform of the TDOS.
A finite TDOS at the origin produces the universal long-memory tail
$K(t)\sim 1/t$, and the corresponding retarded self-energy exhibits the
marginal nonanalytic structure
$\mathrm{Im}\,\Sigma_R(\omega)\propto \mathrm{sgn}(\omega)$ together with a
logarithmic real part.
This makes explicit that memory-dominated infrared dynamics arises from the
self-organized accumulation of slow internal relaxation modes, rather than from
coupling to a fast Markovian bath.

\section{Appendix D: Wilsonian renormalization with a full-kernel shell
in the relaxation-spectrum effective theory}
\label{app:full_kernel_rg}

In this Appendix we formulate the Wilsonian renormalization-group (RG)
analysis of the relaxation-spectrum effective theory using a shell
defined directly by the full inverse propagator.
This construction is more natural than a pure frequency-shell procedure
when the infrared dynamics is governed by a nontrivial dynamical kernel
$\Gamma_{\rm dyn}(\omega)$, because the distinction between fast and
slow modes is then determined not by $\omega$ alone, but by the full
quadratic kernel
\begin{equation}
\Gamma_2(\mathbf q,\omega)
=
r + c q^2 + \Gamma_{\rm dyn}(\omega).
\label{eq:full_kernel_gamma2}
\end{equation}
Here $\Gamma_{\rm dyn}(\omega)$ denotes the dynamical part of the
quadratic inverse kernel. In the main text this role is played by the
retarded Cooper (polarization) kernel $\Pi^R(\omega)$; we adopt the
notation $\Gamma_{\rm dyn}$ in this Appendix in order to emphasize that
the Wilsonian analysis depends only on the general structure of the
dynamical kernel, rather than on its specific microscopic origin.

The purpose of the present Appendix is to derive, in a fully explicit
way, the one-loop renormalization of the mass parameter $r$ and the
quartic coupling $u_4$, and to clarify how the infrared behavior is
controlled by the geometry of the equal-kernel shell.

\vspace{10pt}
\noindent
\emph{(I) Effective action and Gaussian propagator.}

We begin from the effective collective action
\begin{equation}
S[\phi] = S_2[\phi] + S_4[\phi],
\end{equation}
with quadratic part
\begin{equation}
S_2[\phi]
=
\frac{1}{2}
\int \frac{d^d q}{(2\pi)^d}\frac{d\omega}{2\pi}\;
\phi(-\mathbf q,-\omega)\,
\Gamma_2(\mathbf q,\omega)\,
\phi(\mathbf q,\omega),
\label{eq:appD_S2}
\end{equation}
and quartic interaction
\begin{equation}
S_4[\phi]
=
\frac{u_4}{4!}
\int d^d x\, d\tau\;
\phi^4(x,\tau).
\label{eq:appD_S4_realspace}
\end{equation}
The corresponding Gaussian propagator is
\begin{equation}
G(\mathbf q,\omega)
=
\frac{1}{r + c q^2 + \Gamma_{\rm dyn}(\omega)}.
\label{eq:appD_Gaussian_propagator}
\end{equation}

Throughout this Appendix we assume that \(\Gamma_{\rm dyn}(\omega)\) is
an even function of \(\omega\), and that for \(\omega>0\) it is
monotonic in the infrared.
This allows the shell constraint to be solved for \(\omega\) at fixed
\(\mathbf q\).

\vspace{10pt}
\noindent
\emph{(II) Full-kernel shell definition.}

Instead of classifying fast modes by a pure frequency shell
\(\Lambda/b < |\omega| < \Lambda\), we define the Wilsonian shell by the
full inverse propagator:
\begin{equation}
\Lambda e^{-dl}
<
r + c q^2 + \Gamma_{\rm dyn}(\omega)
<
\Lambda,
\qquad dl \ll 1.
\label{eq:appD_full_shell_definition}
\end{equation}
Equivalently, if we introduce
\begin{equation}
F(\mathbf q,\omega)
\equiv
r + c q^2 + \Gamma_{\rm dyn}(\omega),
\label{eq:appD_X_definition}
\end{equation}
the shell is the thin layer
\begin{equation}
\Lambda e^{-dl} < F(\mathbf q,\omega) < \Lambda.
\end{equation}

This definition is physically natural because the distinction between
fast and slow collective modes is controlled by the full quadratic
kernel.
In particular, when the dynamical term is nonanalytic or singular in the
infrared, a classification based on \(\omega\) alone obscures the actual
geometry of the low-energy sector.
By contrast, Eq.~(\ref{eq:appD_full_shell_definition}) directly follows
the equal-kernel contours of the effective propagator.

\vspace{10pt}
\noindent
\emph{(III) Wilsonian mode decomposition.}

We decompose the field into slow and fast components,
\begin{equation}
\phi = \phi_< + \phi_>,
\end{equation}
where \(\phi_>\) contains modes satisfying the shell condition
(\ref{eq:appD_full_shell_definition}), while \(\phi_<\) contains all
remaining lower-kernel modes.
The effective action for the slow sector is defined by
\begin{equation}
e^{-S_{\rm eff}[\phi_<]}
=
\int D\phi_>\;
e^{-S[\phi_<+\phi_>]}.
\label{eq:appD_Seff_definition}
\end{equation}

As usual in Wilsonian RG, the fast modes are integrated out first, while
the slow field is held fixed as an external background.
Only after this partial integration does one rescale momenta,
frequencies, and fields in order to restore the cutoff.

\vspace{10pt}
\noindent
\emph{(IV) One-loop correction to the quadratic term.}

We now derive the renormalization of the mass parameter.
Expanding the quartic interaction yields
\begin{equation}
(\phi_<+\phi_>)^4
=
\phi_<^4
+
4\phi_<^3\phi_>
+
6\phi_<^2\phi_>^2
+
4\phi_<\phi_>^3
+
\phi_>^4.
\label{eq:appD_phi4_expansion}
\end{equation}
Under Gaussian averaging over the fast shell, the odd terms vanish.
The leading correction to the quadratic slow-mode action therefore comes
from the term
\begin{equation}
6\phi_<^2\phi_>^2.
\end{equation}
Substituting this into the interaction and averaging over the fast modes
gives
\begin{equation}
\delta S_2
=
\frac{u_4}{4!}
\int d^d x\, d\tau\;
6\,\phi_<^2(x,\tau)\,
\langle \phi_>^2(x,\tau)\rangle_>.
\label{eq:appD_deltaS2_realspace_step1}
\end{equation}
Using \(6/4! = 1/4\), this becomes
\begin{equation}
\delta S_2
=
\frac{u_4}{4}
\int d^d x\, d\tau\;
\phi_<^2(x,\tau)\,
\langle \phi_>^2(x,\tau)\rangle_>.
\label{eq:appD_deltaS2_realspace_step2}
\end{equation}

By translational invariance, the coincident fast-mode correlator is
independent of \(x\) and \(\tau\):
\begin{equation}
\langle \phi_>^2(x,\tau)\rangle_>
=
\int_{\rm shell}
\frac{d^d q}{(2\pi)^d}\frac{d\omega}{2\pi}\;
G(\mathbf q,\omega).
\label{eq:appD_fast_correlator_shell}
\end{equation}
Therefore
\begin{equation}
\begin{aligned}
\delta S_2
=
&\frac{u_4}{4}
\left[
\int_{\rm shell}
\frac{d^d q}{(2\pi)^d}\frac{d\omega}{2\pi}\;
\frac{1}{r+cq^2+\Gamma_{\rm dyn}(\omega)}
\right]
\\
&\times \int d^d x\, d\tau\;
\phi_<^2(x,\tau).
\label{eq:appD_deltaS2_realspace_step3}
\end{aligned}
\end{equation}

The original quadratic action contains the mass term
\begin{equation}
S_2 \supset
\frac{1}{2}
\int d^d x\, d\tau\;
r\,\phi_<^2(x,\tau).
\end{equation}
Comparing with Eq.~(\ref{eq:appD_deltaS2_realspace_step3}), we identify
the one-loop mass correction as
\begin{equation}
\delta r
=
\frac{u_4}{2}
\int_{\rm shell}
\frac{d^d q}{(2\pi)^d}\frac{d\omega}{2\pi}\;
\frac{1}{r+cq^2+\Gamma_{\rm dyn}(\omega)}.
\label{eq:appD_delta_r_shell_general}
\end{equation}
The overall numerical prefactor depends on the normalization convention
for the quartic vertex.
In what follows we absorb such factors into a dimension-dependent
constant when discussing scaling behavior.

\vspace{10pt}
\noindent
\emph{(V) Thin-shell representation by a delta function.}

We now convert the shell integral into a more useful form.
For an infinitesimal shell \(dl \ll 1\), the width in the variable
\(F(\mathbf q,\omega)\) is
\begin{equation}
dF = \Lambda - \Lambda e^{-dl} \simeq \Lambda\, dl.
\label{eq:appD_shell_width_X}
\end{equation}
Hence, for any smooth integrand \(G(\mathbf q,\omega)\),
\begin{equation}
\begin{aligned}
&\int_{\rm shell} d^d q\, d\omega\; G(\mathbf q,\omega)
\\
& \quad \quad \simeq
\Lambda\, dl
\int d^d q\, d\omega\;
\delta\!\Bigl(F(\mathbf q,\omega)-\Lambda\Bigr)\,
G(\mathbf q,\omega).
\label{eq:appD_coarea_shell}
\end{aligned}
\end{equation}
Applying this identity to Eq.~(\ref{eq:appD_delta_r_shell_general}),
with
\begin{equation}
G(\mathbf q,\omega)
=
\frac{1}{r+cq^2+\Gamma_{\rm dyn}(\omega)}
=
\frac{1}{F(\mathbf q,\omega)},
\end{equation}
we obtain
\begin{align}
\delta r
&\simeq
\frac{u_4}{2}\,\Lambda\, dl
\int
\frac{d^d q}{(2\pi)^d}\frac{d\omega}{2\pi}\;
\delta\!\Bigl(F(\mathbf q,\omega)-\Lambda\Bigr)\,
\frac{1}{F(\mathbf q,\omega)}
\nonumber\\[4pt]
&=
\frac{u_4}{2}\,dl
\int
\frac{d^d q}{(2\pi)^d}\frac{d\omega}{2\pi}\;
\delta\!\Bigl(r+cq^2+\Gamma_{\rm dyn}(\omega)-\Lambda\Bigr),
\label{eq:appD_delta_r_delta_shell}
\end{align}
because the delta function enforces \(F=\Lambda\).

Equation~(\ref{eq:appD_delta_r_delta_shell}) is the central full-kernel
shell formula.
It shows that the one-loop mass correction is determined not directly by
the propagator amplitude, but by the density of modes lying on the
equal-kernel contour
\begin{equation}
r+cq^2+\Gamma_{\rm dyn}(\omega)=\Lambda.
\label{eq:appD_equal_kernel_contour}
\end{equation}

\vspace{10pt}
\noindent
\emph{(VI) Explicit evaluation of the shell density.}

We now perform the \(\omega\)-integral in
Eq.~(\ref{eq:appD_delta_r_delta_shell}).
At fixed \(\mathbf q\), let \(\omega_*(q)\) denote the positive
solution of
\begin{equation}
r + c q^2 + \Gamma_{\rm dyn}(\omega_*) = \Lambda.
\label{eq:appD_omega_star_definition}
\end{equation}
Because \(\Gamma_{\rm dyn}(\omega)\) is assumed even, there are in
general two solutions, \(\omega=\pm \omega_*(q)\).
Using the identity
\begin{equation}
\delta(f(\omega))
=
\sum_i
\frac{\delta(\omega-\omega_i)}{|f'(\omega_i)|},
\label{eq:appD_delta_of_function}
\end{equation}
with
\begin{equation}
f(\omega)=r+cq^2+\Gamma_{\rm dyn}(\omega)-\Lambda,
\end{equation}
we obtain
\begin{equation}
\int d\omega\;
\delta\!\Bigl(r+cq^2+\Gamma_{\rm dyn}(\omega)-\Lambda\Bigr)
=
\sum_{\omega=\pm \omega_*(q)}
\frac{1}{|\partial_\omega \Gamma_{\rm dyn}(\omega)|}.
\end{equation}
Therefore the one-loop mass correction becomes
\begin{equation}
\delta r
=
A_d\,u_4\, dl
\int_0^{q_{\rm max}}
dq\; q^{d-1}
\sum_{\omega=\pm\omega_*(q)}
\frac{1}{|\Gamma'_{\rm dyn}(\omega)|},
\label{eq:appD_delta_r_surface_measure}
\end{equation}
where \(A_d\) absorbs the angular factor \(\Omega_d/(2\pi)^{d+1}\), and
\(q_{\rm max}\) is determined by the condition that the shell equation
(\ref{eq:appD_omega_star_definition}) admit a real solution.
Equivalently, \(q_{\rm max}\) is fixed by
\begin{equation}
r + c q_{\rm max}^2 = \Lambda,
\qquad
q_{\rm max} = \sqrt{\frac{\Lambda-r}{c}},
\label{eq:appD_qmax}
\end{equation}
provided \(\Lambda>r\).

Equation~(\ref{eq:appD_delta_r_surface_measure}) makes the geometric
content of the full-kernel shell completely explicit:
the fluctuation correction is weighted by the inverse slope
\(|\Gamma'_{\rm dyn}|^{-1}\), which measures how much frequency-space
volume is associated with a fixed interval \(dX\) of the inverse kernel.

\vspace{10pt}
\noindent
\emph{(VII) Canonical scaling and general RG equation for \(r\).}

After integrating out the shell, one rescales
\(\mathbf q\), \(\omega\), and \(\phi\) so as to restore the cutoff.
The mass parameter has canonical scaling dimension \(2\), hence the
tree-level contribution is
\begin{equation}
\left.\frac{dr}{dl}\right|_{\rm tree} = 2r.
\end{equation}
Combining this with Eq.~(\ref{eq:appD_delta_r_surface_measure}), the
general full-kernel shell RG equation for the mass parameter is
\begin{equation}
\frac{dr}{dl}
=
2r
+
A_d\,u_4
\int_0^{q_{\rm max}}
dq\; q^{d-1}
\sum_{\omega=\pm\omega_*(q)}
\frac{1}{|\Gamma'_{\rm dyn}(\omega)|}.
\label{eq:appD_rg_r_general_fullkernel}
\end{equation}
This is the natural analogue, in the present relaxation-spectrum
framework, of the standard one-loop mass flow equation in Wilsonian RG.

\vspace{10pt}
\noindent
\emph{(VIII) One-loop renormalization of the quartic interaction.}

We next derive the renormalization of the quartic coupling.
The effective action for the slow modes is obtained from the cumulant
expansion
\begin{equation}
S_{\rm eff}[\phi_<]
=
S_2[\phi_<]
+
\langle S_4 \rangle_>
-
\frac{1}{2}
\Bigl(
\langle S_4^2\rangle_>
-
\langle S_4\rangle_>^2
\Bigr)
+\cdots.
\label{eq:appD_cumulant_expansion}
\end{equation}
The first cumulant produces the mass correction derived above.
The quartic coupling is renormalized by the second cumulant, which
connects two quartic vertices by two fast propagators.

Schematically, this generates
\begin{equation}
\delta S_4
\sim
- u_4^2
\int d^d x\, d\tau\;
\phi_<^4(x,\tau)\,
\langle \phi_>(x,\tau)\phi_>(x,\tau)\rangle_>^2.
\label{eq:appD_deltaS4_schematic}
\end{equation}
In momentum-frequency space the correction takes the form
\begin{equation}
\delta u_4
=
- B_d\,u_4^2
\int_{\rm shell}
\frac{d^d q}{(2\pi)^d}\frac{d\omega}{2\pi}\;
\frac{1}{\left(r+cq^2+\Gamma_{\rm dyn}(\omega)\right)^2},
\label{eq:appD_delta_u_shell_general}
\end{equation}
where \(B_d\) is a positive numerical factor depending on combinatorics
and vertex normalization.

Using the same thin-shell identity as before,
\begin{align}
\delta u_4
&\simeq
- B_d\,u_4^2\,
\Lambda\, dl
\int
\frac{d^d q}{(2\pi)^d}\frac{d\omega}{2\pi}\;
\delta\!\Bigl(F(\mathbf q,\omega)-\Lambda\Bigr)\,
\frac{1}{F(\mathbf q,\omega)^2}
\nonumber\\[4pt]
&=
- B_d\,u_4^2\,
\frac{dl}{\Lambda}
\int
\frac{d^d q}{(2\pi)^d}\frac{d\omega}{2\pi}\;
\delta\!\Bigl(r+cq^2+\Gamma_{\rm dyn}(\omega)-\Lambda\Bigr).
\label{eq:appD_delta_u_delta_shell}
\end{align}
Performing the \(\omega\)-integral as above yields
\begin{equation}
\delta u_4
=
- B_d\,u_4^2\,
\frac{dl}{\Lambda}
\int_0^{q_{\rm max}}
dq\; q^{d-1}
\sum_{\omega=\pm\omega_*(q)}
\frac{1}{|\Gamma'_{\rm dyn}(\omega)|}.
\label{eq:appD_delta_u_surface_measure}
\end{equation}
Including the canonical scaling term, one finds
\begin{equation}
\begin{aligned}
\frac{du_4}{dl}
=
&(4-d-z)\,u_4
\\
&-
B_d\,u_4^2\,
\frac{1}{\Lambda}
\int_0^{q_{\rm max}}
dq\; q^{d-1}
\sum_{\omega=\pm\omega_*(q)}
\frac{1}{|\Gamma'_{\rm dyn}(\omega)|}.
\end{aligned}
\label{eq:appD_rg_u_general_fullkernel}
\end{equation}

The structure of Eqs.~(\ref{eq:appD_rg_r_general_fullkernel}) and
(\ref{eq:appD_rg_u_general_fullkernel}) is transparent.
The mass correction contains one fast propagator and is therefore
proportional to the shell density itself, whereas the quartic correction
contains two fast propagators and is suppressed by one additional power
of the running cutoff \(\Lambda\).

\vspace{10pt}
\noindent
\emph{(IX) Power-law dynamical kernel: \(\Gamma_{\rm dyn}(\omega)\sim A|\omega|^\alpha\), \(\alpha>0\).}

We now specialize to a low-frequency power-law kernel
\begin{equation}
\Gamma_{\rm dyn}(\omega)=A|\omega|^\alpha,
\qquad \alpha>0.
\label{eq:appD_powerlaw_kernel_positive}
\end{equation}
The shell condition becomes
\begin{equation}
r+cq^2+A|\omega_*|^\alpha=\Lambda,
\end{equation}
hence
\begin{equation}
|\omega_*|
=
\left(
\frac{\Lambda-r-cq^2}{A}
\right)^{1/\alpha}.
\label{eq:appD_omega_star_powerlaw_positive}
\end{equation}
Moreover,
\begin{equation}
|\Gamma'_{\rm dyn}(\omega_*)|
=
A\alpha |\omega_*|^{\alpha-1}.
\end{equation}
Therefore
\begin{equation}
\frac{1}{|\Gamma'_{\rm dyn}(\omega_*)|}
=
\frac{1}{A\alpha}
|\omega_*|^{1-\alpha}
=
\frac{1}{A\alpha}
\left(
\frac{\Lambda-r-cq^2}{A}
\right)^{\frac{1-\alpha}{\alpha}}.
\label{eq:appD_inverse_slope_powerlaw_positive}
\end{equation}
Substituting into Eq.~(\ref{eq:appD_rg_r_general_fullkernel}) yields
\begin{equation}
\frac{dr}{dl}
=
2r
+
\widetilde A_d\,u_4
\int_0^{\sqrt{(\Lambda-r)/c}}
dq\; q^{d-1}
\left(
\Lambda-r-cq^2
\right)^{\frac{1-\alpha}{\alpha}},
\label{eq:appD_rg_r_powerlaw_positive}
\end{equation}
where \(\widetilde A_d\) absorbs constants.
Near criticality, \(r\ll \Lambda\), setting
\begin{equation}
q=\sqrt{\frac{\Lambda}{c}}\,x,
\qquad
0\le x\le 1,
\end{equation}
gives
\begin{equation}
\frac{dr}{dl}
=
2r
+
\widetilde A_d\,u_4\,
\Lambda^{\,\frac{d}{2}+\frac{1}{\alpha}-1}
\int_0^1 dx\;
x^{d-1}(1-x^2)^{\frac{1-\alpha}{\alpha}}.
\label{eq:appD_rg_r_powerlaw_scaling}
\end{equation}
Thus the shell correction scales as
\begin{equation}
\delta r \sim u_4\,dl\,\Lambda^{\,\frac{d}{2}+\frac{1}{\alpha}-1}.
\label{eq:appD_delta_r_scaling_powerlaw_positive}
\end{equation}

This expression has a direct geometric interpretation.
In the full-kernel shell formulation, the exponent \(\alpha\) controls
not directly the size of the propagator, but the density of modes on the
equal-kernel contour through the factor
\(|\Gamma'_{\rm dyn}|^{-1}\).
For \(0<\alpha<1\), the slope \(|\Gamma'_{\rm dyn}|\) diverges as
\(\omega\to0\), so the shell becomes compressed in frequency space and
the fluctuation contribution is correspondingly reduced.
For \(\alpha=1\), the shell density is frequency-independent.
For \(\alpha>1\), the shell expands in frequency space, although the
overall infrared behavior still depends on the additional
\(q\)-phase-space factor.

\vspace{10pt}
\noindent
\emph{(X) Marginal flat-TDOS case: \(\alpha=0\).}

For a flat TDOS, the retarded kernel is logarithmic:
\begin{equation}
\Gamma_{\rm dyn}(\omega)
\sim
\rho_0 \ln\frac{\Lambda_0}{|\omega|}.
\label{eq:appD_log_kernel}
\end{equation}
The shell condition reads
\begin{equation}
r+cq^2+\rho_0\ln\frac{\Lambda_0}{|\omega_*|}=\Lambda,
\end{equation}
which gives
\begin{equation}
|\omega_*|
=
\Lambda_0
\exp\!\left[
-\frac{\Lambda-r-cq^2}{\rho_0}
\right].
\label{eq:appD_omega_star_log}
\end{equation}
The derivative of the kernel is
\begin{equation}
\Gamma'_{\rm dyn}(\omega)
=
-\frac{\rho_0}{\omega},
\end{equation}
hence
\begin{equation}
\frac{1}{|\Gamma'_{\rm dyn}(\omega_*)|}
=
\frac{|\omega_*|}{\rho_0}
=
\frac{\Lambda_0}{\rho_0}
\exp\!\left[
-\frac{\Lambda-r-cq^2}{\rho_0}
\right].
\label{eq:appD_inverse_slope_log}
\end{equation}
Substituting into the flow equation yields
\begin{equation}
\frac{dr}{dl}
=
2r
+
\widehat A_d\,u_4
\int_0^{\sqrt{(\Lambda-r)/c}}
dq\; q^{d-1}
\exp\!\left[
-\frac{\Lambda-r-cq^2}{\rho_0}
\right].
\label{eq:appD_rg_r_log}
\end{equation}

This form differs in appearance from the simpler frequency-shell result,
where the logarithmic kernel appears directly in the propagator
denominator.
In the present full-kernel shell formulation, the same physics is
encoded instead in the exponentially small frequency width of the
equal-kernel shell.
Thus the flat-TDOS case remains marginal, but the shell geometry makes
this marginality manifest in a different representation.

\vspace{10pt}
\noindent
\emph{(XI) Infrared-singular kernel: \(\alpha<0\).}

For an infrared-singular TDOS, the low-frequency kernel behaves as
\begin{equation}
\Gamma_{\rm dyn}(\omega)
\sim
\frac{A}{|\omega|^{|\alpha|}},
\qquad \alpha<0.
\label{eq:appD_powerlaw_kernel_negative}
\end{equation}
The shell condition becomes
\begin{equation}
r+cq^2+\frac{A}{|\omega_*|^{|\alpha|}}=\Lambda,
\end{equation}
which implies
\begin{equation}
|\omega_*|
=
\left(
\frac{A}{\Lambda-r-cq^2}
\right)^{1/|\alpha|}.
\label{eq:appD_omega_star_powerlaw_negative}
\end{equation}
Differentiating the kernel gives
\begin{equation}
|\Gamma'_{\rm dyn}(\omega_*)|
\sim
\frac{A|\alpha|}{|\omega_*|^{|\alpha|+1}},
\end{equation}
so that
\begin{equation}
\frac{1}{|\Gamma'_{\rm dyn}(\omega_*)|}
\sim
|\omega_*|^{|\alpha|+1}
\sim
\left(
\Lambda-r-cq^2
\right)^{-\frac{|\alpha|+1}{|\alpha|}}.
\label{eq:appD_inverse_slope_powerlaw_negative}
\end{equation}
Hence
\begin{equation}
\frac{dr}{dl}
=
2r
+
\bar A_d\,u_4
\int_0^{\sqrt{(\Lambda-r)/c}}
dq\; q^{d-1}
\left(
\Lambda-r-cq^2
\right)^{-\frac{|\alpha|+1}{|\alpha|}}.
\label{eq:appD_rg_r_powerlaw_negative}
\end{equation}

The most important physical point is not the detailed power of the
integrand, but the mechanism by which the shell contribution is
controlled.
Because \(\Gamma_{\rm dyn}(\omega)\) diverges in the infrared, its slope
\(|\Gamma'_{\rm dyn}(\omega)|\) becomes very large at small \(\omega\).
As a result, the equal-kernel shell is strongly compressed in frequency
space.
In the full-kernel shell language, this means that the quartic
fluctuation correction is geometrically suppressed even though slow
modes accumulate in the relaxation spectrum.
This is fully consistent with the conclusion reached in the simpler
frequency-shell treatment: the enhancement of pairing does not come from
a divergence of quartic fluctuation corrections, but from the infrared
structure of the quadratic Cooper kernel itself.

\vspace{10pt}
\noindent
\emph{(XII) Relation to the frequency-shell formulation.}

It is useful to compare the present construction with the simpler
frequency-shell RG used in the main text.
If one classifies fast modes by
\begin{equation}
\Lambda e^{-dl}<|\omega|<\Lambda,
\end{equation}
then the one-loop mass correction takes the approximate form
\begin{equation}
\begin{aligned}
\delta r
\sim
&u_4
\int_{\omega\text{-shell}}
\frac{d^d q}{(2\pi)^d}\frac{d\omega}{2\pi}\;
\frac{1}{r+cq^2+\Gamma_{\rm dyn}(\omega)}
\\
&\sim
u_4\,\Lambda\,dl
\int
\frac{d^d q}{(2\pi)^d}
\frac{1}{r+cq^2+\Gamma_{\rm dyn}(\Lambda)}.
\label{eq:appD_frequency_shell_comparison}
\end{aligned}
\end{equation}
In this representation the shell correction is controlled directly by
the propagator amplitude evaluated at the running frequency cutoff.

By contrast, in the full-kernel shell formulation the same correction is
written as
\begin{equation}
\delta r
\sim
u_4\,dl
\int
\frac{d^d q}{(2\pi)^d}
\sum_{\omega=\pm \omega_*(q)}
\frac{1}{|\Gamma'_{\rm dyn}(\omega)|}.
\label{eq:appD_fullkernel_shell_comparison}
\end{equation}
The correction is therefore controlled by the density of modes on the
equal-kernel contour rather than by the propagator amplitude at a single
running frequency.

The two procedures differ in representation, but they encode the same
infrared physics.
The frequency-shell method is simpler and often more convenient for
explicit estimates.
The full-kernel shell method is more geometrically transparent and more
natural when the dynamical kernel is strongly nonanalytic or singular.
In particular, it makes explicit that the infrared behavior is governed
by the compression or expansion of equal-kernel shells in frequency
space.

\vspace{10pt}
\noindent
\emph{(XIII) Physical interpretation and scope.}

The main lesson of the present full-kernel shell RG analysis is that the
quartic fluctuation corrections are controlled by the geometry of the
quadratic kernel.
The one-loop mass correction is determined by the density of modes on the
surface
\begin{equation}
r+cq^2+\Gamma_{\rm dyn}(\omega)=\Lambda,
\end{equation}
while the quartic correction is governed by the same shell density with
one additional inverse power of the running kernel scale.

This leads to an important conceptual distinction.
The Wilsonian RG of the quartic sector quantifies the consistency of the
effective collective theory and the size of fluctuation corrections.
It does not determine the superconducting transition scale itself.
The pairing instability remains governed by the quadratic Cooper kernel
through the Thouless condition
\begin{equation}
r(T)=\frac{1}{g_{\rm pair}}-\Pi^R(0;T),
\qquad
r(T_c)=0.
\label{eq:appD_Thouless_condition}
\end{equation}
Hence the enhancement of pairing originates from the infrared structure
of the retarded quadratic kernel, whereas the quartic fluctuation sector
produces only subleading corrections.

In this sense, the present full-kernel shell formulation provides a
natural Wilsonian framework adapted to the relaxation-spectrum
description of infrared collective dynamics.
It shows that the appropriate notion of ``fast'' and ``slow'' modes is
set by the full inverse propagator rather than by frequency or momentum
alone, and that the infrared effect of the TDOS exponent is encoded in
the geometry of the equal-kernel shell.


\section{Appendix E: Microscopic origin of the time--scale density of states}
\label{app:TDOS_microscopic}

In this Appendix we derive the time--scale density of states
introduced in the main text directly from the microscopic
momentum--space structure of the fermionic relaxation rates.
We show that the TDOS is not an ad hoc assumption, but follows
systematically from the reorganization of the Cooper-channel kernel
in terms of decay rates rather than quasiparticle energies.

It is important to emphasize that the TDOS itself is defined as an
intrinsic, temperature-independent relaxation spectrum of the system.
Thermal effects enter separately through the infrared cutoff of the
finite-temperature Thouless criterion, rather than through the
definition of the TDOS itself.
The temperature dependence of the superconducting instability is thus
controlled by the lower integration bound
\(
\lambda_{\mathrm{IR}}\sim T
\),
while the TDOS describes the underlying distribution of relaxation
modes.

\vspace{10pt}
\noindent
\emph{(I) Spectral representation of the Cooper kernel.}

Starting from the overdamped Cooper-channel expression,
\begin{equation}
\Pi^R(\Omega)
=
\int \frac{d^d k}{(2\pi)^d}\;
\frac{1}{2\gamma_{\mathbf k}-i\Omega},
\label{eq:PiR_gamma_appendix}
\end{equation}
we define the pair relaxation rate
\begin{equation}
\lambda_{\mathbf k}
\equiv
2\gamma_{\mathbf k}.
\end{equation}
Introducing a spectral decomposition,
Eq.~(\ref{eq:PiR_gamma_appendix})
may be rewritten as
\begin{equation}
\Pi^R(\Omega)
=
\int_0^\Lambda d\lambda\;
\frac{\rho(\lambda)}{\lambda-i\Omega},
\label{eq:PiR_TDOS_appendix}
\end{equation}
where the TDOS is defined by
\begin{equation}
\rho(\lambda)
=
\int \frac{d^d k}{(2\pi)^d}\;
\delta\!\bigl(\lambda-2\gamma_{\mathbf k}\bigr).
\label{eq:TDOS_definition_appendix2}
\end{equation}

Equation~(\ref{eq:TDOS_definition_appendix2})
is formally analogous to the ordinary density of states,
with the energy dispersion replaced by the relaxation-rate dispersion.
Thus, the TDOS counts the number of momentum states contributing to a
given decay rate.

The precise thermal occupation factors discussed in the main text do
not modify this intrinsic relaxation spectrum, but only renormalize
the overall amplitude of the Cooper kernel.
The infrared singular structure and the transition-temperature scaling
are therefore governed by the geometry of
\(
\rho(\lambda)
\)
together with the thermal infrared cutoff.

\vspace{10pt}
\noindent
\emph{(II) Reduction to an isotropic integral.}

Assuming an isotropic relaxation-rate dispersion,
\begin{equation}
\gamma_{\mathbf k}
=
\gamma(k),
\qquad
k=|\mathbf k|,
\end{equation}
the TDOS becomes
\begin{equation}
\rho(\lambda)
=
\int \frac{d^d k}{(2\pi)^d}\;
\delta\!\bigl(\lambda-2\gamma(k)\bigr).
\end{equation}
Using spherical coordinates in $d$ dimensions,
\begin{equation}
d^d k
=
S_{d-1}\,k^{d-1}dk,
\end{equation}
with
\begin{equation}
S_{d-1}
=
\frac{2\pi^{d/2}}{\Gamma(d/2)},
\end{equation}
we obtain
\begin{equation}
\rho(\lambda)
=
\frac{S_{d-1}}{(2\pi)^d}
\int_0^\infty dk\;
k^{d-1}
\delta\!\bigl(\lambda-2\gamma(k)\bigr).
\label{eq:rho_integral_appendix2}
\end{equation}

\vspace{10pt}
\noindent
\emph{(III) Evaluation of the delta function.}

Using the identity
\begin{equation}
\delta(f(k))
=
\sum_i
\frac{\delta(k-k_i)}{|f'(k_i)|},
\end{equation}
where $k_i$ are defined by
\begin{equation}
\lambda
=
2\gamma(k_i),
\end{equation}
we obtain
\begin{equation}
\rho(\lambda)
=
\frac{S_{d-1}}{(2\pi)^d}
\sum_i
\frac{k_i^{\,d-1}}{|2\gamma'(k_i)|}.
\label{eq:rho_general_appendix2}
\end{equation}

This expression provides the exact relation between the TDOS and the
microscopic relaxation-rate dispersion.
The thermal occupation affects only the overall prefactor of the
Cooper response, while the infrared power-law exponent is determined
entirely by the geometry of the relaxation-rate dispersion itself.

\vspace{10pt}
\noindent
\emph{(IV) Infrared scaling from a power-law dispersion.}

To determine the infrared behavior, consider a power-law form near a
soft manifold $k_0$,
\begin{equation}
\gamma(k)
\sim
a\,|k-k_0|^n,
\qquad a>0.
\label{eq:gamma_powerlaw_appendix}
\end{equation}
Then
\begin{equation}
\lambda = 2a\,|k-k_0|^n,
\qquad
|k-k_0|
=
\left(\frac{\lambda}{2a}\right)^{1/n}.
\end{equation}

\vspace{5pt}
\noindent
\emph{Finite-shell case ($k_0\neq 0$).}
When the slow modes are located near a finite momentum shell,
$k_i\simeq k_0$ for $\lambda\to 0$, and thus
\begin{equation}
\rho(\lambda)
\sim
\lambda^{-(n-1)/n}.
\end{equation}
Hence
\begin{equation}
\rho(\lambda)\sim \lambda^\alpha,
\qquad
\alpha = -\frac{n-1}{n}.
\label{eq:alpha_shell_appendix}
\end{equation}

In particular,
\begin{align}
\gamma(k)\sim |k-k_0|
&\Rightarrow
\rho(\lambda)\sim \lambda^0,
\\
\gamma(k)\sim |k-k_0|^2
&\Rightarrow
\rho(\lambda)\sim \lambda^{-1/2}.
\end{align}

\vspace{5pt}
\noindent
\emph{Origin-centered case ($k_0=0$).}
If the slow modes are centered at $k=0$, the phase-space factor
must be retained:
\begin{equation}
\rho(\lambda)
\sim
\lambda^{(d-n)/n}.
\end{equation}
Thus
\begin{equation}
\rho(\lambda)\sim \lambda^\alpha,
\qquad
\alpha = \frac{d}{n}-1.
\label{eq:alpha_origin_appendix}
\end{equation}

Examples include
\begin{align}
\gamma(k)\sim k^2,\ d=3
&\Rightarrow
\rho(\lambda)\sim \lambda^{1/2},
\\
\gamma(k)\sim k^2,\ d=2
&\Rightarrow
\rho(\lambda)\sim \lambda^0.
\end{align}

\vspace{10pt}
\noindent
\emph{(V) Physical interpretation.}

The TDOS is controlled by two independent ingredients:
the scaling of the relaxation rate near its zeros,
and the phase-space structure of the slow modes.

A flat TDOS ($\alpha=0$) arises when slow modes are uniformly
distributed down to vanishing decay rates, while an
infrared-singular TDOS ($\alpha<0$) requires an anomalous
accumulation of slow modes.

\vspace{10pt}
\noindent
\emph{(VI) Relation to the dynamical kernel.}

Substituting $\rho(\lambda)\sim \lambda^\alpha$ into
Eq.~(\ref{eq:PiR_TDOS_appendix}), one obtains
\begin{equation}
\Pi^R(\Omega)\sim |\Omega|^\alpha,
\qquad (\alpha\neq 0),
\end{equation}
while for $\alpha=0$,
\begin{equation}
\Pi^R(\Omega)
\sim
\ln\!\left(\frac{\Lambda}{|\Omega|}\right)
+
i\,\frac{\pi}{2}\,\mathrm{sgn}(\Omega).
\end{equation}
Thus the exponent $\alpha$ directly determines the universality
class of the infrared dynamics.

\vspace{10pt}
\noindent
\emph{Summary.}
We have shown that the TDOS is uniquely determined by the microscopic
relaxation-rate dispersion,
and that its infrared scaling follows from the interplay between the
vanishing of $\gamma_{\mathbf k}$ and the phase-space structure of the
slow modes.  This establishes a direct bridge between microscopic
fermionic dynamics and the TDOS formulation used in the main text.

\section{Appendix F: Microscopic motivation for TDOS from SYK and RVB frameworks}
\label{app:microscopic_tdos}

In this Appendix we present a minimal and controlled route connecting
microscopic strongly correlated models to the relaxation-spectrum
formulation employed in the main text.

The purpose of this Appendix is not to provide a complete microscopic
derivation of the time-scale density of states, but rather to
demonstrate that known strongly interacting systems naturally give rise
to effective dynamical structures consistent with a broad distribution
of relaxation rates.

We focus on two representative settings:
the Sachdev--Ye--Kitaev (SYK) class of models, and the resonating-valence-bond
(RVB) framework of doped Mott insulators.

\vspace{10pt}
\noindent
\emph{(I) General strategy: from operator dynamics to TDOS.}

The central object in the present theory is the retarded kernel
\begin{equation}
\Pi^R(\omega)
=
\int_0^\Lambda d\lambda\;
\frac{\rho(\lambda)}{\lambda - i\omega},
\label{eq:TDOS_representation_appendix}
\end{equation}
which encodes collective dynamics in terms of a distribution of
relaxation rates.

To connect a microscopic model to this representation, one considers a
physically relevant collective operator $O$ and evaluates its retarded
response function $\chi_O^R(\omega)$ in the infrared regime.
If the response exhibits a scaling form
\begin{equation}
\chi_O^R(\omega)\sim |\omega|^{\alpha},
\label{eq:chi_scaling_appendix}
\end{equation}
then it can be represented in the form
Eq.~\eqref{eq:TDOS_representation_appendix} with an effective TDOS
\begin{equation}
\rho_{\mathrm{eff}}(\lambda)\sim \lambda^{\alpha}.
\label{eq:rho_scaling_appendix}
\end{equation}

The power-law frequency dependence implies that the response cannot be
represented by a finite number of isolated poles, and is therefore
naturally captured by an integral representation over a continuum of
relaxation rates.
In this sense, the TDOS representation should be understood as an
effective spectral decomposition of the retarded kernel, rather than a discrete set of microscopic relaxation eigenmodes.

Thus, the TDOS is best interpreted as a \emph{projected spectral density},
defined with respect to a chosen collective operator.

\vspace{10pt}
\noindent
\emph{(II) SYK models: operator scaling and effective TDOS.}

We first consider the SYK class of models, which provide a controlled
example of strongly interacting quantum dynamics without quasiparticles.

At low energies, these systems exhibit emergent conformal invariance.
For an operator $O$ with scaling dimension $\delta_O$, the Euclidean
two-point function behaves as
\begin{equation}
\langle O(\tau) O(0)\rangle
\sim
\frac{1}{|\tau|^{2\delta_O}}.
\label{eq:SYK_time_scaling_appendix}
\end{equation}

Upon Fourier transform and analytic continuation, the retarded response
takes the form
\begin{equation}
\chi_O^R(\omega)
\sim
|\omega|^{2\delta_O - 1}.
\label{eq:SYK_freq_scaling_appendix}
\end{equation}
Comparing with Eq.~\eqref{eq:chi_scaling_appendix}, one identifies
\begin{equation}
\alpha_O = 2\delta_O - 1,
\label{eq:alpha_SYK_appendix}
\end{equation}
which yields the effective TDOS
\begin{equation}
\rho_{\mathrm{eff}}^{(O)}(\lambda)
\sim
\lambda^{2\delta_O - 1}.
\label{eq:rho_SYK_appendix}
\end{equation}

This result demonstrates that SYK models naturally generate a continuum
of effective relaxation rates, with the TDOS exponent directly controlled
by operator scaling dimensions.

In particular, $\delta_O = 1/2$ leads to $\alpha_O = 0$, corresponding
to a flat TDOS and logarithmic response, while $\delta_O < 1/2$
produces an infrared-singular TDOS.
The dependence of $\alpha_O$ on the operator channel reflects the fact
that different collective sectors probe distinct parts of the many-body
spectrum.

\vspace{10pt}
\noindent
\emph{(III) Sachdev--Ye model: logarithmic response and flat TDOS.}

A particularly transparent realization of a flat relaxation spectrum is
provided by the Sachdev--Ye model of random quantum magnets.
In this case, the local spin susceptibility exhibits a logarithmic
infrared behavior
\begin{equation}
\chi_{\mathrm{loc}}^R(\omega)
\sim
\ln\frac{\Lambda}{|\omega|}
+
i\,\frac{\pi}{2}\,\mathrm{sgn}(\omega),
\label{eq:SY_log_kernel_appendix}
\end{equation}
characteristic of a marginal, scale-invariant dynamical regime.

This form is reproduced by the TDOS representation
Eq.~\eqref{eq:TDOS_representation_appendix} with a constant infrared
density of relaxation rates,
\begin{equation}
\rho(\lambda \to 0) \to \rho_0.
\label{eq:SY_flat_TDOS_appendix}
\end{equation}
Physically, this corresponds to a broad continuum of slow spin
rearrangement processes with no intrinsic time scale, leading to
scale-free dynamics and long-time memory.
The Sachdev--Ye model therefore provides a concrete microscopic example
in which a flat TDOS emerges as an effective infrared description.

\vspace{10pt}
\noindent
\emph{(IV) RVB framework: bond fluctuations and relaxation spectrum.}

We next consider the RVB framework arising from the $t$--$J$ model,
where the natural collective variable is the singlet-bond field.
The projected RVB state itself is a static variational ansatz and does
not directly determine a TDOS.
The relevant dynamical quantity is instead the fluctuation spectrum
around the RVB manifold.

Introducing a coarse-grained bond field $B(t)$, one considers the
autocorrelation function
\begin{equation}
E_B(t)
=
\langle B(t) B^\dagger(0)\rangle.
\label{eq:CB_appendix}
\end{equation}
Expanding the field in normal modes,
\begin{equation}
B(t)=\sum_\mu g_\mu \phi_\mu(t),
\label{eq:B_modes_appendix}
\end{equation}
leads to
\begin{equation}
E_B(t)
=
\sum_\mu |g_\mu|^2 e^{-\lambda_\mu t}.
\label{eq:CB_modes_appendix}
\end{equation}
Passing to a continuum representation yields
\begin{equation}
E_B(t)
=
\int_0^\infty d\lambda\;
\rho_B(\lambda)\,e^{-\lambda t},
\label{eq:CB_TDOS_appendix}
\end{equation}
with
\begin{equation}
\rho_B(\lambda)
=
\sum_\mu |g_\mu|^2 \delta(\lambda-\lambda_\mu).
\label{eq:rhoB_appendix}
\end{equation}

Here the exponential form should be understood as an effective
description of coarse-grained dissipative dynamics, arising after
integrating out fast degrees of freedom and environmental couplings.
The function $\rho_B(\lambda)$ defines a bond-weighted relaxation-rate
density, which plays the role of a TDOS in the singlet sector.

The infrared behavior of $\rho_B(\lambda)$ determines long-time
dynamics:
\begin{equation}
\rho_B(\lambda)\sim \lambda^{\alpha}
\quad \Longrightarrow \quad
E_B(t)\sim t^{-(1+\alpha)}.
\label{eq:RVB_scaling_appendix}
\end{equation}
A flat TDOS ($\alpha=0$) yields
\begin{equation}
E_B(t)\sim \frac{1}{t},
\label{eq:RVB_flat_appendix}
\end{equation}
indicating long-time memory of singlet-bond fluctuations.

The RVB framework does not uniquely determine $\rho_B(\lambda)$, but
naturally accommodates a broad distribution of low-energy bond and
gauge modes.

\vspace{10pt}
\noindent
\emph{Interpretation.}
The above constructions clarify the physical meaning of the TDOS.
In SYK-type systems, the TDOS emerges from the scaling structure of
operator dynamics in a strongly interacting conformal regime.
In RVB systems, it arises from the fluctuation spectrum of a constrained
manifold of singlet-bond configurations.
In both cases, the TDOS should be viewed as an effective infrared
description of a large set of slow collective processes, rather than
a microscopic input.
The detailed form of $\rho(\lambda)$ remains model-dependent and
requires further microscopic analysis.
Nevertheless, these results demonstrate that a broad relaxation-rate
spectrum, including flat and infrared-singular TDOS, arises naturally
from strongly correlated many-body dynamics.

\section{Appendix G: Possible connections between flat TDOS and scale--free dynamical phenomena}

In this Appendix we briefly discuss a broader dynamical perspective
suggested by the relaxation--spectrum framework developed in the main
text. The central object of the theory is the density of relaxation
rates associated with the Liouvillian generator of the coarse--grained
dynamics.

Near dynamical critical regimes the spectrum of relaxation rates
naturally develops a dense continuum extending toward vanishing decay
rates.  In particular an important limiting case is a finite infrared
density of slow modes,

\begin{equation}
\rho(\lambda\rightarrow 0)=\rho_0 \neq 0 .
\label{eq:flat_tdos_appendix}
\end{equation}

This condition corresponds to a situation in which relaxation processes
exist across arbitrarily long time scales.
The system therefore cannot be characterized by a single relaxation
time, and long--time memory effects emerge naturally.

\vspace{10pt}
\noindent
\emph{(I) Long--time memory kernel.}

The dynamical memory kernel associated with the relaxation spectrum can
be written as
\begin{equation}
K(t)
=
\int_0^{\Lambda} d\lambda \,
\rho(\lambda)\, e^{-\lambda t},
\label{eq:memory_kernel_appendix}
\end{equation}
where $\Lambda$ denotes the ultraviolet cutoff of the relaxation
spectrum.

If the TDOS approaches a constant in the infrared,
$\rho(\lambda)\approx\rho_0$ for $\lambda\ll\Lambda$, one obtains
\begin{equation}
K(t)
\approx
\rho_0
\int_0^{\Lambda} d\lambda\, e^{-\lambda t}
=
\frac{\rho_0}{t}\left(1-e^{-\Lambda t}\right).
\end{equation}
For long times $t\gg \Lambda^{-1}$ this yields the algebraic decay
\begin{equation}
K(t)\sim \frac{\rho_0}{t},
\label{eq:memory_tail_appendix}
\end{equation}
indicating the emergence of long--time memory.

\vspace{10pt}
\noindent
\emph{(II) Relation to scale--free relaxation statistics.}

Distributions of relaxation times with power--law statistics are widely
encountered in complex dynamical systems, including models of
self--organized criticality (SOC).
In many SOC models the lifetime distribution of events takes the
approximate form
\begin{equation}
P(\tau)\sim \tau^{-2}.
\label{eq:scale-free}
\end{equation}

The relation between lifetime $\tau$ and relaxation rate $\lambda$ is
\begin{equation}
\tau=\lambda^{-1}.
\end{equation}
Since the number of modes must be invariant under a change of variables,
the densities satisfy
\begin{equation}
\rho(\lambda)\,d\lambda
=
P(\tau)\,d\tau .
\end{equation}
Using
\begin{equation}
\lambda=\tau^{-1},
\qquad
\left|\frac{d\lambda}{d\tau}\right|=\tau^{-2},
\end{equation}
one obtains
\begin{equation}
P(\tau)
=
\rho(\tau^{-1})\,\tau^{-2}.
\label{eq:Ptau_relation}
\end{equation}

Thus a flat infrared TDOS,
$\rho(\lambda\rightarrow0)=\rho_0$,
implies
\begin{equation}
P(\tau)\sim \rho_0\,\tau^{-2},
\end{equation}
consistent with Eq.~(\ref{eq:scale-free}).
A scale--free lifetime distribution therefore corresponds to a
relaxation spectrum that is approximately constant near $\lambda=0$.

\vspace{10pt}
\noindent
\emph{(III) Implications for low--frequency noise.}

A broad distribution of relaxation times is also known to generate
$1/f$--type noise spectra.
Within the present framework the power spectrum associated with a
superposition of relaxational modes takes the form
\begin{equation}
S(\omega)
\sim
\int_0^\Lambda d\lambda\,
\frac{\rho(\lambda)}
{\lambda^2+\omega^2}.
\end{equation}

Changing variables to the lifetime $\tau=1/\lambda$ and using
Eq.~(\ref{eq:Ptau_relation}) gives
\begin{equation}
S(\omega)
\sim
\int_0^\infty d\tau\,
P(\tau)\,
\frac{\tau^2}{1+(\omega\tau)^2}.
\label{eq:Somega_tau}
\end{equation}
For a scale--free distribution $P(\tau)\sim\tau^{-2}$ one finds
\begin{equation}
S(\omega)
\sim
\int d\tau
\frac{1}{1+(\omega\tau)^2}.
\end{equation}
Introducing the dimensionless variable $x=\omega\tau$ yields
\begin{equation}
S(\omega)
\sim
\frac{1}{\omega}
\int dx\,\frac{1}{1+x^2}.
\end{equation}
Since the remaining integral is finite, the low--frequency scaling is
\begin{equation}
S(\omega)\sim \frac{1}{|\omega|}.
\end{equation}
Thus a flat TDOS naturally produces a $1/f$--type noise spectrum.

\vspace{10pt}
\noindent
\emph{(IV) Relation to the infrared spectral function.}

A flat TDOS produces a characteristic
logarithmic structure in the retarded susceptibility,
\begin{equation}
\chi_R^{-1}(\omega,q)
=
M_q
+
\rho_0 \ln\!\left(\frac{\Lambda}{|\omega|}\right)
+
i\Gamma\,\mathrm{sgn}(\omega),
\end{equation}
where
\begin{equation}
M_q\equiv r+cq^2,
\qquad
\Gamma=\frac{\pi}{2}\rho_0,
\end{equation}

The corresponding spectral function,
\begin{equation}
\mathcal A(\omega,q)
=
\frac{2\Gamma}
{\left[M_q+\rho_0\ln(\Lambda/|\omega|)\right]^2+\Gamma^2},
\end{equation}
describes a broad continuum rather than a sharp quasiparticle peak.

\vspace{10pt}
\noindent
\emph{(V) Consequences for superconducting pairing.}

In the main text we have shown that superconducting pairing is governed
by the \emph{retarded} Cooper-channel kernel entering the Thouless
criterion.
Within the relaxation-spectrum framework the pairing susceptibility is
written as
\begin{equation}
\chi_\Delta^{-1}(\omega)
=
\frac{1}{g}-\Pi^R(\omega),
\end{equation}
where the retarded polarization kernel takes the form
\begin{equation}
\Pi^R(\omega)
=
\int_0^\Lambda d\lambda\,
\frac{\rho(\lambda)}{\lambda-i\omega}.
\end{equation}

For a flat infrared TDOS,
$\rho(\lambda\rightarrow0)=\rho_0$,
the real part of the retarded kernel exhibits logarithmic scaling,
\(
\mathrm{Re}\,\Pi^R(\omega)
\sim
\rho_0 \ln\!\left(\frac{\Lambda}{|\omega|}\right).
\)
Evaluating at the thermal scale $\omega\sim T$ yields the
BCS--like transition condition $1 - g\,\rho_0 \ln\frac{\Lambda}{T_c}=0$,
leading to
\begin{equation}
T_c \sim \Lambda
\exp\!\left(-\frac{1}{g\,\rho_0}\right).
\end{equation}

Thus, in the flat TDOS regime the slow-mode reservoir produces strong
infrared fluctuations and long-time memory, but the superconducting
transition remains marginal, retaining a BCS-like exponential form.
Only when the TDOS becomes infrared-singular,
$\rho(\lambda)\sim\lambda^{\alpha}$ with $\alpha<0$,
does the retarded kernel itself acquire a power-law divergence,
leading to an algebraically enhanced transition scale.

\vspace{10pt}
\noindent
\emph{Summary.}
These results indicate that the infrared organization of the relaxation
spectrum provides a unified dynamical framework for a wide range of
scale--free phenomena observed in correlated systems.

A flat infrared TDOS generates long--time memory,
$K(t)\sim 1/t$, scale--free lifetime distributions,
$P(\tau)\sim\tau^{-2}$, and $1/f$ noise,
reflecting the presence of a dense continuum of slow relaxation modes.

At the same time, the superconducting instability is governed by the
retarded response of this spectrum.
In the flat TDOS regime the Cooper channel exhibits logarithmic
infrared enhancement, leading to a BCS--like exponential transition
scale controlled by the infrared spectral weight.
A qualitatively new regime emerges only when the TDOS becomes
infrared-singular, where the retarded kernel itself becomes
power-law divergent and superconductivity is algebraically enhanced.

Within this perspective, superconductivity and anomalous dynamics arise
from a common origin: the infrared spectral structure of the
Liouvillian relaxation spectrum, rather than from the exchange of a
specific bosonic mediator.

\begin{figure*}[t]
\centering
\includegraphics[width=0.9\textwidth, trim=0.5cm 13.3cm 0cm 0cm]{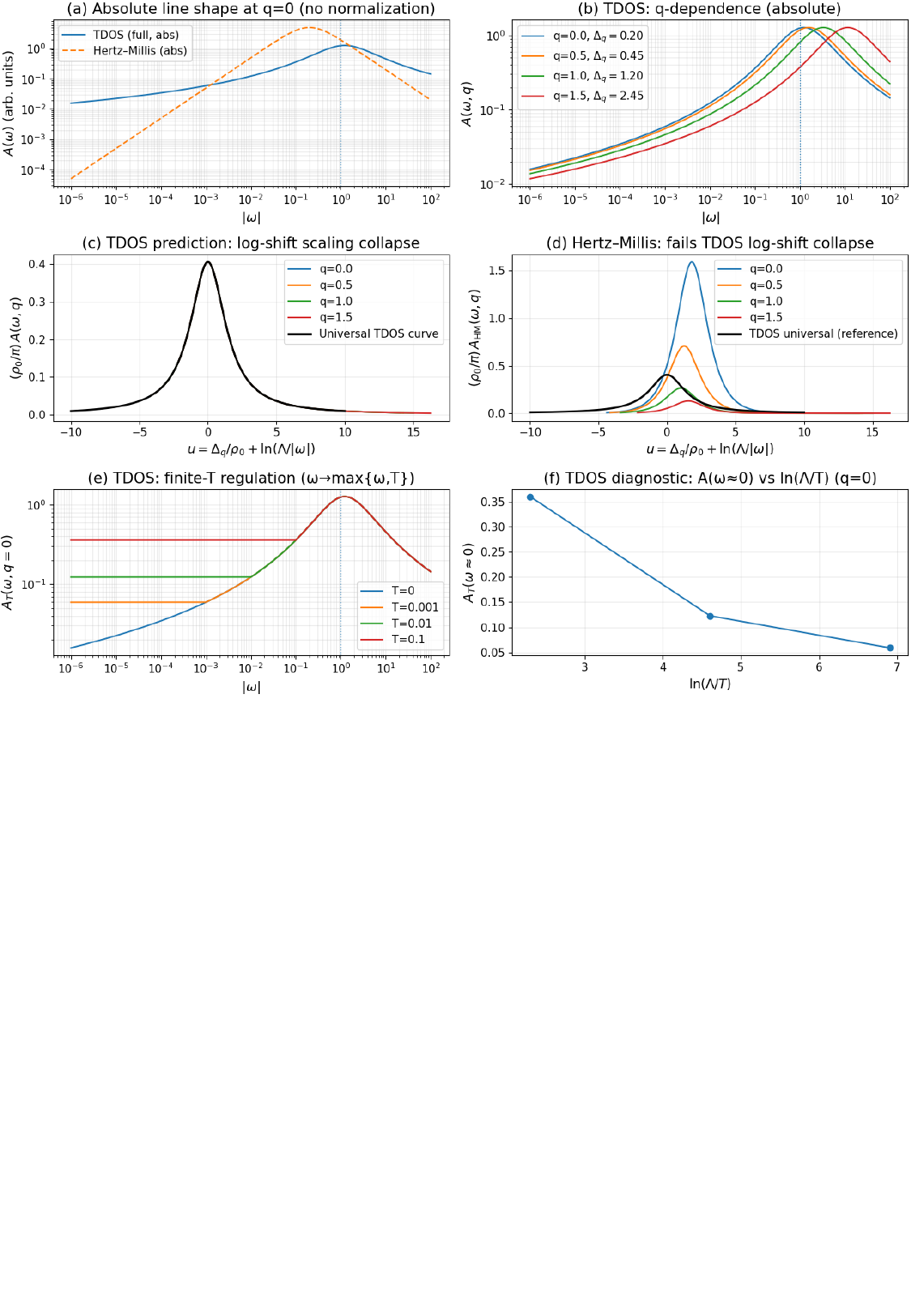}
\caption{Experimental diagnostics distinguishing the memory-dominated TDOS regime from Hertz--Millis (HM) dynamics.
(a) Absolute spectral line shape at $q=0$: the TDOS theory yields a broad continuum, whereas HM produces a sharper peak.
(b) Momentum dependence of the TDOS spectral function: varying $q$ shifts the hump through $M_q=r+cq^2$.
(c) \textit{TDOS prediction (collapse):} when plotted versus the scaling variable
$u=M_q/\rho_0+\ln(\Lambda/|\omega|)$, the TDOS spectra collapse onto a universal curve,
$A(\omega,q)=\frac{\pi}{\rho_0}\,[u^2+(\pi/2)^2]^{-1}$.
(d) \textit{Control test:} HM spectra do not exhibit the same log-shift collapse.
(e) Finite-temperature regulation implemented by $|\omega|\rightarrow \max\{|\omega|,cT\}$ yields infrared saturation.
(f) Corresponding $T$-diagnostic at $q=0$: $A_T(\omega\!\to\!0)$ varies systematically with $\ln(\Lambda/T)$, providing a practical fitting protocol.
All curves are shown in arbitrary units with consistent prefactor conventions.}
\label{fig:tdos_hm_diagnostics}
\end{figure*}


\section{Appendix H: Spectral function generated by memory--dominated self--energy}
\label{app:spectral_function}

In this Appendix we derive the real--frequency spectral function associated
with the collective retarded susceptibility in the memory--dominated critical
regime.  We show that a flat slow--mode TDOS produces a broad non--Lorentzian
continuum with logarithmic infrared scaling and a universal collapse form.

\vspace{10pt}
\noindent
\emph{(I) Spectral function in the memory--dominated regime.}

In the infrared regime controlled by a flat TDOS,
the retarded susceptibility takes the form
\begin{equation}
\chi_R(\omega,\mathbf q)
=
\frac{1}
{M_{\mathbf q}
+
\rho_0\ln(\Lambda/|\omega|)
+
i\Gamma\,\mathrm{sgn}(\omega)} .
\label{eq:chiR_memory}
\end{equation}

The spectral function of collective fluctuations is defined by
$A(\omega,\mathbf q)
\equiv
-2\,\mathrm{Im}\,\chi_R(\omega,\mathbf q)$,
hence
\begin{equation}
A(\omega,\mathbf q)
=
\frac{2\Gamma}
{\left[M_{\mathbf q}
+\rho_0\ln(\Lambda/|\omega|)\right]^2+\Gamma^2}.
\label{eq:spectral_memory_exact}
\end{equation}

\vspace{10pt}
\noindent
\emph{(II) Infrared scaling at criticality.}

Along the critical manifold $M_{\mathbf q}=0$,
\begin{equation}
A(\omega,\mathbf q=0)
=
\frac{2\Gamma}
{\rho_0^2\ln^2(\Lambda/|\omega|)+\Gamma^2}.
\end{equation}
For $|\omega|\ll\Lambda$ the logarithm dominates and the universal infrared tail is
\begin{equation}
A(\omega)\xrightarrow{|\omega|\to0}
\frac{2\Gamma}{\rho_0^2}\,
\frac{1}{\ln^2(\Lambda/|\omega|)} .
\label{eq:infrared_tail}
\end{equation}
Thus the spectral weight is suppressed only logarithmically toward zero
frequency, reflecting the accumulation of memory across many time scales.

\vspace{10pt}
\noindent
\emph{(III) Peak / hump scale away from criticality.}

For finite $M_{\mathbf q}$, Eq.~(\ref{eq:spectral_memory_exact}) is maximized
when
\begin{equation}
M_{\mathbf q}
+
\rho_0\ln\!\left(\frac{\Lambda}{|\omega_*|}\right)=0,
\end{equation}
which yields
\begin{equation}
|\omega_*|
=
\Lambda\exp\!\left(-\frac{M_{\mathbf q}}{\rho_0}\right).
\label{eq:omega_peak}
\end{equation}
For $M_{\mathbf q}>0$ this scale lies above the infrared window
($|\omega_*|\gtrsim\Lambda$) and no low--energy peak develops, whereas for
$M_{\mathbf q}<0$ a pronounced low--frequency hump appears at
$|\omega_*|\ll\Lambda$.

\vspace{10pt}
\noindent
\emph{(IV) Universal scaling collapse.}

Define the scaling variable
\begin{equation}
u(\omega,\mathbf q)
\equiv
\frac{M_{\mathbf q}}{\rho_0}
+
\ln\!\left(\frac{\Lambda}{|\omega|}\right).
\label{eq:scaling_variable}
\end{equation}
Using $\Gamma=(\pi/2)\rho_0$, Eq.~(\ref{eq:spectral_memory_exact}) reduces to the
universal form
\begin{equation}
A(\omega,\mathbf q)
=
\frac{\pi}{\rho_0}\,
\frac{1}{u(\omega,\mathbf q)^2+(\pi/2)^2},
\label{eq:spectral_scaling_form}
\end{equation}
so that all momentum and tuning dependence enters only through $u$, implying a
collapse of line shapes onto a single curve.

\vspace{10pt}
\noindent
\emph{(V) Finite--temperature infrared cutoff.}

At finite temperature the logarithmic growth is regulated by the thermal scale.
A minimal implementation consistent with the Matsubara-frequency
structure is
\begin{equation}
|\omega|\;\rightarrow\;\max\{|\omega|,cT\},
\qquad c=O(1),
\end{equation}
yielding
\begin{equation}
A_T(\omega,\mathbf q)
=
\frac{2\Gamma}
{\left[M_{\mathbf q}
+\rho_0\ln(\Lambda/\max\{|\omega|,cT\})\right]^2+\Gamma^2}.
\label{eq:spectral_finiteT}
\end{equation}
For $|\omega|\ll T$ the logarithm saturates, producing a finite infrared spectral
weight rather than the zero--temperature logarithmic suppression.

\vspace{10pt}
\emph{Physical interpretation.} 
Unlike conventional overdamped criticality, where $\mathrm{Im}\,\Sigma^R(\omega)$
vanishes linearly with $|\omega|$, the memory--dominated regime exhibits
frequency--independent damping ($\propto \mathrm{sgn}(\omega)$) accompanied by a
logarithmically singular reactive part.  The resulting spectral function is
therefore a broad continuum with slowly varying infrared tails rather than a
sharp Lorentzian quasiparticle peak.

\vspace{10pt}
\noindent
\emph{(VI) Experimental diagnostics and scaling tests.}

Figure~\ref{fig:tdos_hm_diagnostics} summarizes a set of practical
diagnostics that distinguish the memory--dominated TDOS regime from
conventional overdamped critical dynamics such as the Hertz--Millis
(HM) theory.

Panel (a) compares the absolute spectral line shapes at $q=0$ predicted
by the two frameworks.
The HM theory produces a relatively narrow Lorentzian-like peak,
reflecting dynamics controlled by a small number of overdamped bosonic
modes.
In contrast, Eq.~(\ref{eq:spectral_memory_exact}) yields a much broader
continuum with slowly varying logarithmic tails,
arising from the collective response of an extensive continuum of
slow relaxation modes.

Panel (b) illustrates the predicted momentum dependence of the TDOS
spectral function.
Because the tuning parameter enters through
$M_{\mathbf q}=r+cq^2$, varying the momentum effectively shifts
the spectral hump according to the scale
$|\omega_*|=\Lambda\exp(M_{\mathbf q}/\rho_0)$ derived in
Eq.~(\ref{eq:omega_peak}).
This produces a characteristic logarithmic displacement of the hump
position as $q$ changes.

Panels (c) and (d) demonstrate the most direct experimental test of the
TDOS framework.
Introducing the scaling variable
\begin{equation}
u(\omega,\mathbf q)
=
\frac{M_{\mathbf q}}{\rho_0}
+
\ln\!\left(\frac{\Lambda}{|\omega|}\right),
\end{equation}
all TDOS spectra collapse onto the universal curve
(\ref{eq:spectral_scaling_form}).
This collapse reflects the fact that the full line shape depends only
on the combined variable $u$.
By contrast, spectra generated by the conventional Hertz--Millis form
do not collapse under the same transformation, providing a clear
diagnostic distinction.

Panel (e) illustrates the finite--temperature regulation discussed in
Eq.~(\ref{eq:spectral_finiteT}).
Replacing $|\omega|$ by $\max\{|\omega|,cT\}$ cuts off the logarithmic
infrared growth and produces saturation of the low--frequency spectral
weight.
The resulting spectra evolve smoothly from the zero--temperature form
to a temperature--dominated plateau.

Finally, panel (f) shows a convenient experimental diagnostic at
$q\simeq0$.
In the regime $|\omega|\ll T$, Eq.~(\ref{eq:spectral_finiteT}) predicts
that the infrared spectral weight varies systematically with
$\ln(\Lambda/T)$.
Measuring $A_T(\omega\!\rightarrow0)$ as a function of temperature
therefore provides a direct way to extract the TDOS parameter
$\rho_0$ and test the predicted logarithmic scaling.

Taken together, these diagnostics provide a practical experimental
protocol for identifying the memory--dominated TDOS universality class.
In particular, the logarithmic peak shift and the universal collapse of
line shapes constitute strong signatures of collective dynamics
governed by an extensive slow--mode reservoir rather than by a single
overdamped critical mode.

\end{document}